\documentclass[a4paper,11pt]{article}
\usepackage{jheppub}
\usepackage{slashed}
\usepackage{bbm,mathtools}
\usepackage{multirow,booktabs}
\usepackage{colortbl}
\usepackage{subcaption}
\usepackage{array,arydshln}
\usepackage[usenames,dvipsnames,svgnames,table]{xcolor}
\usepackage{enumitem}
\usepackage{comment}
\usepackage{tikz}
\usetikzlibrary{patterns}

\definecolor{nicered}{rgb}{0.6,0.1,0.1}
\definecolor{nicegreen}{rgb}{0.1,0.5,0.1}
\definecolor{mediumcandyapplered}{rgb}{0.99, 0.12, 0.07}
\definecolor{red}{rgb}{1.0, 0, 0}
\hypersetup{colorlinks,citecolor= nicered,linkcolor= nicered}

\def\Tr{\mbox{Tr}\,}
\def\tr{\mbox{Tr}\,}

\def\ie{\hbox{\it i.e.}{}}
\def\eg{\hbox{\it e.g.}{}}

\newcommand{\U}{\mathbf{U}}
\newcommand{\V}{\mathbf{V}}
\newcommand{\T}{\mathbf{T}}
\newcommand{\Lag}{\mathcal{L}}
\newcommand{\F}{\mathcal{F}}
\newcommand{\cP}{\mathcal{P}}
\newcommand{\cO}{\mathcal{O}}
\newcommand{\de}{\partial}

\newcommand{\gCoupl}{\kappa}
\newcommand{\xsCount}{\alpha}
\newcommand{\meCount}{N}
\newcommand{\meLabel}{\mathcal{M}}
\newcommand{\xsLamP}{\xsCount_\Lambda^p}
\newcommand{\xsLamG}{\xsCount_\Lambda^\gCoupl}
\newcommand{\xsFpP}{\xsCount_{4\pi}^p}
\newcommand{\xsFpG}{\xsCount_{4\pi}^\gCoupl}
\newcommand{\xsChiP}{\xsCount_\chi^p}
\newcommand{\xsChiG}{\xsCount_\chi^\gCoupl}

\title{
The Art of Counting:\\ a reappraisal of the HEFT expansion
}

\author[a]{Ilaria Brivio,}
\author[b]{Ramona Gr\"{o}ber,}
\author[b]{Konstantin Schmid}

\emailAdd{ilaria.brivio@unibo.it}
\emailAdd{ramona.groeber@pd.infn.it}
\emailAdd{konstantin.schmid@pd.infn.it}

\affiliation[a]{Dipartimento di Fisica e Astronomia, Universit\`a di Bologna, and Istituto Nazionale di Fisica Nucleare, Sezione di Bologna, I-40126 Bologna, Italy}
\affiliation[b]{Dipartimento di Fisica e Astronomia ``G. Galilei", Universit\`a di Padova, and Istituto Nazionale di Fisica Nucleare, Sezione di Padova, I-35131 Padova, Italy}

\abstract{
We revisit the power counting of the Higgs Effective Field Theory (HEFT) from first principles, by requiring that predictions for physical observables follow a series expansion in small, dimensionless quantities. 
Depending on whether HEFT is formulated in terms of a unique low-energy scale $v$ or in terms of two scales $v<f$, this approach identifies two viable power counting rules that can accommodate any operator normalization choice. 
We provide quantitative prescriptions for the consistent truncation of HEFT operators, amplitudes and observable contributions and we illustrate our arguments with a number of examples. 

}

\date{}

\subheader{\hfill\rm COMETA-2025-51}

\begin{document} 

\maketitle

\section{Introduction}
Effective Field Theories (EFTs) have emerged as indispensable tools for interpreting the increasingly precise data collected at the Large Hadron Collider (LHC) and at other high-energy physics experiments. In particular, the so-called Standard Model EFT (SMEFT)~\cite{Buchmuller:1985jz,Grzadkowski:2010es} and Higgs EFT (HEFT)\footnote{HEFT should not be confused with the EFT obtained in the large top quark mass limit, and it is also known as Electroweak Chiral Lagrangian with a light Higgs.}~\cite{Alonso:2012px,Buchalla:2013rka,Brivio:2013pma,Gavela:2014vra,Brivio:2016fzo,Sun:2022ssa,Sun:2022snw} provide model-independent parameterizations of non-resonant beyond-Standard Model (BSM) signals, that could be potentially generated by heavy new particles and that can be searched for in precisely measured observables~\cite{Brivio:2017vri, Isidori:2023pyp}. Both EFTs extend the Standard Model (SM) by introducing classically non-renormalizable interactions.
They differ in the representation of the four scalar fields of the theory: the SMEFT adopts a linear doublet representation of the $SU(2)_L$ gauge symmetry, that contains both the physical Higgs boson and the three would-be Goldstone bosons of electroweak symmetry breaking (EWSB). The HEFT, that descends from technicolor \cite{Susskind:1978ms, Dimopoulos:1979es, Dimopoulos:1981xc} and composite Higgs models theories~\cite{Kaplan:1983sm, Kaplan:1983fs, Georgi:1984af,  Agashe:2004rs, Contino:2006qr}, adopts a non-linear representation for the three Goldstones, which are embedded in a bi-doublet of the global $SU(2)_L\times SU(2)_R$~\cite{Appelquist:1980vg,Longhitano:1980iz,Longhitano:1980tm,Feruglio:1992wf,Appelquist:1993ka,Buchalla:2012qq}, and adds the physical Higgs boson as a pure gauge singlet.

The relationship between SMEFT and HEFT has long been the subject of theoretical and phenomenological studies, aimed at establishing whether the two formalisms are actually independent, whether one of the two is more general than the other, and also whether the two EFTs could be disentangled experimentally, and what such a signature would imply in terms of viable BSM scenarios.
Recent studies have established that the HEFT framework offers a more general description than the SMEFT, as certain BSM scenarios preclude a consistent matching onto the SMEFT due to the presence of singularities at the $H^\dag H=0$ point~\cite{Falkowski:2019tft}, typically arising in theories involving \emph{loryons} ~\cite{Banta:2021dek,Crawford:2024nun}, or due to the existence of BSM sources of EWSB~\cite{Cohen:2020xca}. These conclusions can be achieved employing geometrical methods~\cite{Alonso:2015fsp,Alonso:2016oah,Alonso:2023upf} designed to circumvent the complications arising from the fact that the two EFTs are related by a non-linear field redefinition, and therefore hold independently of how the EFT series are truncated.

HEFT appears to be more general (i.e.~it contains more free parameters) than SMEFT in order-by-order comparisons as well, which has also been shown via on-shell amplitude techniques~\cite{Dong:2022jru,Liu:2023jbq,Goldberg:2024eot, Grober:2025vse}. Most notably, the doublet structure adopted in the SMEFT enforces correlations among interactions involving different powers of $h$, which are only broken by going to higher orders in the SMEFT expansion. By contrast, the singlet representation adopted in HEFT allows for arbitrary Higgs couplings already at leading order (LO), which generally ensures better convergence properties compared to SMEFT~\cite{Cohen:2020xca}. Moreover, interactions involving longitudinally polarized gauge bosons typically appear at lower orders in HEFT compared to SMEFT, and the same is true for operators that violate the custodial symmetry: these breakings are unsuppressed in HEFT, while in SMEFT they require insertions of the Higgs doublet, which increase the operator dimension. This effect leads, for instance, to neutral anomalous quartic gauge couplings appearing at much lower order in HEFT compared to SMEFT. Chirality-flipping fermionic operators also appear at lower orders in HEFT for the same reason. The order-by-order comparison of the two EFTs therefore predicts distinctive HEFT signals in di-Higgs and multi-Higgs production~\cite{Gomez-Ambrosio:2022qsi,Delgado:2023ynh,Davila:2023fkk,Anisha:2024ljc,Anisha:2024ryj, Anisha:2024xxc,Bhardwaj:2024lyr,Englert:2025xrc, Domenech:2025gmn}, Higgs decays~\cite{Isidori:2013cga,Isidori:2013cla,Buchalla:2013mpa}, Higgs and diboson production~\cite{Brivio:2013pma,Gavela:2014vra,Brivio:2016fzo,Eboli:2021unw}, vector boson scattering~\cite{Delgado:2014jda,Eboli:2023mny,Mahmud:2025wye}, as well as lower energy flavor processes~\cite{Alonso:2012jc,Alonso:2012pz,Gavela:2014vra,Fortuna:2024rqp}. It is understood that these signatures can disentangle the two EFTs only if contributions from the respective higher orders are negligible.

SMEFT is by far the most commonly adopted framework for BSM searches. It represents the standard EFT for the interpretation of LHC data, widely employed by both the theory community and the experimental collaborations. Its fast advancement over the last decade was fostered by the development of a number of tools, that almost fully automate the matching-running-simulation chain and allow the combination of large numbers of measurements in statistical global analyses, see \eg~\cite{Aebischer:2023nnv} for an overview.
HEFT has been adopted less frequently in the literature (see~\cite{Corbett:2015mqf,Brivio:2016fzo,Eboli:2021unw} for global fits), but interest in this EFT has been growing again recently, mainly as a framework for exploring BSM scenarios whose signatures are not satisfactorily captured by the dimension-six SMEFT. This happens in the presence of new physics that cannot be matched onto SMEFT, as mentioned above, or whenever the SMEFT expansion, while possible, is poorly convergent, as is often the case in the matching to composite Higgs models. In this approach, HEFT can be employed as a framework to test the validity of the SMEFT expansion, similarly to the dimension-8 SMEFT.

One of the main challenges in adopting HEFT descends from its power counting. 
In this work, we will use the term ``power counting" to denote the set of rules that sort in a coherent way (i) EFT operators, (ii) Feynman diagrams and (iii) observables (\eg\ cross section contributions) by expected importance, and that allow to implement a valid EFT truncation at any of these orders.
While for SMEFT the power counting boils down to expanding in powers of the cutoff scale $\Lambda$ at each step, the HEFT case is less obvious and generates recurring confusion in the literature.
The origin of this complication lies in the fact that HEFT merges the formalism of chiral Lagrangians for Goldstone bosons, that are known to follow a counting ``in derivatives"~\cite{Weinberg:1978kz,Gasser:1983yg}, 
with gauge and fermions fields that transform linearly under the symmetries and with a scalar singlet that -- in general -- does not have a Goldstone nature. Moreover, all the masses in the theory originate from the spontaneous breaking of the electroweak (EW) group, and correspondingly share the same scale parameter $v$. As will be shown later, an important factor in the definition of the power counting is that, while in SMEFT $v$ emerges as the vacuum expectation value (VEV) of the Higgs field, in HEFT it is introduced as a free parameter. This difference is directly related to a well-known, fundamental distinction between the two theories, namely that
while SMEFT is naturally written as an expansion around the EW-preserving field configuration $\langle H\rangle= 0$ such that the EW-symmetry breaking (EWSB) mechanism is explicitly modeled in the Lagrangian, HEFT can only be written directly in the broken electroweak phase~\cite{Alonso:2016oah,Cohen:2020xca}. 

The issue of power counting within the HEFT framework has been previously studied in the literature, see Refs.~\cite{Buchalla:2012qq,Buchalla:2013eza,Gavela:2016bzc,Buchalla:2016sop,Brivio:2016fzo}. However, previous works mostly discussed the classification of HEFT operators, proposing various criteria to generalize the power counting of chiral perturbation theory ($\chi$PT)~\cite{Weinberg:1978kz,Gasser:1983yg,Manohar:1983md}. To our knowledge, a consistent set of rules taking the user from the Lagrangian to the cross section has never been fully spelled out. In particular, the interplay between the organization of the operator series and the expansion of observable predictions in powers of some small parameter, that ultimately retains the physical meaning of an EFT, has not been examined in detail.  Clarifying these aspects is clearly indispensable in order to build a consistent program of HEFT searches comparable to the SMEFT one, and it is a necessary step towards the potential automation of simulation tools. Moreover, the power counting of the EFT is closely related to the algorithms for BSM-HEFT matching, that also present some ambiguities and are the subject of ongoing research~\cite{Dawson:2023oce,Dawson:2023ebe,Buchalla:2023hqk,Song:2024kos,Alonso:2025ksv}.

In this work, we revisit the power counting of HEFT by requiring, as a guiding principle, that the classification of EFT operators and Feynman diagrams mirrors the expansion of \emph{observable} predictions in powers of small, dimensionless physical quantities, including $(p/\Lambda), (g/4\pi)$, etc. In doing so, we will heavily rely on existing results, particularly on Naive Dimensional Analysis (NDA)~\cite{Manohar:1983md,Cohen:1997rt,Luty:1997fk,Gavela:2016bzc} and on well-known diagrammatic relations (see App.~\ref{app.diagrammatics}). Although our reasoning shares several arguments with previous analyses, we find that it yields a clearer picture of the HEFT power counting: it removes ambiguities in the operator classification and it provides a simple set of rules for the coherent organization of the Lagrangian, diagrams and observable series. More specifically, we find that two alternative power counting rationales apply, depending on whether the EW scale $v$ is the only mass-dimensionful scale present in the theory, or an independent quantity $f$ is introduced to suppress scalar insertions. Interestingly, the sets of rules we identify present some differences compared to previously proposed rationales.

This work is structured in the following way. We begin, in Section~\ref{sec.formalism}, with a definition of the SMEFT and HEFT frameworks and a brief review of the two EFTs. Section~\ref{sec.pc} presents a self-contained discussion of power counting in general EFTs, reviewing in particular how insertions of small expansion parameters propagate from the Lagrangian to amplitudes and observables. 
Section~\ref{sec.SMEFT_PC} briefly covers the power counting of SMEFT, specializing the discussion of Section~\ref{sec.pc} to this well-known case. The main results of the paper are presented in Section~\ref{sec.HEFT_PC}, where the power counting of HEFT is studied in detail, and the proposed set of expansion rules is derived. 
Section~\ref{sec.examples} contains a number of examples showcasing the rationale behind the proposed power counting and its application in concrete processes. Finally, in Section~\ref{sec.con} we conclude. 
A self-contained summary of the power counting rules can be found in App.~\ref{app.nutshell}.

\section{EFT extensions of the SM}\label{sec.formalism}

\subsection{Standard Model Effective Field Theory}

The SMEFT is built out of the same fields and symmetries as the SM. In particular, the four scalar degrees of freedom are embedded in a complex $SU(2)_L$ doublet
\begin{align}
H  =\frac{1}{\sqrt{2}}\binom{\phi_2+i\phi_1}{\phi_0-i\phi_3} \,,  
\end{align}
such that, upon EWSB, $\phi_0 = v+h$, with $v=\langle\phi_0\rangle\simeq 246~{\rm GeV}$ the vacuum expectation value of the field and $h$ the physical Higgs excitation.
The fields  $\phi_I$, $I=1,2,3$ are instead the Goldstone bosons (GBs) of the spontaneous symmetry breaking, that are reabsorbed into the longitudinal components of the $W^\pm,Z$ bosons.

The SM Lagrangian is written as\footnote{We omit the QCD theta term as its role is beyond the scope of this work. We also neglect the presence of neutrino masses.}
\begin{align}\label{eq.LagSM_SMEFT}
\Lag_{\mathrm{SM}} =&\, -\frac{1}{4}B_{\mu\nu}B^{\mu\nu} - \frac{1}{4}W_{\mu\nu}^I W^{I\mu\nu} - \frac{1}{4}G_{\mu\nu}^A G^{A\mu\nu}+D_\mu H^\dag D^\mu H -V(H)
\nonumber\\
&+\sum_{\psi} \bar\psi i\slashed{D}\psi-\left[\bar Q_L Y_d H d_R+ \bar Q_L Y_u \tilde H u_R + \bar L_L Y_e H e_R+\text{h.c.}\right]\,,\\
V(H) &= -\frac{m_h^2}{2} H^\dag H +\lambda (H^\dag H)^2\,,
\end{align}
where $B_\mu, W^I_\mu, G^A_\mu$ are the gauge fields of the $U(1)_Y\times SU(2)_L\times SU(3)_c$ gauge groups. The respective coupling constants will be denoted by $g',g,g_s$. The index $\psi$ runs over the 5 fermion fields of the SM: 
\begin{align}
      Q_L &= (u_L, d_L)\,,
      &
      L_L &= (\nu_L, e_L)\,,
      &
      u_R\,,& 
      &
      d_R\,,& 
      & 
      e_R\,.
\end{align}
Each of them carries flavor indices which are implicitly contracted in Eq.~\eqref{eq.LagSM_SMEFT}, where the Yukawa couplings $Y_u,Y_d,Y_e$ are $3\times3$ complex matrices in flavor space.
We have also introduced the dual Higgs doublet
\begin{align}
\tilde H = \varepsilon H^* =\frac{1}{\sqrt{2}}\binom{\phi_0+i\phi_3}{i\phi_1 - \phi_2} \,,
\end{align}
which has opposite hypercharge with respect to $H$.
The scalar potential $V(H)$ enjoys an exact $O(4)$ global symmetry, which is due to
\begin{equation}
   H^\dag H =\frac12\left( \phi_0^2 + \phi_1^2+ \phi_2^2 + \phi_3^2\right)\,,
\end{equation}
being invariant under rotations of the four real scalars. This symmetry is locally isomorphic to a global $SU(2)_L\times SU(2)_R$:
one defines the bidoublet field 
\begin{equation}\label{eq.Sigma_definition}
\Sigma = (\tilde H \; H)
\end{equation}
such that  $\Tr(\Sigma^\dag \Sigma) = 2H^\dag H$. Since $\Sigma$ transforms as $\Sigma \mapsto \Omega_L \Sigma \Omega_R^\dag$ under generic $SU(2)_{L,R}$ transformations  $\Omega_{L,R}$, it is easy to see that $\Tr(\Sigma^\dag \Sigma)$ is manifestly invariant.
Upon EWSB, the global symmetry is spontaneously broken down to the so-called \emph{custodial symmetry}, i.e. the diagonal subgroup $SU(2)_V$ (or the $O(3)$ rotating only the GB fields).
The gauging of the hypercharge ($g'\neq 0$) is a source of explicit breaking of the global symmetry, which can be seen identifying $U(1)_Y$ with the group generated by the third generator of $SU(2)_R$.\footnote{This group actually corresponds to a linear combination of $U(1)_{Y}$ and $U(1)_{B-L}$~\cite{Graf:2022rco}.\label{note.T3}}

\subsubsection*{SMEFT power counting}
In SMEFT, BSM effects are parameterized by an infinite tower of effective operators that are sorted by their canonical dimension $d$:
\begin{align}
\Lag_{\mathrm{SMEFT}} &= \Lag_{\mathrm{SM}} 
+ \frac{C^{(5)}}{\Lambda} \mathcal{O}^{(5)} 
+ \sum_i \frac{C_i^{(6)}}{\Lambda^2} \mathcal{O}_i^{(6)} 
+ \sum_i \frac{C_i^{(7)}}{\Lambda^3} \mathcal{O}_i^{(7)} 
+ \sum_i \frac{C_i^{(8)}}{\Lambda^4} \mathcal{O}_i^{(8)} + \dots
\label{eq:SMEFTL},
\end{align}
where the sums run over complete and non-redundant bases of operators, indicated as $\mathcal{O}_i^{(d)}$, that contain only SM fields and respect the SM gauge symmetries.
The quantities $C_i^{(d)}$ are the Wilson coefficients, and $\Lambda$ is a constant with the dimensions of a mass, that is introduced in order to work with dimensionless $C_i^{(d)}$.
The dots stand for terms of higher canonical dimension. 
Complete and non-redundant bases of SMEFT operators have been constructed up to dimension 12~\cite{Weinberg:1979sa,Grzadkowski:2010es,Lehman:2014jma,Henning:2015alf,Li:2020gnx,Murphy:2020rsh,Li:2020xlh,Liao:2020jmn,Harlander:2023psl}, and the number of independent parameters at each order is known through Hilbert series calculations~\cite{Henning:2015alf}. Dimension-5 contains only one term, yielding Majorana masses for the three neutrinos, and all operators of odd dimension violate the conservation of the baryon and/or the lepton number~\cite{Kobach:2016ami}. For this reason, phenomenological studies focusing on $B$ and $L$-conserving phenomena only consider operators of even dimension. 

The power counting rules for SMEFT are well-known, and ultimately amount to an order-by-order expansion in inverse powers of $\Lambda$ at every stage of the calculation: the leading ($B,L$ conserving) corrections to SM predictions are given by terms of order $\mathcal{O}(\Lambda^{-2})$, the subleading ones by terms of $\mathcal{O}(\Lambda^{-4})$ and so on.
The perturbative expansion is orthogonal to the EFT expansion in SMEFT, in the sense that, at each order in $\Lambda$, observable predictions can be truncated at any desired order in the perturbative parameters of the SM ($g/4\pi$ etc.) without stumbling upon formal inconsistencies. 
As will be reviewed in the next sections, this seemingly natural fact is actually a direct consequence of the power counting in canonical dimensions, which also ensures that the SMEFT is renormalizable order-by-order in its expansion. 

\subsection{Higgs Effective Field Theory}\label{sec.HEFT}
The HEFT (also known as Electroweak Chiral Lagrangian with a light Higgs) relaxes the assumption that the physical Higgs boson $h$ is part of a $SU(2)_L$ doublet and rather treats it independently of the three Goldstone bosons. The latter are incorporated in the dimensionless $2\times2$ complex field
\begin{align}
    \U = \exp \left( \frac{i\,\pi_I \sigma^I}{v}\right)\,,
\end{align}
where $\pi^I$ are the GBs in a nonlinear representation, and $\sigma^I$ are the Pauli matrices.
$\U$~transforms as a bidoublet under the global $SU(2)_L \times SU(2)_R$ symmetry. Its covariant derivative can be written as
\begin{align}
    D_{\mu} \U = \partial_{\mu} \U + \frac{i g}{2} W_{\mu}^I\, \sigma^I \U - \frac{i g'}{2} B_{\mu}\, \U \,\sigma^3\,,
\end{align}
and it is convenient to define the objects
\begin{align}
    \V_{\mu} &= (D_{\mu} \U) \U^{\dagger}\,,
    \label{eq.def_VT}
    &
    \T &= \U \sigma^3\U^\dagger 
    \,,
\end{align}
which both transform in the adjoint representation of $SU(2)_L$: $\V_{\mu} \rightarrow \Omega_L \V_{\mu} \Omega_L^{\dagger}$ and analogously for $\T$. While $\V_\mu$ is a singlet under $SU(2)_R$, $\T$ is only invariant under $U(1)_Y\subset SU(2)_R$. In fact, it constitutes a \emph{spurion} that injects an explicit breaking of the custodial symmetry.\footnote{ Due to the caveat pointed out in footnote~\ref{note.T3}, the use of $\T$ as a building block is compatible with the preservation of QED only for operators that conserve $B-L$~\cite{Graf:2022rco}. In this work we will restrict to the sector respecting this condition.}

The physical Higgs boson $h$ is introduced as a gauge singlet, which therefore can be arbitrarily coupled to any operator. It is customary to package the infinite series of allowed $h$ insertions into dimensionless functionals~\cite{Grinstein:2007iv}
\begin{align}
\label{eq.general_flare_function}
    \F_i(h) &= \sum_{n=0}^{\infty} a_{i,n} \left(\frac{h}{v}\right)^n\,,
    &
    a_{i,n}&\in \mathbb{C}\,,
\end{align}
that can be treated as building blocks for the Lagrangian. The coefficients $a_{i,n}$ are complex in general, but, if $\F_i$ is inserted in a hermitian operator, only their real parts are physical.

Finally, because the scalar fields representation makes the global $SU(2)_L\times SU(2)_R$ manifest, it is convenient to adopt a consistent notation for the fermions, introducing right-handed doublets:
\begin{align}
    Q_R &= (u_R, d_R)\,, & L_R &= (0, e_R)\,.
\end{align}
Using this formalism, the SM Lagrangian can be written as~\cite{Feruglio:1992wf} 
\begin{align}
    \Lag_{\mathrm{SM}} = 
    &- \frac{1}{4} G_{\mu \nu}^a G^{a \mu \nu} - \frac{1}{4} W^I_{\mu \nu} W^{I \mu \nu} - \frac{1}{4} B_{\mu \nu} B^{\mu \nu} 
    \nonumber\\
    &
    + \frac{1}{2} \partial_{\mu} h \partial^{\mu} h - \frac{(v+h)^2}{4} \mathrm{Tr}\left(\V_{\mu} \V^{\mu}\right)  - V(h) 
    \nonumber\\
    &+ i \overline{Q}_L \slashed{D} Q_L + i \overline{Q}_R \slashed{D} Q_R + i \overline{L}_L \slashed{D} L_L + i \overline{L}_R \slashed{D} L_R 
    \nonumber\\
    &- \frac{v+h}{\sqrt{2}} \left(\overline{Q}_L \U Y_Q Q_R + \mathrm{h.c.} \right) - \frac{v+h}{\sqrt{2}} \left(\overline{L}_L \U Y_L L_R + \mathrm{h.c.} \right),
\label{eq.lagrangian_HEFT_SM}
\end{align}
with
\begin{align}
V(h) &= \lambda v^4\left[\frac{h^2}{v^2} + \frac{h^3}{v^3}+\frac{1}{4} \frac{h^4}{v^4}\right]\,,
&
Y_Q &= \begin{pmatrix}Y_u&\\&Y_d\end{pmatrix}\,,
&
Y_L&= \begin{pmatrix}0&\\&Y_e\end{pmatrix}\,.
\end{align}
A mapping from SMEFT to HEFT can always be identified by replacing\footnote{Note that, although we use $h$ to denote the Higgs excitation in both cases, the SMEFT fields ($\phi_i,h$) and the HEFT ones ($\pi_i,h$) are related to each other by a nonlinear field redefinition~\cite{Alonso:2016oah}.} 
\begin{align}
H &\to \frac{v+h}{\sqrt2} \U \binom{0}{1}\,,
&
\tilde H &\to \frac{v+h}{\sqrt2} \U \binom{1}{0}\,,
\label{eq.SMEFT_to_HEFT}
\end{align}
while the inverse
\begin{align}
\label{eq.HEFT_to_SMEFT}
\U&\to \frac{(\tilde H \; H)}{\sqrt{H^\dag H}}\,,
&
h&\to \sqrt{2H^\dag H}-v\,,
\end{align}
is physical \emph{iff} HEFT admits an $O(4)$ fixed point at $h=-v$ (or equivalently at $H^\dag H= 0$) and it is possible to analytically expand the theory around it~\cite{Alonso:2015fsp,Alonso:2016oah,Falkowski:2019tft,Cohen:2020xca}.

\subsubsection*{HEFT power counting: state of the art}
In the previous paragraphs we have introduced the HEFT formalism, but we purposely avoided giving an expression for $\Lag_{\rm HEFT}$. Indeed, 
Eq.~\eqref{eq.lagrangian_HEFT_SM} provides the SM Lagrangian in HEFT notation, which does not necessarily coincide with the LO of the HEFT expansion, that we discuss in the following.

When examining the HEFT power counting, one typically starts from the pure scalar sector, which has two characterizing features: first of all,
the building blocks $\U$ and $\F(h)$ are dimensionless and contain infinite series of scalar field insertions. 
In the case of $\U$, all terms in the series are needed in order to preserve its bi-doublet transformation property, implying that a gauge-invariant power counting must weigh all series terms equally. This observation clearly rules out a naive use of canonical dimensions. The second important feature of the scalar sector of HEFT is that the EW scale $v$ does not  appear dynamically (\ie\ as a VEV) but it is introduced as an explicit suppression scale in the expressions for $\U$ and $\F(h)$.

The close resemblance between the pure scalar sector of HEFT and chiral perturbation theory suggests that it should naturally follow a counting “in derivatives”~\cite{Weinberg:1978kz,Gasser:1983yg}, such that one would write \eg
\begin{align}
\Lag_{0\de}&=
-\lambda v^4 \mathcal{V}(h)\\
\Lag_{2\de}&=    
\frac{1}{2} \de_\mu h\de^\mu h + \frac{v^2}{4}\tr[\V_\mu\V^\mu]\F_C(h) + C_T\, \frac{v^2}{4}\tr[\T\V_\mu]^2\F_T(h)
\label{eq.scalar_L2}
\\
\Lag_{4\de}&=    
C_1 \,\Tr[\V_\mu\V^\mu]^2\F_1(h) 
+ C_2 \,\Tr[\V_\mu\V_\nu]^2\F_2(h) + \dots
\label{eq.scalar_L4}
\\
\Lag_{6\de}&=
\frac{C_3}{\Lambda^2}\, \tr[\V_\mu\V^\mu]^3 \F_3(h) + 
\frac{C_4}{\Lambda^2}\, \tr[\V_\mu\V_\nu]^2 \tr[\V_\rho\V^\rho]\F_4(h)
+\dots
\end{align}
where the dots stand for other operators with 4 or 6 derivatives and the two functionals $\mathcal{V}(h),\F_C(h)$ take the form
\begin{align}
\mathcal{V}(h) &= \frac{h^2}{v^2} + a_{V,3} \frac{h^3}{v^3} + a_{V,4}  \frac{1}{4}\frac{h^4}{v^4} + \sum_{n=5}^{\infty} a_{V,n} \left(\frac{h}{v}\right)^n, 
\label{eq.Vh}
\\
 \F_C(h) &= 1 + \sum_{n=1}^{\infty} a_{C,n} \left(\frac{h}{v}\right)^n, 
 \label{eq.FCh}
\end{align}
to assign a canonical kinetic term to the GB fields, preserve  $m_h^2=2\lambda v^2$ and avoid a $h$ tadpole. The SM limit is recovered for $a_{V,3}=a_{V,4}=a_{C,2}=1$, $a_{C,1}=2$ and $a_{V,(n\geq 5)}=a_{C,(n\geq 3)}=0$.

The classification of the interaction terms in derivatives is clearly compatible with the gauge invariance requirement. Moreover, it is intuitively associated to an expansion  of the scattering amplitudes in $(p/\Lambda)$, where $p$ represents generic momenta of the external state particles.\footnote{The arguments of Section~\ref{sec.pc} will clarify why the expansion is in $(p/\Lambda)$ and not $(p/v)$. Moreover, when present, mass insertions are treated as $m^2\sim p^2$ to ensure homogeneous equations of motion and propagators.}
More precisely, as we will see in Section~\ref{sec.pc}, a $L$-loop diagram with $S$ external scalar legs, whose $i$-th vertex is extracted from an operator with $q_i$ derivatives, scales with $(p/\Lambda)^{\alpha}$, being\footnote{This discussion is still restricted to the pure scalar case. The inclusion of fermions requires a generalization of the expressions, that can be found in Section~\ref{sec.pc}.} 
\begin{equation}
\alpha = 2L + S-2 + \sum_i (q_i-2)\,,
\label{eq.chiPT_count}
\end{equation}
where the sum is over all vertices. Note that this quantity differs from the usual order $\mathcal{D}$ defined in~\cite{Weinberg:1978kz} (see also \eg~\cite{Pich:1995bw,Buchalla:2013eza}) as
\begin{align}
\label{eq.D_vs_alpha}
\alpha = \mathcal{D} +S-4\,.
\end{align}
As will be shown later, the last two terms avoid counting powers of $p$ that are associated to the dimensions of the external legs, and that have no net physical impact at observables level, due to a cancellation against phase space factors~\cite{Gavela:2016bzc}. A well-known consequence of the derivative expansion is that one-loop diagrams containing only operators with two derivatives are renormalized by counterterms belonging to the $4\de$ class, and similarly for higher orders. In fact, considering that the counterterm must give a tree-level contact diagram with the same $\alpha$ and $S$ as the loop, Eq.~\eqref{eq.chiPT_count} gives 
\begin{equation}
q_{\rm CT} = 2L +2 + \sum_i (q_i-2) = \mathcal{D}\,.
\end{equation}
This well-known result expresses the common lore that adding a perturbative order requires adding a derivative order and vice versa. Thus, the perturbative and EFT expansions are tied together, differently from what occurs with a power counting in canonical dimensions.

Let us now address the full HEFT, where the chiral scalar sector is coupled to gauge and fermion fields. The simultaneous presence of these sectors generates ambiguities in the definition of a unified power counting. To give some very naive examples: a gauge field strength $X_{\mu\nu}\sim [D_\mu,D_\nu]$ could be considered as a two-derivative object. However, this would shift the gauge kinetic terms to $\Lag_{4\de}$. Similarly, operators of class $X^3$, \ie\ formed by three field strengths, could either belong to $\Lag_{6\de}$, or to $\Lag_{4\de}$, if we consider the kinetic terms as LO and note that $X^3$ have two derivatives more than those. Further, if $\U$ insertions are unsuppressed, operators such as $B_{\mu\nu}\tr[\T W^{\mu\nu}]$ (contributing to the $S$ parameter) or $\tr[\T W_{\mu\nu}]^2$ (the $U$ parameter) are of the same order as the gauge kinetic terms, which can seem incongruous. The presence of fermions leads to similar conundrums: for instance, as discussed further below, four-fermion operators can be classified as either $2\de$ or $4\de$ depending on the rules adopted.

The issue of identifying a unified power counting for the HEFT Lagrangian has been addressed in the literature drawing on a variety of logical arguments.  
Building upon the conventional coupling of $\chi$PT to QED~\cite{Urech:1994hd,Knecht:1999ag}, Refs.~\cite{Buchalla:2012qq,Buchalla:2013eza,Buchalla:2016sop} define HEFT as an expansion in $(1/16\pi^2)$, which is equated to $(v^2/\Lambda^2)$ or $(f^2/\Lambda^2)$ factors arising from NDA. Within this approach, the LO of the HEFT Lagrangian cannot be fixed unambiguously by a counting rule, but must be selected by the EFT practitioner. The choice is subject to the constraint that all LO terms must have the same "chiral dimension" $d_\chi=2$, which is a quantity that generalizes the number of derivatives. Once the LO is fixed, the NLO Lagrangian is determined by selecting the operator classes that contain the counterterms required for the 1-loop renormalization of the LO. The definition of $d_\chi$ is such that all NLO operators obtained in this way have $d_\chi=4$, which maintains a similarity with the derivative expansion.
Ref.~\cite{Gavela:2016bzc}, on the other hand, proposes to sort HEFT operators by their "primary dimension" $d_p$, which is defined as the \emph{lowest} canonical dimension among the terms resulting from the series expansion of $\U,\,\F(h),\,\V_\mu$ etc.
This choice is motivated by considerations on the scaling of observable predictions with powers of $(p/\Lambda)$, and it seeks an EFT organization that resembles more closely the SMEFT expansion, while remaining compatible with the gauge invariance constraints. 

None of these rationales appears to be fully satisfactory: for instance, the classification based on chiral dimensions assigns
the operator classes $\psi^4$, $X^2\U$ and the custodial-breaking operator $\tr[\T\V_\mu]^2$ to the LO Lagrangian ($d_\chi=2$), which was deemed unrealistic in Refs.~\cite{Buchalla:2013eza,Buchalla:2016sop}. The authors argued that the chiral counting should be supplemented with further considerations, that motivate pushing those operators to the NLO sector. Analogous arguments were adopted in~\cite{,Brivio:2013pma,Brivio:2016fzo} as well. For instance, the 2-derivative custodial violating operator can be assumed to be loop-suppressed\footnote{By ``loop-suppressed" here we mean that, if its coefficient was set to 0, the operator would be generated by loop corrections within the EFT, pretty much in the same way as in the SM. This is distinct from assuming that the operator is generated at 1-loop in the matching to BSM models.}  based on experimental observations. Four-fermion operators and  operators containing gauge field strength are understood to carry a suppression due to these fields being expected to couple weakly to the UV sector.

Another unappealing feature of the chiral counting is that it assigns a finite chiral dimension to coupling constants, in particular to the gauge ($g,g',g_s$) and Yukawa couplings ($y_\psi$)~\cite{Buchalla:2013eza,Buchalla:2016sop}. This feature is analogous to the treatment of the electromagnetic coupling~$e$ in $\chi$PT~\cite{Urech:1994hd}, and it is understood as a convenient formal prescription, while a fundamental physics principle supporting it is not identified. Nevertheless, the dimensionality of the coupling constant plays a relevant role in the classification of HEFT operators by the chiral rationale, as $g,g'\dots$ factors increase the chiral dimension of HEFT operators. On one hand, this can be seen as introducing ambiguities in the counting. On the other, coupled with UV assumptions, it motivates the order increase for certain operators. 
For instance, if gauge fields are weakly coupled in the UV, then $B_{\mu\nu}$ insertions \emph{must} be accompanied by $g'$, and analogously for the other gauge groups. Similar effects can emerge \eg\ from arguments based on the assumption of Minimal Flavor Violation (MFV)~\cite{Chivukula:1987py,Hall:1990ac,DAmbrosio:2002vsn}, that requires the insertion of Yukawa couplings in some fermionic currents.

At the same time, the classification based on primary dimensions is independent of the number of loops and coupling insertions, but it leads to an organization of the EFT Lagrangian that does not reproduce the counting in derivatives for the pure scalar sector~\cite{Gavela:2016bzc}. Moreover, as $d_p$ only provides the \emph{minimum} order at which an operator can contribute, it cannot be used to determine unambiguously the suppression order of a Feynman diagram. The latter needs to be computed on a case by case basis.
Finally, a somewhat unappealing feature is that, while it justifies suppressing four-fermion operators, the primary dimension does not treat $X^2\U$ operators any differently from the gauge kinetic terms. 

In fact, HEFT operator classes are generally assigned to different orders depending on the selected counting rationales. Further examples are dipole operators (class $\psi^2X$), that have lower $d_p=5$ compared to four-fermion operators ($d_p=6$), but are considered NNLO in Ref.~\cite{Buchalla:2013rka}, as the one-loop diagrams generating them from LO couplings are finite. Classes $\psi^2 D\U$ and $\psi^2 D^2\U$ are both considered NLO in~\cite{Buchalla:2013eza,Buchalla:2013rka}, but have different primary dimensions (5 and 7 respectively), and so on.

In the absence of an unambiguous criterion to sort HEFT operators, the "standard" HEFT Lagrangian is usually defined as
\begin{equation}
  \Lag_{\rm HEFT} = \Lag_{\rm LO} + \Lag_{\rm NLO} + \Lag_{\rm NNLO}+\dots \,. 
\end{equation}
The LO term is conventionally taken to be~\cite{Contino:2010mh,Azatov:2012bz}\footnote{The insertion of $\F(h)$ in $\partial_{\mu} h \partial^{\mu}h$ is redundant and can be removed by field redefinitions. The same holds for the fermionic kinetic terms.} 
\begin{align}
    \Lag_{\rm LO} = &
    - \frac{1}{4} G_{\mu \nu}^a G^{a \mu \nu} - \frac{1}{4} W^I_{\mu \nu} W^{I \mu \nu} - \frac{1}{4} B_{\mu \nu} B^{\mu \nu} 
    \nonumber\\
    &
    + \frac{1}{2} \partial_{\mu} h \partial^{\mu} h - \frac{v^2}{4} \mathrm{Tr}\left(\V_{\mu} \V^{\mu}\right) \mathcal{F}_C(h) - \lambda v^4\mathcal{V}(h) 
    \nonumber\\
    &+ i \overline{Q}_L \slashed{D} Q_L + i \overline{Q}_R \slashed{D} Q_R + i \overline{L}_L \slashed{D} L_L + i \overline{L}_R \slashed{D} L_R 
    \nonumber\\
    &- \frac{v}{\sqrt{2}} \left(\overline{Q}_L \U \mathcal{Y}_Q(h) Q_R + \mathrm{h.c.} \right) - \frac{v}{\sqrt{2}} \left(\overline{L}_L \U \mathcal{Y}_L(h) L_R + \mathrm{h.c.} \right)\,,
\label{eq.lagrangian_HEFT_LO}
\end{align}
where the functionals $\mathcal{V}(h), \F_C(h)$ take the form in Eqs.~\eqref{eq.Vh},~\eqref{eq.FCh} and similarly, in the Yukawa terms:
\begin{align}
    \mathcal{Y}_Q(h) &= {\rm diag}\left(\mathcal{Y}_U(h),\, \mathcal{Y}_{D}(h)\right)\,,
    \qquad
    \mathcal{Y}_L(h) = {\rm diag}\left(0,\, \mathcal{Y}_{E}(h)\right)\,,
    \\
    \mathcal{Y}_{U}(h) &= Y_{u}\left(\mathbbm{1}+\sum_{n=1}^\infty a_{u,n}\left(\frac{h}{v}\right)^n\right)\,.
    \label{eq.YU_expansion}
\end{align}
The expansions of $\mathcal{Y}_D$ and $\mathcal{Y}_E$ are analogous to Eq.~\eqref{eq.YU_expansion}, with $Y_d,Y_e$ Yukawa matrices and $a_{d,n},a_{e,n}$ coefficients respectively.
In general, the quantities $a_{\psi,n}$ ($\psi=u,d,e$) are $3\times 3$ complex matrices in flavor space. The SM limit is recovered for $
 a_{\psi,1} =\mathbbm{1},\, a_{\psi,(n\geq 2)} =0$.
This choice of $\Lag_{\rm LO}$ evidently aims at having the LO Lagrangian as close as possible to the SM. It is easy to check that Eq.~\eqref{eq.lagrangian_HEFT_LO} generalizes Eq.~\eqref{eq.lagrangian_HEFT_SM} simply by allowing for arbitrary $h$ interactions.

The custodial breaking operator
\begin{align}
\label{eq.PT_def}
 \mathcal{P}_T &= \frac{v^2}{4}\Tr[\T\V_\mu]^2\F_T(h)\,,
\end{align}
where $\F_T(h)$ is parameterized as in~\eqref{eq.general_flare_function},
can be assigned to $\Lag_{\rm LO}$ (by virtue of its 2 derivatives) or to $\Lag_{\rm NLO}$, as discussed above. 

The NLO Lagrangian $\Lag_{\rm NLO}$ is typically chosen to  include operators of classes:
\begin{align}
\label{eq.HEFT_L4_classes}
  &d^4\,,  
  & 
  &d^2X\,,
  &
  &X^2\U\,,
  &
  &X^3\,,
  &
  &d\psi^2\U\,,
  &
  &d^2\psi^2\,,
  &
  &X\psi^2\,,
  &
  &\psi^4\,,
\end{align}
where $d$ stands for any insertion of $D_\mu$ or $\V_\mu$, $X$ represents a gauge field strength and $\psi$ a generic fermion field. The addition of a $\U$ indicates that the class \emph{only} contains operators with scalar field insertions, which distinguishes \eg\ $B_{\mu\nu}^2$ and $\bar\psi i\slashed{D}\psi$ (which are of course LO) from $B_{\mu\nu}^2 h$ and $\bar\psi\gamma_\mu \V^\mu\psi$. In all other classes, arbitrary scalar insertions can be added without affecting the counting. Therefore they are left unspecified. 

All the classes in Eq.~\eqref{eq.HEFT_L4_classes} 
were identified as NLO in~\cite{Buchalla:2013eza,Buchalla:2013rka}, except $\psi^2 X,\,X^3$ that were considered NNLO. At the same time, this selection completes the HEFT Lagrangian up to $d_p=8$~\cite{Gavela:2016bzc}, and it provides effective interactions that are not too different from those in $d=6$ SMEFT.  In this sense, the "conventional" $\Lag_{\rm NLO}$  constitutes a reasonable compromise, compatible with all the relevant physics arguments, that can be readily used in phenomenological studies. However, the fact that it was not unambiguously defined by a  power counting clouds its physical interpretation, and can generate confusion in its use in scattering amplitudes, observable predictions and even in the matching to BSM models. While in the SMEFT case we have a very clear physical correspondence between the expansions performed at the level of $\Lag_{\rm SMEFT}$ and at the level of diagrams and observables, the prescription to truncate HEFT predictions is more blurry. The picture worsens further when going to NNLO and higher orders, for which power counting considerations were not spelled out precisely and that have not been systematically deployed in phenomenological studies verifying their impact. For instance, none of the power counting studies presented so far has spelled out the full list of operator classes pertaining to NNLO according to their rationale\footnote{It is of course possible to determine which operator classes have higher values of $d_\chi$ or $d_p$. However, as seen above, this is not sufficient to determine the NNLO unambiguously: the conventional NLO basis contains terms with $2\leq d_\chi\leq4$ and $5\leq d_p\leq 8$, and additional physics arguments were employed in its definition, whose generalization to higher orders has not been worked out to the best of our knowledge.}, and it is not obvious that a customary set compatible with all of them could be defined.

Complete bases for the NLO operator classes in Eq.~\eqref{eq.HEFT_L4_classes} were built in~\cite{Buchalla:2013rka,Brivio:2016fzo,Sun:2022ssa}, with~\cite{Sun:2022ssa} clarifying in particular the flavor structure. Ref.~\cite{Sun:2022snw} presented a basis for $\Lag_{\rm NNLO}$, which was defined by assigning $d_\chi=2$ to gauge field strengths and fermion bilinears (under the assumption that they would only appear in $g X_{\mu\nu}$, $g\bar\psi\psi$ or $y\bar\psi\psi$ building blocks) and selecting operator classes with overall $d_\chi=5, 6$ (while the NLO basis in~\cite{Sun:2022ssa} included classes with $d_\chi=3,4$ in the same conventions).
The number of independent parameters in $\Lag_{\rm NLO}$ was computed with Hilbert series methods as well~\cite{Graf:2022rco,Sun:2022aag,Alonso:2024usj}, see also~\cite{Graf:2020yxt}. These works classified HEFT operators based on their mass dimensions (computed assigning \eg\ $d=1$ to $\V_\mu$ and $d=3/2$ to $\psi$), but their results can be easily specialized to the operator classes in Eq.~\eqref{eq.HEFT_L4_classes}~\cite{Graf:2022rco}.

Finally, the parameter
\begin{equation}
\xi = \frac{v^2}{f^2}\;\in[0,1]\,,
\label{eq.xi_def}
\end{equation}
is often introduced in the construction of HEFT operator bases, see \eg~\cite{Alonso:2012px,Buchalla:2013rka,Brivio:2013pma,Brivio:2016fzo}.
The constant $f\geq v$ in Eq.~\eqref{eq.xi_def} is an additional energy scale: its introduction yields a non-minimal version of HEFT, where $f$ replaces $v$ as the suppression scale for scalar insertions ($\pi^I/f$, $h/f$) and  factors of $\xi$ are introduced such that $v$ controls the size of the SM masses. 

This construction was originally inspired by composite Higgs models~\cite{Kaplan:1983sm,Georgi:1984af,Dugan:1984hq,Contino:2003ve,Agashe:2004rs,Contino:2006qr,Gripaios:2009pe}, in which the four SM scalars emerge as pseudo-Goldstone bosons with decay constant $f$, and the condition $f>v$ can be ensured \eg\ via vacuum misalignment~\cite{Kaplan:1983fs,Banks:1984gj}. When matching composite Higgs models to the HEFT, one typically finds that the matching expressions for the Wilson coefficients scale with some power of $\sqrt{\xi}$ and that the operators (in particular the $\F_i(h)$ structures) bear an implicit non-linear dependence on $\xi$~\cite{Alonso:2014wta,Hierro:2015nna,Gavela:2016vte,Qi:2019ocx} (see also~\cite{Contino:2010mh,Azatov:2012bz,Buchalla:2013rka} for LO matchings). 
Depending on the UV assumptions, $\xi$ can take values anywhere in the range $[0,1]$. The limit $\xi\to 1$ ($f\sim v$) corresponds to scenarios in which non-linearities are most relevant: the couplings of the Higgs boson largely deviate from their SM values, and unitarity is restored by the exchange of composite resonances, which must be light: $M\sim 4\pi v\sim 3$~TeV.  On the other hand, for $\xi\to 0$ ($v\ll f$) one recovers the \emph{decoupling limit}, in which the composite sector is much heavier than the EW scale. In this case, the $\xi$-dependence hidden in the HEFT operators can be expanded out, and $\xi$ can be interpreted as inducing a new expansion on top of the one "in derivatives".
An important observation is that, in this scenario, the expansion in $\xi$ is expected to reproduce the SMEFT expansion: intuitively,  expanding in $\xi$ is equivalent to expanding in $(H^\dag H)/f^2$ at the vacuum. Moreover, composite Higgs models contain the $SU(2)_L$ scalar doublet, which is embedded within a larger representation of the coset symmetry, and they do not induce singularities at the $H^\dag H=0$ point. The equivalence of $\xi$ and SMEFT expansions in composite Higgs models has been checked by verifying that the matching procedure assigns a leading scaling in $\xi^n$ to HEFT operators that, via Eq.~\eqref{eq.HEFT_to_SMEFT}, map to $d=4+2n$ SMEFT operators~\cite{Alonso:2014wta,Hierro:2015nna,Qi:2019ocx}. 
The Strongly-Interacting Light Higgs
(SILH) Lagrangian~\cite{Giudice:2007fh} constitutes a well-known realization of this picture~\cite{Buchalla:2014eca}, in which the EFT of a light composite Higgs is directly parametrized by dimension-6 SMEFT operators.

These arguments motivated the introduction of the $\xi$ parameter in the "bottom-up" construction of the HEFT Lagrangian. In this approach, HEFT operators are assigned a (leading) $\xi$ order on top of their primary classification into LO, NLO, etc. discussed above. The $\xi$ order is determined by identifying, for each operator, the lowest canonical dimension among its SMEFT "siblings"\footnote{The SMEFT-HEFT mapping was actually performed using Eq.~\eqref{eq.SMEFT_to_HEFT}, assigning a certain $\xi$ order to all the HEFT operators produced in the expansion of a SMEFT one. 
}~\cite{Alonso:2012px,Buchalla:2013rka,Gavela:2014vra,Brivio:2013pma}. This classification has been mostly used as a marker of the SMEFT expansion, rather than as a power-counting criterion, as large values of $\xi$ remain allowed in principle.
Nevertheless, Refs.~\cite{Buchalla:2014eca,Brivio:2016fzo} argued that the HEFT Lagrangian can be organized as a dual expansion in $\sqrt{\xi}$ and $(p/\Lambda)$, with Ref.~\cite{Brivio:2016fzo}
indicating that counting both with the same weight is effectively equivalent to the power counting in primary dimensions. This argument was laid out qualitatively in Ref.~\cite{Brivio:2016fzo}, and compared to the conventional definition of $\Lag_{\rm NLO}$ presented earlier in this section, which identifies the operator classes in Eq.~\eqref{eq.HEFT_L4_classes}.

\vskip 1em

In the next sections we revisit the power counting of HEFT from first principles. Our goal is twofold: on one hand we would like to identify "clean" physical arguments and theoretical constraints that apply to the EFT in its own merit, removing as much as possible theoretical prejudice and references to UV assumptions, which can always be introduced at a second stage in order to specialize further the EFT structure. On the other hand, we seek a quantitative, unambiguous criterion for sorting HEFT operators, that can be associated to a prescription for the truncation of scattering amplitudes and observable predictions, analogous to the expansion in inverse powers of $\Lambda$ in SMEFT.

\section{Power counting in Effective Field Theories} \label{sec.pc}
\subsection{General aspects}\label{sec.pc_general}

The fundamental idea of EFTs is that they approximate another theory within a specific parametric or kinematic regime, where observables predictions can be meaningfully expanded in powers of some small quantity $\delta\ll 1$. 
Depending on the desired accuracy, predictions can then be truncated at a given order $\delta^n$, leaving a finite theoretical error of order $\delta^{n+1}$. 
The corresponding EFT is a theory that is manifestly organized as an expansion at the Lagrangian level, such that predictions of observables in this framework are directly expressed as a series expansion in $\delta$. 

The EFT can be built "from the bottom up" 
as an independent, self-consistent theory,
by specifying its
field content, symmetries (exact or approximate) and \emph{power counting}, \ie\ a set of rules for sorting and truncating
\begin{enumerate}[label=(\roman*)]
\item the infinite tower of allowed EFT operators in the Lagrangian \label{item.ops}
\item the set of Feynman diagrams (of any loop order) contributing to a given process\label{item.M}
\item the contributions to given observable (e.g. cross section, decay rate) obtained squaring the scattering amplitudes\label{item.xs}
\end{enumerate}
These criteria have to be mutually consistent, and such that the final result, \ie\ the sorting and truncation of the contributions to an observable, matches the expansion in the desired quantity $\delta$.

In the SMEFT case, \ref{item.ops} operators are sorted by their canonical dimension $d$, which is associated with a $\Lambda^{4-d}$ suppression factor as in Eq.~\eqref{eq:SMEFTL}. The Lagrangian truncation follows the same rationale, \ie\ it should be performed retaining all the terms up to the desired order in $\Lambda$. The same holds for \ref{item.M} Feynman diagrams and \ref{item.xs} observable contributions. In particular, the well-known power counting of SMEFT tells us that diagrams with one insertion of $d=6$ operators (of order $\Lambda^{-2})$ are less suppressed than diagrams with two $d=6$ insertions or one $d=8$ insertion, which are of the same order~$\Lambda^{-4}$. Similarly, the interference between a SM and a dimension-six diagrams  enters at a lower order\footnote{Here we are only interested in suppressions from the EFT power counting:  we do not consider kinematic suppressions, which often affect interference terms and occur on a process-dependent basis.
} ($\Lambda^{-2}$) compared to the interference between SM and dimension-eight or to the square of a dimension-six diagram, both scaling with $\Lambda^{-4}$, and so on. 
At the end of the calculation of \eg\ a scattering cross section in SMEFT, it becomes manifest that this prescription realizes a series expansion of the cross section in powers of $(p/\Lambda)\sim (v/\Lambda)$, where the $\sim$ indicates that both factors are counted as equivalent suppressions.

It is important to stress that the expansion obtained at the observables level is the only one that carries a physical meaning. 
Lagrangian terms, Feynman rules and scattering amplitudes are not observable quantities, so the  prescription for those intermediate objects can be viewed merely as an algorithmic rule designed to guarantee a consistent expansion of the final results.  
When adopting SMEFT, we assume that new physics is sufficiently decoupled from the SM, such that \emph{observables} can be expanded in  $\delta = (p,v/\Lambda)\ll 1$. The well-known power counting prescription can be derived from here by working backwards through the calculation -- tracing the amplitude, Feynman rules and operators steps -- and introducing at each stage a sorting rule that matches the expansion in~$\delta$. 
In the SMEFT case, this "backward projection" of the observables expansion to amplitudes and operators is trivial. However, this is not always the case. For instance, in the case of $\chi$PT,  the powers of $(p/\Lambda)$ in an observable depend on both the number loops of the contributing diagrams and the number of derivatives of the operators inserted, as in Eq.~\eqref{eq.chiPT_count}. Therefore the operators and diagrams series must be organized in a way that accounts for both features.

The rest of this work aims at applying this rationale to the HEFT, that is: we approach the power counting question by requiring that a HEFT calculation yields observable predictions that are organized according to some physically meaningful expansion, and we analyze the implications of this condition on the organization of the Lagrangian and amplitudes series. Although the conclusions are expected to contain similarities with previously proposed power counting rationales, our approach is completely independent of those adopted in the literature and reviewed in Section~\ref{sec.HEFT}.
Our hope is that it helps pinpointing model-independent consistency conditions that can inform a more rigorous organization of $\Lag_{\rm HEFT}$.

Before addressing the HEFT case, it is useful to derive a number of results that establish the relation between the suppression factors in observable predictions and the properties of effective operators and Feynman diagrams in general EFTs. Section~\ref{sec.Lag_to_obs} introduces our notation and it presents a self-contained discussion of such results, which draws heavily from previous literature, in particular from Ref.~\cite{Gavela:2016bzc}.

\subsection{Lagrangian, amplitudes and observables expansions}\label{sec.Lag_to_obs}
In general, predictions of observables in QFT can depend on various sorts of "small parameters" worth expanding in. A series-expanded cross section prediction, for instance, will have the form
\begin{align}
\sigma = \sigma_0 \,\delta_1^{\alpha_1}\dots \delta_n^{\alpha_n}
\end{align}
where we take  $\delta_i\ll 1$ to be a set of small \emph{dimensionless} quantities, and $\alpha_i$ are the powers with which they appear in the final expression.
The parameters $\delta_i$ can be either:
\begin{enumerate}
\item dimensionful parameters appearing in the cross section, normalized to a reference quantity that qualitatively defines their maximum value. This category includes \eg\ $(p/\Lambda)$ and the expansion parameters of the perturbative series $(g/4\pi)$.
\item dimensionless quantities bringing model-dependent suppressions, such as spurions introducing small breakings of an approximate symmetry.
\end{enumerate}
In this work we will only consider suppressions of  category 1, whose insertions are controlled by dimensional analysis. Power counting prescriptions for parameters of category 2 can usually be added trivially on top of the dimensional ones. An example was already given in Section~\ref{sec.HEFT}, when discussing the possibility of pushing the 2-derivative operator $\cP_T$ in Eq.~\eqref{eq.PT_def} to the NLO HEFT Lagrangian, due to its breaking of the custodial symmetry. 

\subsubsection{Lagrangian}
The scaling of dimensionful quantities at the Lagrangian level is determined by
Naive Dimensional Analysis (NDA)~\cite{Manohar:1983md,Cohen:1997rt,Luty:1997fk,Gavela:2016bzc}. Keeping $c=\hbar$
while allowing $\hbar\neq 1$ dictates that any physical quantity can be expressed in units of mass and $\hbar$. For instance, the effective action $S$ has dimension $\hbar$, while the Lagrangian has dimension $\hbar \times ({\rm length})^{-4} = \hbar M^4$.\footnote{The assumption $c=\hbar$ is distinct from $c=1$ when $\hbar\neq 1$. For instance,  a Lagrangian's dimension is $\hbar\times({\rm length})^{-4} = \hbar\times (Mc/\hbar)^4$. Choosing $c=1$ would give $M^4/\hbar^3$, while $c=\hbar$ leads to the desired $\hbar M^4$.}
The well-known NDA master formula introduces two reference objects carrying these dimensions: $\Lambda$, which is a mass, and $(4\pi)$ which can be interpreted as having dimension $\hbar^{-1/2}$ in the usual conventions adopted in QFT calculations.
A generic Lagrangian interaction is then cast in a form where any dimensionful quantity (field or constant) is normalized by appropriate factors of $\Lambda$ and $4\pi$ according to:
\begin{equation}
 \Lag \supset \frac{\Lambda^4}{(4\pi)^2}
 \left(\frac{\de_\mu}{\Lambda}\right)^q   
 \left(\frac{4\pi\phi}{\Lambda}\right)^s   
 \left(\frac{4\pi\psi}{\Lambda^{3/2}}\right)^f   
 \left(\frac{g}{4\pi}\right)^{n_g}
 \left(\frac{y}{4\pi}\right)^{n_y}
 \left(\frac{\lambda}{(4\pi)^2}\right)^{n_\lambda}
 \left(\frac{4\pi v}{\Lambda}\right)^{n_v}\,.
\label{eq.NDA_MF}
\end{equation}
Each bracketed quantity in Eq.~\eqref{eq.NDA_MF} is dimensionless: $\phi$ represents any scalar or gauge field insertion, while $\psi$ is a generic fermion.  The constant $v$ is defined to have the same dimensions as a scalar field, potentially representing a non-zero VEV. $g,y,\lambda$ are defined as the coupling constants of the interactions $A_\mu\bar\psi\gamma^\mu\psi$, $\phi\bar\psi\psi$ and $\phi^4$, respectively. 

Since NDA only addresses the dimensional scaling of an interaction, it only depends on the number of fields ($s$, $f)$, derivatives $q$ and constants ($n_g,n_y,n_\lambda,n_v$) appearing in the Lagrangian term, and not on the details of the Lorentz structure. 
For economy of notation, Eq.~\eqref{eq.NDA_MF} shows only one representative for each class of dimensionful objects appearing in the SM Lagrangian. The full NDA master formula for the SM (and its EFT extansions) contains an independent factor for each gauge coupling constant $g,g',g_s$, and Yukawa coupling $y_\psi$. Gauge fields have the same dimensions as scalars, so the factor $(4\pi\phi/\Lambda)$ could be split into powers of $(4\pi\phi/\Lambda)$ and $(4\pi A_\mu/\Lambda)$, powers of $(4\pi X_{\mu\nu}/\Lambda^2)$ could be added for field strengths, and the notation could be further specialized to each gauge group. The generalization to EFTs with different field content and/or coupling constants is straightforward.
Finally, if the EFT Lagrangian is defined with dimensionful\footnote{From here on, "dimensionful/less" means with/without mass \emph{or} $\hbar$ dimensions.} Wilson coefficients, additional factors such as $(C_i\Lambda^2/(4\pi)^2)$ should be included in Eq.~\eqref{eq.NDA_MF}.
In the following we will adopt the NDA normalization because it is most convenient for the purpose of discussing EFT power counting, but we will come back to discussing alternative operator normalizations for the specific cases of SMEFT and HEFT in the next sections.  

NDA satisfies a number of important properties:
\begin{itemize}
\item it assigns homogeneous scalings to the terms inside a covariant derivative, and to those inside a scalar acquiring a non-vanishing VEV
\begin{align}
\frac{\de_\mu}{\Lambda}&\sim \frac{g}{4\pi}\frac{4\pi A_\mu}{\Lambda} = \frac{g A_\mu}{\Lambda}\,,
&
\frac{4\pi \phi}{\Lambda} &\sim \frac{4\pi h}{\Lambda} + \frac{4\pi v}{\Lambda}
\end{align}
\item it assigns homogeneous scalings to all terms in a propagator, i.e. to momenta and masses, including those that are induced by spontaneous symmetry breaking:
\begin{align}
\frac{p}{\Lambda}&\sim \frac{g}{4\pi}\frac{4\pi v}{\Lambda} = \frac{g v}{\Lambda} \sim \frac{m_V}{\Lambda}\,,
&&\text{and analogously for }g\leftrightarrow y\,,\label{eq:pmscaling}
\\
\frac{p}{\Lambda}&\sim \frac{\sqrt{\lambda}}{4\pi}\frac{4\pi v}{\Lambda} \sim \frac{m_h}{\Lambda}\,,
\end{align}
\item it is consistent diagrammatically (see below)
\item it makes perturbativity bounds trivial. For instance, if we define SMEFT operators with NDA normalization
\begin{align}
\Lag_{\rm SMEFT} &\supset \frac{(4\pi)^2}{\Lambda^2}  C_{4\psi}\; (\bar\psi\psi)^2 + \frac{(4\pi)^2}{\Lambda^2} C_{HG} \;G_{\mu\nu}^A G^{A\mu\nu} H^\dag H
+\dots
\end{align}
then the Wilson coefficients are dimensionless, and they must satisfy
\begin{align}
|C_i |\leq 1\,,
\end{align}
in order for the perturbative series expansion to be valid, \ie\ for the size of loop diagrams not to exceed that of tree-level ones, and so on.
This property follows from the choice of $4\pi$ as normalization for $\hbar$ dimensions, and it holds for general EFTs.
\end{itemize}

What we need for the power counting discussion is to keep track of the powers of $\Lambda$ and $4\pi$ appearing in an \emph{interaction vertex}, \ie\ a Lagrangian term defined \emph{after} expanding out covariant derivatives, field strengths, and scalars taking a VEV. Applying the NDA formula, we can write any interaction schematically  as
\begin{equation}
\label{eq.generic_interaction}
\Lag_i = 
\frac{1}{\Lambda^{N_{\Lambda,i}}}\frac{1}{(4\pi)^{N_{4\pi,i}}} \; \gCoupl_i \; \de^{q_i}\,\phi^{s_i}\, \psi^{f_i}\,,
\end{equation}
where, as above, $\phi$ represents both scalars and gauge fields, $\psi$ represents  fermions, and $\gCoupl_i$ denotes a generic product of dimensionful constants.  This expression could be multiplied by arbitrary factors containing dimensionless constants, such as Wilson coefficients $C$, which are irrelevant for the rest of the discussion, and will therefore be omitted. The index $i$ is a label for the interaction vertex.

We defined $N_{\Lambda,i}, N_{4\pi,i}$ to count \emph{inverse} powers of $\Lambda$ and $4\pi$, as eventually we will be interested in suppression factors in the observables predictions.
It is convenient to split them into powers $N^\gCoupl$ brought by the dimensionful constants ($v,g,y,\lambda,\dots)$ and powers $N^p$ brought by fields and momenta:
\begin{align}
\label{eq.Ni_lag}
N_{\Lambda,i} &= N_{\Lambda,i}^p + N_{\Lambda,i}^\gCoupl\,,
&
N_{4\pi,i} &= N_{4\pi,i}^p + N_{4\pi,i}^\gCoupl\,,
\end{align}
where $N_{\Lambda,i}^p,N_{4\pi,i}^p$ can be read off from Eq.~\eqref{eq.NDA_MF}
\begin{align}
\label{eq.Nip_lag}
N_{\Lambda,i}^p &= -4 +q_i +s_i +\frac{3}{2}f_i\,,
&
N_{4\pi,i}^p &= 2-s_i-f_i\,,
\end{align}
and $N_{\Lambda,i}^\gCoupl, N_{4\pi,i}^\gCoupl$ are obtained summing the powers brought by constants, according to the following table:
\begin{center}
\renewcommand{\arraystretch}{1.2}
\begin{tabular}{>{$}c<{$}cc}
\hline
& $N_{\Lambda}^\gCoupl$& $N_{4\pi}^\gCoupl$
\\\hline
g, g', g_s, y_i& 0& 1\\
\lambda& 0& 2\\
v& 1& $-1$
\\\hline
\end{tabular}
\vspace{1em}
\end{center}
Insertions of the electromagnetic coupling $e = gg'/\sqrt{g^2+g^{\prime2}}$ behave identically to $g,g'$ insertions. 
Insertions of (trigonometric functions of) the weak mixing angle $\theta = g'/g$ are dimensionless and therefore do not contribute to any of the countings. 

Table~\ref{tab.operator_counting_examples} provides some examples, showcasing alternative definitions of the same interactions, as well as multiple interactions that in SMEFT/HEFT would stem from the same operator. It is worth noting that: 
\begin{description}
\item[$N_{\Lambda,i}^p,N_{4\pi,i}^p$] only depend on the field and derivative content of the interactions. They are independent of arbitrary normalizations chosen for the operators, but are different for each interaction produced by a gauge invariant operator (\eg\ $\bar \psi_L \psi_R h$ vs $\bar \psi_L \psi_R h^2$). In particular, $N_{4\pi,i}^p$ is  different for the terms inside a covariant derivative $D_\mu = \de_\mu + i g A_\mu$.

\item[$N_{\Lambda,i}^\gCoupl,N_{4\pi,i}^\gCoupl$] obviously depend on the dimensionful constants inserted, which makes them dependent on arbitrary normalizations chosen for the operators (\eg\ $G_{\mu\nu}G^{\mu\nu} H^\dag H$ vs. $g_s^2 G_{\mu\nu} G^{\mu\nu} H^\dag H$). They are also different for the various interactions produced by a single operator, and in particular for the terms inside a covariant derivative.
\item[$N_{\Lambda,i},N_{4\pi,i}$] are homogeneous for all interactions produced by an operator, but depend on the overall normalization chosen in terms of dimensionful constants.
\end{description}
By construction, all SM interactions defined as in Eqs.~\eqref{eq.LagSM_SMEFT}, or~\eqref{eq.lagrangian_HEFT_SM}
have ${N_{\Lambda,i}=0=N_{4\pi,i}}$.

A short remark on the Higgs mass term is in order: in Eq.~\eqref{eq.LagSM_SMEFT} it was introduced as $(m_h^2/2) H^\dag H$. The NDA scaling of the interaction would be $\Lambda^2 H^\dag H$, which is consistent with the fact that -- dimensionally -- $m_h$ is a mass. The parameter $m_h$ was introduced as an independent quantity in Eq.~\eqref{eq.LagSM_SMEFT}. However, it is convenient to consider it as a function of $v$ and $\lambda$, and to retain only the latter two as independent parameters. In this way, the treatment of $m_h$ is identical to the treatment of fermion and gauge boson masses.

\begin{table}[t]\centering
\renewcommand{\arraystretch}{1.3}
\begin{tabular}{>{$}r<{$}@{ }>{$}l<{$}|*3{>{$}c<{$}}|*3{>{$}c<{$}}|*3{>{$}c<{$}}}
\hline
\multicolumn{2}{l}{$\Lag$ Interaction}& 
N_{\Lambda,i}^p &
N_{\Lambda,i}^\gCoupl &
N_{\Lambda,i} &
N_{4\pi,i}^p &
N_{4\pi,i}^\gCoupl&
N_{4\pi,i} &
N_{\chi,i}^p &
N_{\chi,i}^\gCoupl&
N_{\chi,i}
\\\hline
g& \bar \nu \gamma^\mu l \, W^+_\mu  &
0& 0& 0&
-1& 1& 0&
-1& 1& 0
\\
y_\psi& \bar \psi_L \psi_R h&
0& 0& 0&
-1& 1& 0&
-1& 1& 0
\\
e& W^+_\mu W^-_\nu A^{\mu\nu}&
0& 0& 0& 
-1& 1& 0& 
-1& 1& 0
\\
g^2v& W^+_\mu W^-_\nu h&
-1& 1& 0& 
-1& 1& 0& 
-2& 2& 0
\\
\lambda v& h^3 & 
-1& 1& 0&
-1& 1& 0&
-2& 2& 0
\\
\hdashline
v& (\de_\mu G_\nu^A)^2 h&
1& 1& 2&
-1& -1& -2&
0& 0& 0
\\
&(\de_\mu G_\nu^A)^2 h^2&
2& 0& 2&
-2& 0& -2&
0& 0& 0
\\
g_s^2& (\de_\mu G_\nu^A)^2 (h/v)&
1& -1& 0&
-1& 3& 2&
0& 2& 2
\\
g_s^2& (\de_\mu G_\nu^A)^2 (h/v)^2&
2& -2& 0&
-2& 4& 2&
0& 2& 2
\\
\hdashline
& W^{+\mu}_{\nu} W^{-\nu}_\rho A^\rho_\mu&
2& 0& 2&
-1& 0& -1&
1& 0& 1
\\
g^2& W^{+\mu} W^-_{\nu} W^{+\nu} W^-_\rho A^\rho_\mu&
2& 0& 2&
-3& 2& -1&
-1& 2& 1
\\
e & W^{+\mu}_{\nu} W^{-\nu}_\rho A^\rho_\mu&
2& 0& 2&
-1& 1& 0&
1& 1& 2
\\
\hdashline
v& \bar t_L \sigma^{\mu\nu}T^A t_L \de_\mu G_\nu^A& 
1& 1& 2&
-1& -1& -2&
0& 0& 0
\\
&\bar t_L \sigma^{\mu\nu}T^A t_L \de_\mu G_\nu^A h& 
2& 0& 2&
-2& 0& -2&
0& 0& 0
\\
vg_s& \bar t_L \sigma^{\mu\nu}T^A t_L G_\mu^B G_\nu^C f^{ABC}& 
1& 1& 2&
-2& 0& -2&
-1& 1& 0
\\
g_s& \bar t_L \sigma^{\mu\nu}T^A t_L \de_\mu G_\nu^A& 
1& 0& 1&
-1& 1& 0&
0& 1& 1
\\
g_s& \bar t_L \sigma^{\mu\nu}T^A t_L \de_\mu G_\nu^A (h/v)& 
2& -1& 1&
-2& 2& 0& 
0& 1& 1
\\
g_s^2& \bar t_L \sigma^{\mu\nu}T^A t_L G_\mu^B G_\nu^C f^{ABC}& 
1& 0& 1&
-2& 2& 0&
-1& 2& 1
\\
\hdashline
v^2&\bar \psi_L \psi_R h &
0& 2& 2& 
-1& -2& -3&
-1& 0& -1
\\
v&\bar \psi_L \psi_R h^2 &
1& 1& 2& 
-2& -1& -3&
-1& 0& -1
\\
y_\psi v&\bar \psi_L \psi_R (h/v)^2 &
1& -1& 0& 
-2& 2& 0&
-1& 1& 0
\\
\hline
\end{tabular}
\caption{Examples of $N_{\Lambda,4\pi,\chi}^{p,\gCoupl}$ countings for Lagrangian interaction vertices. The table includes alternative normalizations of the same interactions and different interactions stemming from the same gauge-invariant operator. $A_{\mu\nu} = \partial_{\mu} A_{\nu} -\partial_{\nu} A_{\mu}$ denotes the photon field strength.}\label{tab.operator_counting_examples}
\end{table}

\subsubsection{Chiral dimension}\label{sec.chiraldim} Before moving on to diagrammatic rules, it is useful to introduce the concept of "chiral dimension". Rather than borrowing definitions from previous literature, here we define it as a physical quantity, on the same footing as a length or a power. As any other physical quantity, the chiral dimension can be expressed as a composition of mass and $\hbar$ dimensions when $c=\hbar\neq 1$. Specifically
\begin{equation}
\label{eq.dchi_def}
d_\chi = d_M -2d_\hbar \,,
\end{equation}
is the combination that leaves scalar fields chirally dimensionless, \ie\ that assigns them $d_\chi=0$. With this definition, we can easily incorporate the chiral dimension in the dimensional analysis. In terms of the $N$ countings, Eq.~\eqref{eq.dchi_def}  translates into  
\begin{equation}
\label{eq.Nchi_def}
\framebox{$N_\chi = N_\Lambda + N_{4\pi}$}\,,   
\end{equation}
\ie\ we can define a quantity $N_\chi$ that counts the total number of suppression factors from powers of either $\Lambda$ or $4\pi$, without distinguishing between them.
This composition rule holds in general, at the Lagrangian, diagrams or observables levels, and it applies to the breakdown in $N^p+N^\gCoupl$ contributions, \ie\ $N_\chi^p = N_\Lambda^p + N_{4\pi}^p$, $N_\chi^\gCoupl = N_\Lambda^\gCoupl + N_{4\pi}^\gCoupl$ etc.
In particular, for a generic Lagrangian interaction we can define
\begin{align}
\label{eq.Nchi_amplitude}
N_{\chi,i} &= N_{\chi,i}^p + N_{\chi,i}^\gCoupl\,,
&
N_{\chi,i}^p &= -2 + q + \frac{f}{2}\,,
\end{align}
which trivially follow from Eqs.~\eqref{eq.Ni_lag} and~\eqref{eq.Nip_lag}.
In the same way, $N_{\chi,i}^\gCoupl$ can be computed remembering that $g,g',g_s,y_t$ contribute $N_\chi^\gCoupl=1$, and $\lambda$ contributes $N_\chi^\gCoupl=2$.

The values of the chiral countings for example interactions were reported in the last three columns of Table~\ref{tab.operator_counting_examples}. $N_{\chi,i}^p,N_{\chi,i}^\gCoupl,N_{\chi,i}$ behave analogously to the $\Lambda,4\pi$ countings, with the additional specification that they are insensitive to the presence of scalars, gauge fields, or VEV insertions. It is also worth noting that all SM interactions have $N_{\chi,i}=0$.

With the definition given here, the chiral dimension of the various quantities in the Lagrangian (\ie\ $d_\chi[\psi]=1/2$, $d_\chi[g]=1$ etc.) agrees with the assignment in Ref.~\cite{Buchalla:2013eza,Buchalla:2016sop}. 
The chiral order $``d = 2L+2"$ of an effective interaction defined in~\cite{Buchalla:2013eza} is related to $N_{\chi,i}$ by $N_{\chi,i} = d-2=2L_i$, with $L_i$ defined in the same reference.
The quantity ``$N_\chi$" defined in Ref.~\cite{Gavela:2016bzc}, coincides with $(N_\chi^p+2)$ in our definition.

\subsubsection{Amplitudes} \label{sec_amplitudes} Let us now consider Feynman diagrams of any loop order, contributing to a certain process. Each diagram will be proportional to some powers of $\Lambda$ and $4\pi$: powers of $\Lambda$ can only come from the Feynman rules of the vertices composing the diagram, while powers of $4\pi$ can come from both Feynman rules and loop integrals. The overall scaling of a diagram can be related to the scaling of the individual vertices and to the diagram topology using well-known results that are collected in App.~\ref{app.diagrammatics}.

Besides powers of $\Lambda$ and $4\pi$, we are interested in keeping track of the powers of momenta $p$ appearing in an amplitude. Momenta can come from vertices, internal propagators, loop integrals, and external fermionic spinors: $u(p)\sim \sqrt{p}$. 
Importantly, we will not distinguish between momenta and masses of the dynamical degrees of freedom: our analysis restricts to power counting rationales that treat $p/\Lambda$ and $m/\Lambda$ suppressions on the same footing. 
Dropping this assumption would imply that the various terms appearing in equations of motion, propagators or generic kinematic functions (including loop functions) can belong to different EFT orders, which is clearly problematic.
Classification algorithms that treat differently $p$ and $m$ can be constructed in some cases, see \eg\ the recent proposal for SMEFT in Ref.~\cite{Assi:2025zmp}. However, they are necessarily basis-dependent (because of the inhomogeneity of equations of motion) and they only hold in specific kinematic regimes, such as the tails of kinematic distributions where the $p\gg m$ condition is realized. As such, they cannot be adopted as general principles for the organization of an EFT Lagrangian.

\vskip 1em
Let us consider an amplitude $\meLabel$ with $n$ external legs, out of which $S$ scalar or gauge fields and $F$  fermions ($n=S+F$), $L$ loops and $P$ powers of momenta stemming from the amputated diagram (\ie\ counting all $p$ sources except the external spinors). We can define countings for the full amplitude, which are analogous to those for interactions:
\begin{align}
\label{eq.me_NLamP_def}
\meCount_{\Lambda,\meLabel} &=
\meCount_{\Lambda,\meLabel}^p + \meCount_{\Lambda,\meLabel}^\gCoupl
&
\meCount_{\Lambda,\meLabel}^p &= \sum_{i\in {\rm vert}} N_{\Lambda,i}^p
= -4 + P +S+\frac{3}{2}F
\\
\meCount_{4\pi,\meLabel} &=
\meCount_{4\pi,\meLabel}^p + \meCount_{4\pi,\meLabel}^\gCoupl
&
\meCount_{4\pi,\meLabel}^p  &= 2L + \sum_{i\in {\rm vert}}  N_{4\pi,i}^p  = 2-n
\\
\meCount_{\chi,\meLabel} &= \meCount_{\chi,\meLabel}^p + \meCount_{\chi,\meLabel}^\gCoupl
&
\meCount_{\chi,\meLabel}^p  &= 2L  + \sum_{i\in{\rm vert}} N_{\chi,i}^p = -2 + P +\frac{1}{2}F
\label{eq.me_NChiP_def}
\end{align}
where the sums run over the vertices composing the diagram, and the last equalities on the right-hand sides can be derived using the diagrammatic relations in App.~\ref{app.diagrammatics}. 
The fact that the expressions for the $N^p$'s are formally identical to those in Eqs.~\eqref{eq.Nip_lag} and~\eqref{eq.Nchi_amplitude} represents the diagrammatic consistency of NDA mentioned above: the scaling in powers of $\Lambda$ and $4\pi$ of an arbitrary diagram is the same as the scaling of a vertex with the same external legs and overall powers of momenta and coupling constants. In particular, divergent loop diagrams have the same NDA weight as their counterterms.
\\
For the coupling constant components we have, trivially
\begin{align}
\label{eq.me_NG_def}
\meCount_{\Lambda,\meLabel}^\gCoupl &= \sum_{i\in {\rm vert}} N_{\Lambda,i}^\gCoupl
&
\meCount_{4\pi,\meLabel}^\gCoupl &= \sum_{i\in {\rm vert}} N_{4\pi,i}^\gCoupl
&
\meCount_{\chi,\meLabel}^\gCoupl &= \sum_{i\in {\rm vert}} N_{\chi,i}^\gCoupl\,.
\end{align}

The $N$ countings can be of course expressed in other forms, whenever convenient. For instance the powers of $p$ can be written as:
\begin{align}
\meCount_{\Lambda,\meLabel}^p &= \meCount_{\chi,\meLabel}^p - \meCount_{4\pi,\meLabel}^p = 2L + \sum_{i\in {\rm vert.}} \left(q_i + \frac{f_i}{2}-2\right) +n -2\,, 
\end{align}
which reduces to Eq.~\eqref{eq.chiPT_count} in theories without fermions. This expression 
generalizes Weinberg's derivative counting~\cite{Weinberg:1978kz}, with the addition of the term $(n-4)$ already noted in Eq.~\eqref{eq.D_vs_alpha}.
The chiral order/dimension of a diagram defined in~\cite{Buchalla:2013eza} corresponds to $(N_{\chi,\meLabel} + 2) = (N_{\Lambda,\meLabel} + 4-n)$ in our notation, and the composition rules~\eqref{eq.me_NChiP_def},~\eqref{eq.me_NG_def} are consistent with their Eq.~(28). Equating the chiral order to $(2L+2)$ is only valid if
all vertices entering the diagram have $N_{\chi,i}=0$.

As shown in App.~\ref{app.diagrammatics}, once the momentum dependence from external spinors is added in, one finds that an amplitude $\mathcal{M}$ scales with
\begin{align}
 \mathcal{M} &\sim
 p^{2+\meCount_{\chi,\meLabel}^p} \, 
 \Lambda^{-\meCount_{\Lambda,\meLabel}}\, 
 (4\pi)^{-\meCount_{4\pi,\meLabel}}
 \\
 &= p^{4-n} (4\pi)^{n-2}\; 
 \left(\frac{p}{\Lambda}\right)^{\meCount_{\Lambda,\meLabel}^p} 
 \left(\frac{1}{\Lambda}\right)^{\meCount_{\Lambda,\meLabel}^\gCoupl}
 \left(\frac{1}{4\pi}\right)^{\meCount_{4\pi,\meLabel}^\gCoupl}\,,
 \label{eq.M_scaling_general}
\end{align}
where, in the second line, we have explicitly pulled out a factor that depends only on the total number of external legs $n$ and will be canceled out, when moving to observables, by phase space contributions.
If the only mass-dimensionful constants are scalar VEVs, as in SMEFT/HEFT with NDA-normalized effective operators, then the scaling can be further rearranged into
\begin{align}
\label{eq.M_scaling_vevs}
\mathcal{M} &\sim p^{4-n} (4\pi)^{n-2}\; 
 \left(\frac{p}{\Lambda}\right)^{\meCount_{\Lambda,\meLabel}^p} 
 \left(\frac{4\pi}{\Lambda}\right)^{\meCount_{\Lambda,\meLabel}^\gCoupl}
 \left(\frac{1}{4\pi}\right)^{\meCount_{\chi,\meLabel}^\gCoupl}\,,
\end{align}
where we have essentially moved the $4\pi$ factors brought by $v$ from the last factor to the second-to-last one.

To illustrate the meaning of Eqs.~\eqref{eq.M_scaling_general} and~\eqref{eq.M_scaling_vevs}, let us consider for concreteness an EFT whose only dimensionful constants are $g,y,\lambda,v$. In this case, Eq.~\eqref{eq.me_NG_def} gives $\meCount_{\Lambda,\meLabel}^\gCoupl = N_{v,\meLabel}\,,\;\meCount_{\chi,\meLabel}^\gCoupl = N_{g,\meLabel} + N_{y,\meLabel} + 2N_{\lambda,\meLabel}$, where $N_{g,\meLabel}$ counts the overall powers of $g$ appearing in the diagram and analogously for the others.  Eq.~\eqref{eq.M_scaling_vevs} can be specialized to this case by writing it as 
\begin{align}
\label{matrix_element}
\mathcal{M}&\sim
 p^{4-n} (4\pi)^{n-2}\; 
 \left(\frac{p}{\Lambda}\right)^{\meCount_{\Lambda,\meLabel}^p} 
 \left(\frac{4\pi v}{\Lambda}\right)^{\meCount_{v,\meLabel}}
 \left(\frac{g}{4\pi}\right)^{\meCount_{g,\meLabel}}
 \left(\frac{y}{4\pi}\right)^{\meCount_{y,\meLabel}}
 \left(\frac{\lambda}{(4\pi)^2}\right)^{\meCount_{\lambda,\meLabel}}\,.
\end{align}

\subsubsection{Observables}
At the final step of a calculation, an observable prediction is obtained squaring the amplitude and integrating over the phase space. The amplitudes product scales as
\begin{align}
\mathcal{M}_a \mathcal{M}^\dag_b &\sim
p^{8-2n} (4\pi)^{2n-4}\; 
 \left(\frac{p}{\Lambda}\right)^{\xsLamP} 
 \left(\frac{1}{\Lambda}\right)^{\xsLamG}
 \left(\frac{1}{4\pi}\right)^{\xsFpG}\,.
\end{align}
where
\begin{align}
\label{eq.relevant_alphas_def}
\xsLamP &= \meCount_{\Lambda,a}^p + \meCount_{\Lambda,b}^p   
&
\xsLamG &= \meCount_{\Lambda,a}^\gCoupl + \meCount_{\Lambda,b}^\gCoupl
&
\xsFpG &= \meCount_{4\pi,a}^\gCoupl + \meCount_{4\pi,b}^\gCoupl\,.
\end{align}
We can also define:
\begin{align}
\label{eq.alphafp_vs_n}
 \xsFpP &= N_{4\pi,a}^p + N_{4\pi,b}^p = 2(2-n)\,,
 \\
 \label{eq.alphachi_vs_n}
 \xsChiP &= N_{\chi,a}^p + N_{\chi,b}^p =  \xsLamP + 2(2-n)\,,
 \\
 \xsChiG &= N_{\chi,a}^\gCoupl + N_{\chi,b}^\gCoupl = \xsLamG+\xsFpG\,,
 \label{eq.alpha_chi_G_def}
\end{align}
in an analogous way, and 
\begin{align}
\label{eq.other_alpha_defs}
\xsCount_\Lambda &= \xsLamP + \xsLamG\,,
&
\xsCount_{4\pi} &= \xsFpP + \xsFpG\,,
&
\xsCount_\chi &= \xsChiP + \xsChiG\,.
\end{align}
A $k$-body phase space brings additional factors:\footnote{The NDA convention $(4\pi)\sim \hbar^{-1/2}$ descends from the fact that the angular integration of loop momenta in dimensional regularization yields factors $(4\pi)^{d/2}$ in $d$ dimensions. Although they are not introduced through dimensional arguments, $(4\pi)$ factors from phase space integration fundamentally have the same angular origin, and therefore count towards the overall $(4\pi)$ scaling of observables. }
\begin{align}
\int d{\rm PS}_k = 
\int  \prod_{j=1}^k \frac{dq_j}{(2\pi)^32E_j} q_j^2 d\Omega_j
(2\pi)^4\delta^4\left(q_{\rm init} - {\textstyle\sum}_n q_n\right)
\quad\sim\quad
p^{2k-4} (4\pi)^{3-2k}\,.
\end{align}
where $q_{\rm init}$ is the sum of initial state momenta.
Therefore a $1\to (n-1)$ decay rate $\Gamma$ and a $2\to (n-2)$ scattering cross section $\sigma$ scale respectively as
\begin{align}
 \label{eq.w_expansion_MF}
 \Gamma 
 \sim \frac{1}{p}\int d{\rm PS}_{n-1}\,
 \mathcal{M}_a\mathcal{M}_b^\dag
 &
 \;\sim\; p\, (4\pi)  \; 
 \left(\frac{p}{\Lambda}\right)^{\xsLamP} 
 \left(\frac{1}{\Lambda}\right)^{\xsLamG}
 \left(\frac{1}{4\pi}\right)^{\xsFpG}
 \\
 \sigma \sim \frac{1}{p^2}\int d{\rm PS}_{n-2} \,\mathcal{M}_a\mathcal{M}_b^\dag&
 \;\sim\; p^{-2} (4\pi)^3  \; 
 \left(\frac{p}{\Lambda}\right)^{\xsLamP} 
 \left(\frac{1}{\Lambda}\right)^{\xsLamG}
 \left(\frac{1}{4\pi}\right)^{\xsFpG}\,.
 \label{eq.xs_expansion_MF}
\end{align} 
Crucially, the $p$ and $4\pi$ factors from phase space, flux factor, and $n$-dependent terms in the squared amplitude always recombine among themselves to give a universal factor with the dimensions of the observable, which is independent of the diagram topology and of the theory adopted. The remaining powers of $\Lambda$ and $4\pi$ enter as suppression of dimensionful quantities in the cross section, forming dimensionless ratios. Arguably, only these should be considered as introducing EFT suppressions~\cite{Gavela:2016bzc}. This is a conceptual difference compared to the naive counting in derivatives, which is reflected in the $(n-4)$ correction noted in~\eqref{eq.D_vs_alpha}. Remarkably, Eq.~\eqref{eq.relevant_alphas_def} indicates that the powers $\xsLamP,\xsLamG,\xsFpG$ with which such ratios appear in~\eqref{eq.w_expansion_MF},~\eqref{eq.xs_expansion_MF} follow directly, with a trivial sum, from those with which they appeared in the contributing amplitudes $\mathcal{M}_a,\mathcal{M}_b$. They do not receive contributions from any other element in the calculation and, conversely, none of the dimensionless suppression factors in~\eqref{eq.M_scaling_general} gets absorbed elsewhere.

If scalar VEVs are the only constants with dimensions of mass, Eq.~\eqref{eq.xs_expansion_MF} can be rearranged into
\begin{align}
\label{eq.xs_expansion_MF_vevs}
 \sigma &
 \;\sim\; p^{-2} (4\pi)^3  \; 
 \left(\frac{p}{\Lambda}\right)^{\xsLamP} 
 \left(\frac{4\pi}{\Lambda}\right)^{\xsLamG}
 \left(\frac{1}{4\pi}\right)^{\xsChiG}
 \end{align}
and analogously for decay widths.
In the example where $g,y,v,\lambda$ are the only dimensionful constants present, Eq.~\eqref{eq.relevant_alphas_def} gives $\xsLamG=N_v$ and $\xsFpG = N_g + N_y + 2 N_\lambda-N_v$, where $N_g$ are the total powers of $g$ appearing in the cross section etc. Therefore 
Eq.~\eqref{eq.xs_expansion_MF} should be interpreted as
\begin{align}
\label{eq.xs_expansion_MF_explicitconstants}
  \sigma &
 \;\sim\; p^{-2} (4\pi)^3  \; 
 \left(\frac{p}{\Lambda}\right)^{\xsLamP} 
 \left(\frac{4\pi v}{\Lambda}\right)^{N_{v}}
 \left(\frac{g}{4\pi}\right)^{N_g}
 \left(\frac{y}{4\pi}\right)^{N_y}
 \left(\frac{\lambda}{(4\pi)^2}\right)^{N_\lambda}\,.
\end{align}

Eqs.~\eqref{eq.w_expansion_MF} and ~\eqref{eq.xs_expansion_MF} represent our "master formula" for the expansion of observable predictions in series of $p/\Lambda$ and normalized dimensionful constants, in a general EFT.
We argue that any valid power counting prescription based on dimensional analysis
must lead to an expansion of this form, \ie\ 
an expansion in $(p/\Lambda)$ and/or dimensionless combinations $(g/4\pi), (\lambda/16\pi^2)$ etc.
A consistent set of sorting rules can be defined choosing a \emph{linear combination} (with positive coefficients) of the exponents appearing in the observable expansion formula (\eg\ $\xsLamP$ or $2N_\lambda + 4 N_{g_s}$, etc) to define the EFT expansion parameter, and then working backwards through the calculation to define counting rules for diagrams and operators, using Eqs.~\eqref{eq.relevant_alphas_def},~\eqref{eq.me_NLamP_def},~\eqref{eq.me_NG_def} and~\eqref{eq.Nip_lag}.

The results derived in this section ensure that the step from amplitudes to observables is always trivial, as the powers of the expansion parameters $(p/\Lambda),(g/4\pi)$ etc. appearing in Eqs.~\eqref{eq.M_scaling_general} and~\eqref{eq.xs_expansion_MF} are related by a simple sum, see Eq.~\eqref{eq.relevant_alphas_def}. 
This implies that, in any EFT, the sorting rule for Feynman diagrams is always the direct transposition of the sorting rule for observables: for instance, if observables are sorted by $\xsLamP + 3 N_g + 2N_\lambda$, then diagrams can be assigned an order $N_\meLabel = \meCount_{\Lambda,\meLabel}^p+3N_{g,\meLabel}+2N_{\lambda,\meLabel}$. Doing so, the selection of the relevant diagrams can be easily performed remembering that the order of the amplitudes product $\mathcal{M}_a\mathcal{M}_b^\dag$ is the sum of the two orders $N_{\mathcal{M}_a}+N_{\mathcal{M}_b}$. 
Since this last step does not add any physics information, it is equivalent to study the power counting for a given EFT at the level of observables or of amplitudes.

\section{Power counting in SMEFT}\label{sec.SMEFT_PC}
This section specializes the power counting discussion presented in Section~\ref{sec.pc} to the  SMEFT case.
Although  well-known and arguably trivial, examining this familiar example can help clarify the meaning of the notation we introduced and it is a useful exercise in preparation for the HEFT case, that we tackle in Section~\ref{sec.HEFT_PC}.

\vskip 1ex
The construction of SMEFT starts with the choice of expanding observables in 
\begin{equation}
\label{eq.delta_SMEFT}
\delta_{\rm SMEFT} = 
\left(\frac{p}{\Lambda}\right) 
\sim 
\left(\frac{g v}{\Lambda}\right)
\sim 
\left(\frac{g' v}{\Lambda}\right)
\sim 
\left(\frac{y_\psi v}{\Lambda}\right) 
\sim 
\left(\frac{\sqrt{\lambda} v}{\Lambda}\right)\,,
\end{equation}
where we have made explicit that the masses of the SM gauge bosons\footnote{Technically we have $m_W\sim gv$ and $m_Z \sim \sqrt{g^2+g^{\prime 2}}\, v$. In terms of dimensional analysis, expanding in these quantities is equivalent to expanding in $gv$ and $g'v$, so we will use the latter for economy of notation. } ($m_V\sim g v,g'v$), fermions ($m_\psi\sim y_\psi v$) and Higgs boson ($m_h\sim\sqrt{\lambda} v$) must all be treated on the same footing as momenta. For simplicity, we will first work under the assumption that the only dimensionful parameters appearing in the SMEFT Lagrangian are $g,g',g_s,y_\psi,\lambda,v$, while all Wilson coefficients are dimensionless. As per Eq.~\eqref{eq.xs_expansion_MF}, a cross section prediction can then be explicitly expanded as 
\begin{equation}
\label{eq.xs_MF_SMEFT}
\sigma \sim p^{-2}(4\pi)^3 \; 
\left(\frac{p}{\Lambda}\right)^{\xsLamP}
\left(\frac{4\pi v}{\Lambda}\right)^{N_v}
\left(\frac{g_s}{4\pi}\right)^{N_{g_s}}
\left(\frac{g}{4\pi}\right)^{N_g}
\left(\frac{g'}{4\pi}\right)^{N_{g'}}
\left(\frac{y_\psi}{4\pi}\right)^{N_{y_\psi}}
\left(\frac{\lambda}{(4\pi)^2}\right)^{N_\lambda}
C_i^{N_{C_i}}\,.
\end{equation}
Alternative normalization choices for the effective operators will be discussed in Section~\ref{sec.smeft_other_normalizations}.

The expansion parameter choice in Eq.~\eqref{eq.delta_SMEFT} does not uniquely identify a sorting rule for SMEFT contributions to observable predictions: writing the mass factors as 
\begin{equation}
\frac{g v}{\Lambda} = \frac{4\pi v}{\Lambda} \frac{g}{4\pi}\, ,
\end{equation}
and analogously for $g',y_\psi,\sqrt{\lambda}$, it is easy to see that
overall powers of $\delta_{\rm SMEFT}$ could be counted either as $(\xsLamP+N_v)$ or as $(\xsLamP+N_g+N_{g'} + N_{y_\psi}+2N_\lambda)$.
The two options are not physically equivalent, as insertions of $v$ and insertions of the coupling constants $g,g',y_\psi,\lambda$ obviously have different origins in the theory.

SMEFT is defined with the additional requirement that the EFT power counting only depends on mass dimensions carried by momenta and fields (and their VEVs). 
This implies that observable contributions are always sorted by 
\begin{equation}
    N_{\rm SMEFT} = \xsLamP + N_v \,,
 \end{equation}
 which, with the normalization choices made here, coincides with $\xsCount_\Lambda$.
Then one easily finds that the power counting rule reduces to ``expanding in $\Lambda$" at all steps of the calculation. 

At each order in $N_{\rm SMEFT}$, contributions to the cross section can be further sorted by their powers of $(g/4\pi), (g_s/4\pi),(y_\psi/4\pi)$ etc, which reflects the perturbative expansion. 
The fact that the expansion parameters associated to the SMEFT and perturbative series factorize 
ensures that SMEFT and loop orders of a calculation can be selected independently without incurring in theoretical inconsistencies.
Powers of the dimensionless Wilson coefficients $C_i$ will also appear in the observables predictions, as indicated in Eq.~\eqref{eq.xs_MF_SMEFT}. They have no impact on the dimensional counting but they should be treated analogously to powers of $(g/4\pi)$ in the organization of a perturbative expansion. 

Finally, it is worth discussing the consistency of dimensional power counting with gauge invariance. In particular, one could wonder whether the countings $N_g, N_{g'}, N_{g_s}$ could be problematic, since they do not assign homogeneous weights to the components of a covariant derivative $D_\mu = \de_\mu + i gA_\mu$. However, this is actually not an issue because, at the level of scattering amplitudes, gauge invariance is preserved order by order in the coupling. This is a very well-known fact in the Standard Model, and it stays true in both SMEFT and HEFT. 
It is easy to check that none of the other countings could possibly introduce a violation of gauge invariance. For instance, $\xsLamP$ directly descends from $N_{\Lambda,i}^p$ in Eq.~\eqref{eq.Nip_lag}, which is manifestly homogeneous over a covariant derivative $D_\mu = \de_\mu + i g A_\mu$, because $\de_\mu$ and $(g A_\mu)$ have the same mass-dimension.

\subsection{Alternative normalizations of the SMEFT operators}\label{sec.smeft_other_normalizations}
So far, the analysis has been formulated assuming a specific normalization of SMEFT operators, which ensures dimensionless Wilson coefficients. This choice is of course arbitrary, so we now discuss what happens if the Lagrangian is cast into a different form. 

Consider for concreteness the operator $\mathcal{O}_{HG} = G_{\mu\nu}^A G^{A\mu\nu} (H^\dag H)$. We can include it in the Lagrangian in various ways, \eg
\begin{align}
 \Lag &\supset C_{1}\, \frac{(4\pi)^2}{\Lambda^2} \mathcal{O}_{HG} \,, 
 &
 \Lag &\supset C_{2}\, \mathcal{O}_{HG}  \,,
 &
 \Lag &\supset \frac{C_{3}}{\Lambda^2}\, (4\pi)^{2y} \left(\frac{g_s^{}}{4\pi}\right)^{2x} \mathcal{O}_{HG} \,.
\end{align}
The first version is the NDA-normalized one considered so far, so the Wilson coefficient $C_1$ is dimensionless. In the second version $C_2$ has dimensions $M^{-2}\hbar^{-1}$, while in the third $C_3$ has dimensions $\hbar^{-1+y}$. We leave $x,y$ as a free powers, to show explicitly where they enter the counting. The expanded master formula for observable predictions will contain factors\footnote{For economy of notation we behave as if all three parameters could be included simultaneously.\\ Of course in any realistic calculation only one of them will be present.}
\begin{equation}
\sigma \sim \quad
\cdots \quad
C_1^{N_1}
\,
\left(\frac{\Lambda^2}{(4\pi)^2} C_2\right)^{N_2}
\,
\left(\frac{C_3}{(4\pi)^{2(1-y)}}\right)^{N_3}\,,
\end{equation}
where $N_i$ counts the total powers of $C_i$ appearing in the observable expression. Correspondingly,  the $\xsCount^\gCoupl$ countings should be modified to contain
\begin{align}
\xsLamG &= -2N_2 + \dots
&
\xsChiG &= 2(1-y)N_3 + N_{g_s}+ \dots\,,
\end{align}
and $N_{g_s}$ receives a contribution $ 2x N_3$. The dots stand for contributions from other parameters. The counting $\xsLamP$ is unaffected by constant rescalings.

Let us compare the three cases for the simplest observable contribution: the square of the $gg\to hh$ contact term. The scattering amplitudes scale respectively as
\begin{align}
 \mathcal{M}_1 &\sim (4\pi)^2 
 \left(\frac{p}{\Lambda}\right)^{2}
 C_1
\\
 \mathcal{M}_2 &\sim (4\pi)^2 
 \left(\frac{p}{\Lambda}\right)^{2}
 \frac{\Lambda^2}{(4\pi)^2} C_2
 \\
 \mathcal{M}_3 &\sim (4\pi)^2 
 \left(\frac{p}{\Lambda}\right)^{2}
 \frac{C_3}{(4\pi)^{2(1-y)}} \left(\frac{g_s}{4\pi}\right)^{2x}
\end{align}
and the cross section contributions as 
\begin{align}
\sigma_1 &\sim p^{-2}(4\pi)^3 \left(\frac{p}{\Lambda}\right)^{4} C_1^2 
\\
\sigma_2 &\sim p^{-2}(4\pi)^3 \left(\frac{p}{\Lambda}\right)^{4} \left(\frac{\Lambda^2}{(4\pi)^2}C_2\right)^2
\\
\sigma_3 &\sim p^{-2}(4\pi)^3 \left(\frac{p}{\Lambda}\right)^{4} \left(\frac{C_3}{(4\pi)^{2(1-y)}}\right)^2 \left(\frac{g_s}{4\pi}\right)^{4x}\,.
\end{align}
It is easy to verify that all expressions are dimensionally consistent. Moreover:
\begin{itemize}
\item validity of the perturbative expansion requires  
all the quantities in brackets to be smaller than one, except the one involving $C_3$, \ie\
$C_1\leq 1$, $C_2\leq (4\pi/\Lambda)^2$, while $C_3/ (4\pi)^{2(1-y)}\leq (4\pi/g_s)^{2x}$, is allowed to be larger than 1, depending on $g_s,x$ and $y$.

\item  adopting mass-dimensionful Wilson coefficients (\eg\ $C_2$) does not affect the definition of the SMEFT expansion parameter, which is still $\xsLamP+N_v$: all $\mathcal{M}_i$ (and all $\sigma_i$) belong to the same SMEFT order, which, for this operator, is only determined by the powers of $(p/\Lambda)$.
What changes, compared to the NDA-normalized case, is the relation to the total powers of $\Lambda$ ($\xsCount_\Lambda$): omitting other Wilson coefficients, in this case we have $\xsCount_\Lambda - \xsLamP = \xsLamG  = N_v - 2N_2$ and therefore $\xsLamP+N_v=\xsCount_\Lambda+2N_2$.

\item adopting $\hbar$-dimensionful Wilson coefficients (\eg\ $C_2$ and $C_3$ with $y\neq 1$) or, equivalently, adding arbitrary powers of $(4\pi)$ to the operator normalization, does not affect the SMEFT power counting at all. It only has numerical consequences, in the sense that it simply changes the normalization of the Wilson coefficient: for instance,  the perturbative expansion counts powers of $(C_3/(4\pi)^{2(1-y)})$ instead of powers of $C_1$.

\item adding arbitrary powers of dimensionless quantities to the operator normalization (\eg\ powers of $(g_s/4\pi)$ as in $C_3$ with $y=0$) does not affect the SMEFT expansion, but increases the perturbative order of the observable contribution: for instance, for $x=1$, $\sigma_3 = \alpha_s^2\sigma_1 $. 

\item adding arbitrary powers of a dimensionful coupling to the operator normalization (\eg\ powers of $g_s$ as in $C_3$ with $y=x$) does not affect the SMEFT expansion, and leaves the overall powers of $4\pi$ in the amplitude invariant.
This case is equivalent to introducing a dimensionless $(g_s/4\pi)$ and simultaneously making the Wilson coefficient $\hbar$-dimensionful, therefore it changes both the perturbative order of the contributions of the operators and the numerical normalization of the Wilson coefficient.

\end{itemize}

All in all, from the point of view of the EFT power counting, all normalization choices for the operator definitions are allowed. Adopting NDA gives the simplest possible form for the "master formulas" introduced above and it ensures that the SMEFT power counting is equivalent to sorting by powers of $\Lambda$, while the perturbative expansion is marked by powers of $(g/4\pi), (\sqrt{\lambda}/4\pi), C_i$ etc. Adding powers of arbitrary coupling constants (either SM couplings or Wilson coefficients) to the definition of an operator does not affect its SMEFT power counting, but increases the perturbative order of its contributions. Any other rescaling is practically inconsequential.\footnote{We are ignoring the possibility of rescaling by powers of $v$, assuming that in SMEFT $v$ only appears as the Higgs VEV.} 
Ultimately, normalization choices are free for the EFT practitioner to make, and they can be seen as implementations of modeling assumptions. 
In particular, normalization choices for EFT operators are often motivated based on assumptions about UV completions, which determine, to some extent, the scaling of the matching contributions. We will comment further on this in Section~\ref{sec.matching}.

\subsection{On the size of $(4\pi v/\Lambda)$ in SMEFT }\label{sec.vLambda_SMEFT}
Before moving on to the HEFT case, we briefly comment on the expected size of the $(4\pi v/\Lambda)$ factors appearing in Eq.~\eqref{eq.xs_MF_SMEFT}. 
The requirement that SMEFT expands in $p/\Lambda\sim m/\Lambda\leq 1$ gives the constraint
\begin{equation}
    \frac{4\pi v}{\Lambda} \leq \frac{4\pi}{g,g',y_\psi,\sqrt{\lambda}}\,,
\end{equation}
which however is very loose, because it simply implies $\Lambda \geq x v $ with $x=\{g,g',y_\psi,\sqrt\lambda\}\leq 1$. 
More significantly, the fact that this quantity naturally emerges from NDA suggests that the SMEFT indeed expands in $(4\pi v/\Lambda)$, rather than $(v/\Lambda)$ and, therefore, that the SMEFT expansion requires
\begin{align}
  \frac{4\pi v}{\Lambda}&\leq 1  
  &&\Rightarrow&
  \Lambda&\geq 4\pi v\simeq 3~\text{TeV}\,,
\end{align}
which can seem overly strong. To examine this point further, let us first clarify what we mean with the statement above: consider for instance dimension-6 SMEFT corrections to SM vertices in the Warsaw basis~\cite{Grzadkowski:2010es}. If the SMEFT operators are defined with NDA normalization, one always finds that the suppressions are controlled by $(4\pi v/\Lambda)$, \eg
\begin{align}
\frac{[Z\bar t_Lt_L]_{\rm SMEFT}}{[Z\bar t_Lt_L]_{\rm SM}} -1 &\sim  \left(\frac{4\pi v}{\Lambda}\right)^2 \times\left[ C_{Hq}^{(1)}\,, C_{Hq}^{(3)}\,, C_{HWB}\,, C_{HD}\right]\,,
\\
\frac{[hhh]_{\rm SMEFT}}{[hhh]_{\rm SM}} -1 
&\sim  \left(\frac{4\pi v}{\Lambda}\right)^2 \times\left[ C_{HD}\,, C_{H\square}\right]\,,
\\
\frac{[Z W^+W^-]_{\rm SMEFT}}{[Z W^+W^-]_{\rm SM}} -1 
&\sim  \left(\frac{4\pi v}{\Lambda}\right)^2 \times C_{HWB}\,,
\end{align}
where, in these expressions, the Wilson coefficients $C_i$ are exactly dimensionless. Higher order corrections stem from dimension-8 operators carrying an extra insertion of $(H^\dag H)$ and, in NDA normalization, they will scale with 
\begin{equation}
    \left(\frac{4\pi v}{\Lambda}\right)^4 C_8\,,
\end{equation}
and analogously for higher dimensions. Because powers of $4\pi v/\Lambda$ are always accompanied by insertions of the Wilson coefficients, one could argue that perturbativity constraints really require
\begin{align}
\label{eq.perturbativity_weaker}
\left(\frac{4\pi v}{\Lambda}\right)^2 C_6 &\leq 1\,,
&
\left(\frac{4\pi v}{\Lambda}\right)^4 C_8 &\leq 1\,,
\end{align}
for dimension-6 and -8 operators respectively, which is a weaker constraint than demanding separately that $(4\pi v/\Lambda)\leq 1$ \emph{and} $C_6,C_8\leq 1$. Indeed, applying Eq.~\eqref{eq.perturbativity_weaker} one has
\begin{align}
  \Lambda &\geq C_6^{1/2} \, 4\pi v \,,
  &
  \Lambda &\geq C_8^{1/4} \, 4\pi v\,.
\end{align}
For $C_6,C_8$ sufficiently small, one could then have $\Lambda<3$~TeV, while remaining compatible with perturbativity across the whole Lagrangian.
However, for analogous conditions to hold at all SMEFT orders, the Wilson coefficients $C_d$ of dimension-$d$ operators must satisfy
\begin{align}
C_d \leq (\Lambda/4\pi v)^{d-4}\,,
\end{align}
\ie\ they must become smaller and smaller with increasing $d$. Clearly, this implies the presence of suppression factors that are not accounted for in the bottom-up SMEFT construction, where all $C_i$ are dimensionless and \emph{a priori} similar in size.

In practice, this extra suppression can be easily realized, for instance, by the matching to a weakly interacting theory: in this case, the expressions for dimension-8 Wilson coefficients typically contain higher powers of the SM and UV coupling constants compared to those for dimension-6 coefficients, see \eg\ the results in~\cite{Ellis:2023zim,Dawson:2021xei,Cepedello:2024ogz}. If the couplings are perturbative, they effectively bring $(g/4\pi)$ suppressions into $C_i$, which increase with the canonical dimension of the operator. This argument can be compared to the discussion about the convergence of the SMEFT expansion presented in Ref.~\cite{Cohen:2020xca}: taking a scalar singlet extension of the SM as an example, the authors show that the SMEFT series only converges for $
|H^\dag H|$ below a certain threshold. Their bound can be rearranged into $\Lambda \geq \sqrt{\hat\kappa}\, 4\pi v$, where $\Lambda = m$ is the mass of the heavy scalar $S$ and $\kappa = 16\pi^2\hat\kappa$ the coupling of the $H^\dag H S^2$ interaction, with $\hat\kappa\leq 1$.

In conclusion, in the absence of (UV-motivated) assumptions on the expected size of the dimensionless Wilson coefficients, the SMEFT formally expands in $(4\pi v/\Lambda)$, which is therefore required to be $\leq 1$. This is the only statement that can be made based on dimensional analysis. We will maintain it in this work for simplicity and consistency with our arguments, that rely only on NDA. Nevertheless, the reader should keep in mind that this condition is not necessarily enforced in practice, as, in several concrete applications, the requirement is relaxed. This can happen if \eg\ model-dependent suppressions lead to the size of the dimensionless Wilson coefficients decreasing sufficiently fast with the dimension of the corresponding operators.

\section{Power counting in HEFT}\label{sec.HEFT_PC}
What does HEFT expand in, at the level of observables? In the spirit of parameterizing the effects of yet-undiscovered new physics, HEFT should definitely induce an expansion in $p/\Lambda$. Requiring homogeneity of momenta and light masses, that appear together in propagators and kinematic functions, further brings us to guess
\begin{align}
\label{eq.delta_HEFT}
\delta_{\rm HEFT} = 
\left(\frac{p}{\Lambda}\right)
\sim 
\left(\frac{g v}{\Lambda}\right)
\sim 
\left(\frac{g' v}{\Lambda}\right)
\sim 
\left(\frac{y_\psi v}{\Lambda}\right) 
\sim 
\left(\frac{\sqrt{\lambda} v}{\Lambda}\right)\,,
\end{align}
\ie\ $\delta_{\rm HEFT} = \delta_{\rm SMEFT}$. Although it might seem surprising, this statement is actually very sensible, as SMEFT and HEFT are introduced with the same scope, and are only distinguished by their formal implementation.

Indeed the difference between HEFT and SMEFT emerges when translating the expansion in $\delta_{\rm HEFT}$ into a counting of the powers of inserted constants.
The key observation is that \emph{in HEFT we cannot expand in $(4\pi v/\Lambda)$}, and therefore the counting prescription cannot depend on $N_v$. The reason is that, while in SMEFT $v$ emerges as the VEV of the Higgs field, in HEFT it is introduced as \emph{the suppression scale} for $h$ and $\pi_I$ insertions in the $\F,\U$ building blocks. This is a crucial conceptual difference that has a few consequences: most importantly, NDA tells us that~\cite{Manohar:1983md}
\begin{equation}
\label{eq.HEFT_v_inequality}
4\pi v \geq \Lambda
\end{equation}
is required to preserve perturbativity. This constraint is well-known and can be equivalently derived in various ways, such as by requiring unitarity of scattering amplitudes~\cite{Falkowski:2019tft,Cohen:2021ucp}.
Note that Eq.~\eqref{eq.HEFT_v_inequality} is in complete contrast with SMEFT, which by design expands in $(4\pi v/\Lambda) < 1$.
Introducing $v$ as a suppression scale also causes it to appear most often at denominators, which naturally leads to positive powers of $(\Lambda/4\pi v)\leq1$ appearing in observable predictions, as opposed to powers of $(4\pi v/\Lambda)$ which are ubiquitous in SMEFT.
In practice, the results of Section~\ref{sec.pc} would apparently allow an expansion of observables in $(\Lambda/4\pi v)\ll1$. However, this condition clearly doesn't match the phenomenological assumptions we make when working with HEFT,\footnote{Nor is it satisfied in other known theories whose fields naturally come in non-linear representations, such as chiral perturbation theory for pions, where $4\pi f_{\pi}\approx \Lambda_{\rm QCD}$. 
} 
so one is left with two options: 
\begin{enumerate}[label=(\alph*)]
    \item assume that $(4\pi v/\Lambda)\approx 1$ and avoid expanding in this quantity
    \label{option.v}
    \item[\emph{or}]
    \item disentangle the scale $v$ generating masses for SM particles from the scale $f$ suppressing scalars insertions. 
    In this way, one is still left with the constraint $(4\pi f/\Lambda)\geq 1$ but ${\sqrt{\xi} = v/f}$ can be a small quantity, that one could expand in.
    As reviewed in Section~\ref{sec.HEFT}, $\xi$ has been often introduced in the literature, mainly inspired by the matching to composite Higgs models. Here we see that, as an additional motivation, it allows to push the HEFT cutoff scale $\Lambda$ to values higher than $4\pi v\simeq 3$~TeV.
    \label{option.xi}
\end{enumerate}

Option~\ref{option.xi} clearly leads to a non-minimal version of the HEFT Lagrangian, that contains two independent scales $v,f$: both have the dimensions of a scalar field, but they must be treated differently by the power counting. This requires a new set of rules that dictate which scale should appear in a certain HEFT interaction. Because this adds a new layer of complexity, we will address first option~\ref{option.v} and we will come back to~\ref{option.xi} in Section~\ref{sec.xi}.

\subsection{Option (a): $v$ as the only mass-dimensionful parameter}
\label{sec.HEFT_PC_a}
In this section we assume that the only dimensionful parameters in the HEFT Lagrangian are $g,g',g_s,y_\psi,\lambda,v$, and in particular $v$ is the only explicit mass scale present besides $\Lambda$. With this in mind, we go back to Eq.~\eqref{eq.delta_HEFT}. As noted when discussing the SMEFT, powers of $\delta_{\rm SMEFT}=\delta_{\rm HEFT}$ can be counted either as $(\xsLamP + N_v)$ or as $(\xsLamP + N_g + N_{g'} + N_{y_\psi} + 2N_{\lambda})$. While SMEFT is designed to pick the former option, HEFT can only adopt the latter.
As a result, the HEFT and perturbative series are tied together, and observable contributions should be sorted by
\begin{equation}
\label{eq.HEFT_pc}
\framebox{$
  N_{\rm HEFT} \equiv  \xsLamP + N_g + N_{g'} + N_{y_\psi} + 2N_{\lambda}
$}
\end{equation}
Note that because $SU(3)_c$  is an unbroken symmetry and there are no masses given by $g_s v$, the perturbative expansion in $(g_s/4\pi)$ is not subject to the mass-momentum equivalence constraint and can remain independent.\footnote{One could wonder if the combination of $g,g'$ corresponding to the electromagnetic constant $e$ should also be free of the mass-momentum equivalence constraint. However, this is not the case: it is not possible to rotate $(g,g')$ into a pair $(e,G)$ such that only $G$ is associated to the electroweak masses. Equivalently, at LO in the SM, the electromagnetic constant  is
$e = (2m_W/v)\sqrt{1-m_W^2/m_Z^2} = (2m_W/v)\sin\theta$. This expression gives $e\sim (m/v)$, so $(e/4\pi)\sim (m/4\pi v)\sim (p/\Lambda)(\Lambda/4\pi v)\sim (p/\Lambda)$ in the scenario considered. }

\subsubsection{Power counting for HEFT amplitudes}\label{sec.HEFT_pc_a_amplitudes}
As pointed out at the end of Section~\ref{sec.Lag_to_obs}, the sorting criteria for diagrams is necessarily the trivial transposition of the rule for observables, \ie\ HEFT diagrams should be sorted by 
\begin{equation}
\label{eq.HEFT_pc_M}
  N_{\rm HEFT}^\meLabel \equiv  \meCount_{\Lambda,\meLabel}^p + N_{g,\meLabel} + N_{g',\meLabel} + N_{y_\psi,\meLabel} + 2N_{\lambda,\meLabel}\,,
\end{equation}
such that the observable contribution given by a product $\mathcal{M}_a\mathcal{M}_b^\dag$ has order
\begin{align}
    N_{\rm HEFT} =  N_{\rm HEFT}^a +  N_{\rm HEFT}^b\,.
\end{align}
In the next paragraphs, we will sometimes drop the $\meLabel$ label and use $N_{\rm HEFT},N_g$ etc. to indicate the countings for either diagrams or observables. Their meaning can always be inferred from the context. 

\begin{figure}[t] \centering
    \begin{subfigure}[t]{0.29\textwidth}
        \centering
        \captionsetup{labelformat=empty}
        \includegraphics[width=\textwidth]{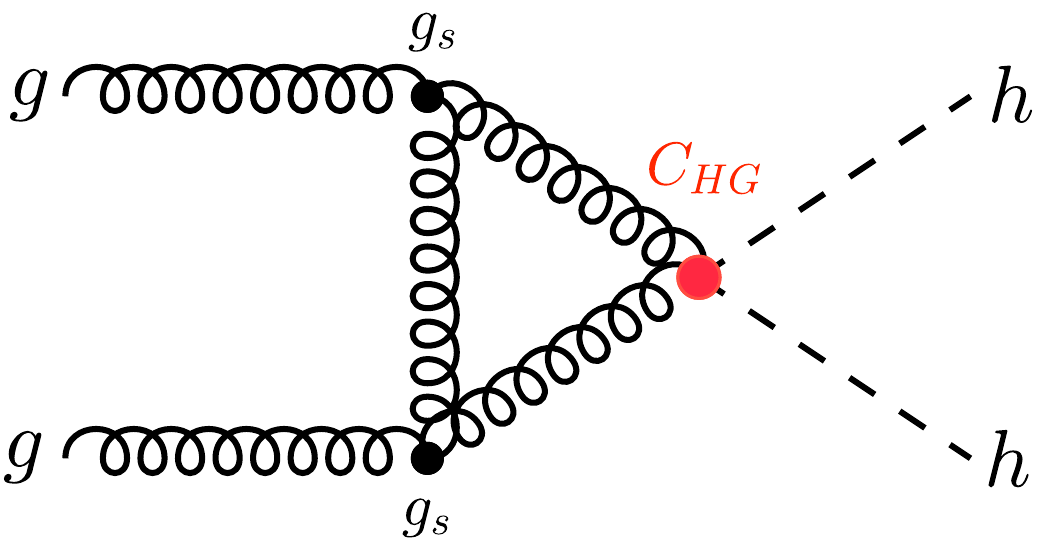}
        \caption{A1}
        \label{fig:gghh1}
    \end{subfigure}\hfill
    \begin{subfigure}[t]{0.31\textwidth}
        \centering
        \captionsetup{labelformat=empty}
        \includegraphics[width=\textwidth]{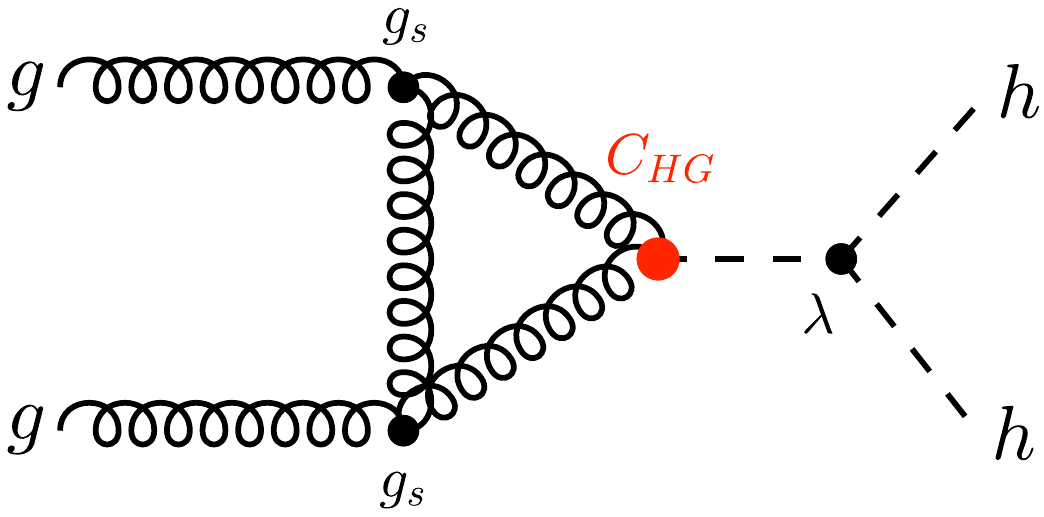}
        \caption{A2}
        \label{fig:gghh2}
    \end{subfigure}\hfill
    \begin{subfigure}[t]{0.27\textwidth}
        \centering
        \captionsetup{labelformat=empty}
        \raisebox{0.25em}{\includegraphics[width=\textwidth]{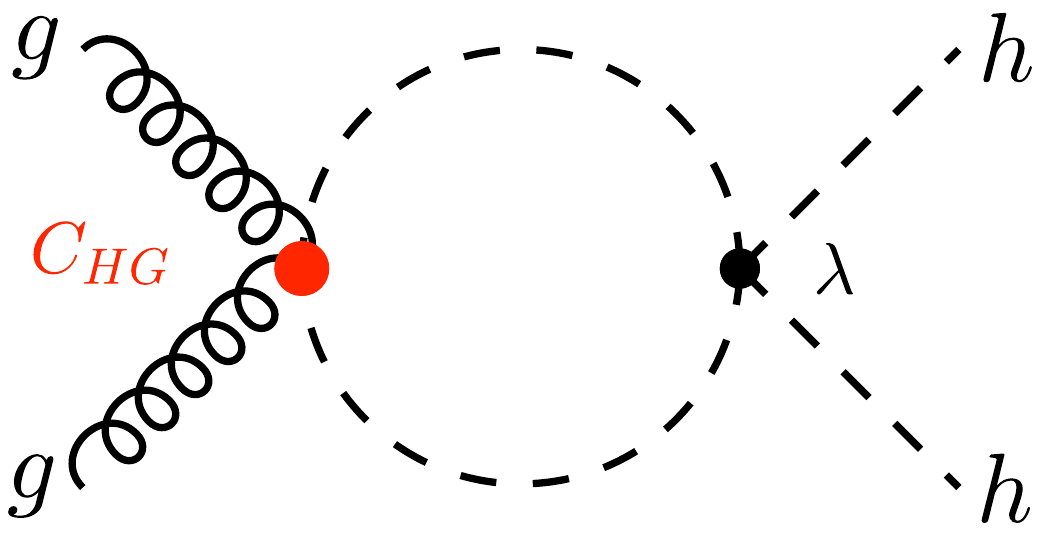}}
        \caption{A3}
        \label{fig:gghh3}
    \end{subfigure}

    \caption{One-loop diagrams contributing to $gg \rightarrow hh$ that contain a $C_{HG}$ coupling.}
    \label{fig.gghh_HEFT_diagrams_scaling}
\end{figure}

The counting rule in Eqs.~\eqref{eq.HEFT_pc} and \eqref{eq.HEFT_pc_M} enjoys an interesting and important property: the individual terms appearing in the sum are inhomogeneous over different diagrams contributing to the same process and containing the same number of insertions of a given operator (or set of operators). However, upon inspection, one finds that almost all inhomogeneities cancel out in the sum. For instance, consider the 1-loop diagrams contributing to $gg\to hh$ shown in Fig.~\ref{fig.gghh_HEFT_diagrams_scaling}.  All of them contain one insertion of the operator $\cP_{HG} = G_{\mu\nu}^A G^{A\mu\nu} (h/v)\,\F_{HG}(h)$. They scale as follows: 
\begin{align}
A_1 &\propto g_s^2\,C_{HG}\,\frac{p^2}{v^2}  
&
&\to&
\meCount_\Lambda^p &= 2\,,
N_\lambda=0\,,
N_{g_s} =2\,,
&
N_{\rm HEFT}&=2\,,
\\[1mm]
A_2 &\propto g_s^2\, C_{HG} \,\lambda
&
&\to&
\meCount_\Lambda^p &= 0\,,
N_\lambda=1\,,
N_{g_s} =2\,,
&
N_{\rm HEFT}&=2\,,
\\
A_3 &\propto C_{HG} \lambda\,\frac{p^2}{v^2} 
&
&\to&
\meCount_\Lambda^p &= 2\,,
N_\lambda=1\,,
N_{g_s} =0\,,
&
N_{\rm HEFT}&=4\,.
\end{align}

Although none of the countings shown in these equations is the same for all three contributions, it is not hard to see that $N_{\rm HEFT} + N_{g_s} = 4$ is constant.
This behavior generalizes to any operator and process, and it can be understood noting that, as long as the HEFT operators are NDA-normalized and the only dimensionful constants are those listed above, then:
\begin{align}
\label{eq.NHEFT_to Nchi_xs}
N_{\rm HEFT} + N_{g_s}
 \;&=\; \xsLamP+\xsChiG 
 \;=\; \xsCount_\chi - \xsFpP 
 \;=\; \xsCount_\chi + 2(n-2) \, ,
 \\
N_{\rm HEFT}^\meLabel + N_{g_s,\meLabel}
 \;&=\; \meCount_{\Lambda,\meLabel}^p+\meCount_{\chi,\meLabel}^\gCoupl
 \;=\; \meCount_{\chi,\meLabel} - \meCount_{4\pi,\meLabel}^p
 \;=\; \meCount_{\chi,\meLabel} + (n-2) \, , 
 \label{eq.NHEFT_to Nchi_me}
\end{align}
where we have explicitly distinguished the observable and amplitude cases, since they differ by the last term. $n$ is the total number of external legs, and $\xsCount_\chi$ ($\meCount_{\chi,\meLabel}$) is the total chiral counting for the cross section contribution (amplitude). We used that $\xsChiP=\xsLamP+\xsFpP$, plus Eqs.~\eqref{eq.other_alpha_defs},~\eqref{eq.alphafp_vs_n} and~\eqref{eq.alphachi_vs_n}, and the analogous relations for the countings in scattering amplitudes. Using Eqs.~\eqref{eq.me_NChiP_def} and~\eqref{eq.me_NG_def}, $\xsCount_\chi$ and $\meCount_{\chi,\meLabel}$ can be computed as
\begin{align}
\xsCount_\chi&=
\meCount_{\chi,a} + \meCount_{\chi,b}
\\[2mm]
\meCount_{\chi,\meLabel} &= -2 + P +\frac{F}{2} + \meCount_{\chi,\meLabel}^{\gCoupl} = 2L + \sum_{i\in {\rm vert}} N_{\chi,i}
\label{eq.alphachi_vs_vertices}
\end{align} 
In Section~\ref{sec.chiraldim} we noted that $N_{\chi,i}$ is always homogeneous across all the interaction vertices stemming from the same gauge-invariant operator and, in particular, it vanishes for all SM interactions. Therefore, Eq.~\eqref{eq.alphachi_vs_vertices} guarantees that $\meCount_{\chi,\meLabel}$ always takes the same value for all diagrams contributing to the same process at the same loop order and with the same operator insertions. In the example above, $\cP_{HG}$ has $N_{\chi,i}=0$. Therefore, for all diagrams shown,  $\meCount_{\chi,\meLabel}=2$ and $N_{\rm HEFT}^\meLabel + N_{g_s,\meLabel} = \meCount_{\chi,\meLabel} + n-2 = 4$.

This result tells us that if we choose to treat $g_s$ on the same footing as the other SM couplings and define an alternative HEFT power counting rule 
\begin{equation}
\label{eq.HEFTs_pc}
\framebox{$N_{\rm HEFT}^s = N_{\rm HEFT} + N_{g_s}$}\,,
\end{equation}
we will find that all contributions from the same gauge-invariant HEFT operator to a fixed process and at the same loop order conveniently share the same $N_{\rm HEFT}^s$.  Moreover, the counting rule for diagrams will be particularly simple and it will depend only on the $N_{\chi,i}$ of the vertices, which is the same as the $N_\chi$ of the operator they stem from, and on the loop order $L$ of the diagram: 
\begin{equation}
\framebox{$\displaystyle
N_{\rm HEFT}^{s,\meLabel} = n-2 + 2L + \sum_{i\in{\rm vert}} N_{\chi,i} $}\,.
\label{eq.NHEFT_to_Nchi}
\end{equation}
If we choose to count in $N_{\rm HEFT}$, on the other hand, we can write, for a diagram $\meLabel$:
\begin{align}
\framebox{$\displaystyle
N_{\rm HEFT}^\meLabel 
\;=\;
N_{\rm HEFT}^{s,\meLabel} - N_{g_s,\meLabel}
\;=\;
n-2+2L + \sum_{i\in {\rm vert}}(N_{\chi,i}-N_{g_s,i}) $}\,,
\label{eq.NHEFT_to_NchiminGs}
\end{align} 
where it is manifest from the sum over $(N_{\chi,i}-N_{g_s,i})$ that, for fixed $L$ and $n$, the same operator will lead to diagrams with different values of $N_{\rm HEFT}^\meLabel$ whenever vertices with different powers of $g_s$ are inserted. 

For the rest of this section, we will consider both $N_{\rm HEFT}$ and $N_{\rm HEFT}^s$ as alternative options, presenting our results in both cases. 
In practice, the "minimal" HEFT power counting for diagrams and observables, obtained by requiring consistency with dimensional analysis and mass-momentum equivalence, is given by $N_{\rm HEFT}$ in Eq.~\eqref{eq.HEFT_pc}. This rule ties together the EFT and perturbative series in all SM parameters except $g_s$, leaving us with two independent expansions: the EFT one, which counts equivalently EFT insertions and EW loops, and the perturbative QCD one, which can be organized independently. 
The expansion in $N_{\rm HEFT}^s$ is a specific linear combination of these two that, on one hand, gives simpler  counting rules for scattering amplitudes, but on the other, is less general, because it leaves us with less freedom, interpreting QCD corrections as increasing the HEFT order.

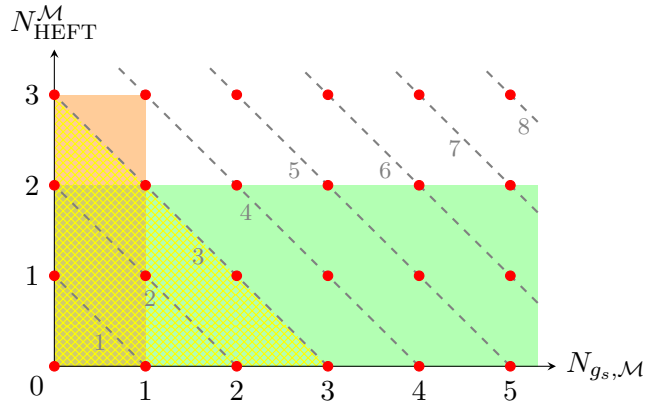
\begin{figure}[t]\centering
\begin{tikzpicture}[>=stealth, scale=1.2]
    \filldraw[fill=green!30, draw=none]
    (0,0) -- (5.3,0) -- (5.3,2) -- (0,2) -- cycle;

    \filldraw[fill=orange, fill opacity=0.4, draw=none] 
     (0,0) -- (1,0) -- (1,3) -- (0,3) -- cycle;

    \filldraw[pattern color=yellow, pattern=crosshatch,  draw=none]
    (0,0) -- (0,3) -- (3,0) -- cycle;
    
    \draw[->] (0,0) -- (5.5,0) node[right] {$N_{g_s,\mathcal{M}}$};
    \draw[->] (0,0) -- (0,3.5) node[above] {$N^{\mathcal{M}}_{\mathrm{HEFT}}$};

    \foreach \x in {1,2,3,4,5}
        \draw (\x,0.05) -- (\x,-0.05) node[below] {\x};

    \foreach \y in {1,2,3}
        \draw (0.05,\y) -- (-0.05,\y) node[left] {\y};
        
    \node[below left] at (0,0) {0};

    \draw[gray, thick, dashed] (1,0) -- (0,1);
    \draw[gray, thick, dashed] (2,0) -- (0,2);
    \draw[gray, thick, dashed] (3,0) -- (0,3);
    \draw[gray, thick, dashed] (4,0) -- (0.7,3.3);
    \draw[gray, thick, dashed] (5,0) -- (1.7,3.3);
    \draw[gray, thick, dashed] (5.3,0.7) -- (2.7,3.3);
    \draw[gray, thick, dashed] (5.3,1.7) -- (3.7,3.3);
    \draw[gray, thick, dashed] (5.3,2.7) -- (4.7,3.3);

    \filldraw[red] (0,0) circle (1.5pt);
    \filldraw[red] (0,1) circle (1.5pt);
    \filldraw[red] (0,2) circle (1.5pt);
    \filldraw[red] (0,3) circle (1.5pt);
    \filldraw[red] (1,0) circle (1.5pt);
    \filldraw[red] (1,1) circle (1.5pt);
    \filldraw[red] (1,2) circle (1.5pt);
    \filldraw[red] (1,3) circle (1.5pt);
    \filldraw[red] (2,0) circle (1.5pt);
    \filldraw[red] (2,1) circle (1.5pt);
    \filldraw[red] (2,2) circle (1.5pt);
    \filldraw[red] (2,3) circle (1.5pt);
    \filldraw[red] (3,0) circle (1.5pt);
    \filldraw[red] (3,1) circle (1.5pt);
    \filldraw[red] (3,2) circle (1.5pt);
    \filldraw[red] (3,3) circle (1.5pt);

    \filldraw[red] (4,0) circle (1.5pt);
    \filldraw[red] (4,1) circle (1.5pt);
    \filldraw[red] (4,2) circle (1.5pt);
    \filldraw[red] (4,3) circle (1.5pt);
    \filldraw[red] (5,0) circle (1.5pt);
    \filldraw[red] (5,1) circle (1.5pt);
    \filldraw[red] (5,2) circle (1.5pt);
    \filldraw[red] (5,3) circle (1.5pt);
    
    {\footnotesize
    \node[gray] at (0.5,0.27) {1};
    \node[gray] at (1.05,0.75) {2};
    \node[gray] at (1.58,1.22) {3};
    \node[gray] at (2.1,1.70) {4};
    \node[gray] at (2.63,2.17) {5};
    \node[gray] at (3.63,2.17) {6};
    \node[gray] at (4.39,2.41) {7};
    \node[gray] at (5.15,2.65) {8};
    }
\end{tikzpicture}
\caption{
HEFT scattering amplitudes can be classified according to a double expansion in QCD ($N_{g_s,\meLabel}$) and $N_{\rm HEFT}^\meLabel$ orders, which can take discrete values represented with red dots. The grey dashed lines are curves of constant $N_{\rm HEFT}^{s\meLabel} = N_{\rm HEFT}^\meLabel+N_{g_s,\meLabel}$, which can be used as an alternative counting. Truncating a calculation at a certain $N_{\rm HEFT}^{s\meLabel}$ selects values within a triangular region, as shown in yellow. Truncating in $N_{\rm HEFT}^\meLabel$ selects a horizontal band (green), while truncating simultaneously in $N_{\rm HEFT}^\meLabel$ and $N_{g_s,\meLabel}$ identifies a rectangular region (orange).
}
\label{fig.M_orders_scheme}
\end{figure}

Fig.~\ref{fig.M_orders_scheme} shows a graphical representation of the classification of HEFT diagrams along the $N_{\rm HEFT}^{\meLabel},N_{g_s,\meLabel}$ axes. These quantities can only take a discrete set of values\footnote{Fig.~\ref{fig.M_orders_scheme} additionally assumes that only integer powers of the coupling constants appear in the Lagrangian, such that the allowed values for $N_{\rm HEFT}^{(s)\meLabel}, N_{g_s,\meLabel}$ must also be integers. This assumption is not strictly required (nothing forbids operator pre-factors containing \eg\
$\sqrt{y_\psi}$) but it is stable under renormalization and consistent with usual conventions.}
such that $N_{\rm HEFT}^\meLabel\geq 0$, as we will prove in Section~\ref{sec.HEFT_counting_operators}.
The order $N_{\rm HEFT}^{s,\meLabel}$ is given by the linear combination along the diagonal, marked by the gray dashed lines. 
Organizing the HEFT expansion based on $N_{\rm HEFT}^{s}$ corresponds to retaining all diagrams falling within one of these diagonals, as shown by the yellow hatched region, while organizing it based on $N_{\rm HEFT}$ corresponds to retaining those below a horizontal cut, as in the green shaded region. The latter technically extends indefinitely to the right, representing the fact that arbitrarily-high-order QCD corrections could be retained. 
In a realistic calculation, though, one will need to truncate the QCD series as well. While with $N_{\rm HEFT}^{s}$ counting the treatment of the QCD series is fixed,  when adopting $N_{\rm HEFT}$, the $N_{g_s}$ cut can be picked independently, resulting in a rectangular selection, as depicted by the orange shaded region. 

The order $N_{\rm HEFT}^{s,\meLabel}$ defined in Eq.~\eqref{eq.NHEFT_to_Nchi} is related to the chiral dimension for scattering amplitudes $[\mathcal{D}]_c$ defined in Refs.~\cite{Buchalla:2013eza,Buchalla:2016sop} by $[\mathcal{D}]_c = N_{\rm HEFT}^{s,\meLabel} + 4-n = N_{\chi,\meLabel}+2$.\footnote{
This equation is a direct generalization to HEFT of the relation identified in Eq.~\eqref{eq.D_vs_alpha} by comparing the total powers of $(p/\Lambda)$ with the derivative counting of Ref.~\cite{Weinberg:1978kz} for chiral perturbation theory.
}
An important aspect we would like to stress is that our "master formulas" Eqs.~\eqref{eq.NHEFT_to_Nchi},~\eqref{eq.NHEFT_to_NchiminGs}  fundamentally emerge by requiring a counting in $N_{\Lambda,\meLabel}^p$ as proposed in Ref.~\cite{Gavela:2016bzc}, and in fact they bear a connection to the primary dimension $d_p$ as well. The relevant relation to understand it is: 
\begin{align}
    N_{\Lambda,\meLabel}^p &= \sum_{i\in\text{vert}} N_{\Lambda,i}^p 
    \label{eq.NLamp_sum}\\
    &= N_{\chi,\meLabel}^p +n-2 \,=\, 2L+\sum_{i\in\text{vert}} N_{\chi,i}^p+n-2\,.
    \label{eq.NLamp_to_Nchip}
\end{align}
The first line tells us that we could have equivalently calculated the first component of $N_{\rm HEFT}^{(s),\meLabel}$ by summing the 
$N_{\Lambda,i}^p$ of each vertex, \ie\ the canonical dimension (carried only by fields and derivatives) of the corresponding interactions, see Eq.~\eqref{eq.Nip_lag}. This choice would have given correct results, but it comes with the drawback that each operator contains potentially infinite interactions, all with different $N_{\Lambda,i}^p$. The idea of the primary dimension is to take\footnote{We use a $\simeq$ here to note that the definition of $d_p$ in~\cite{Gavela:2016bzc} requires some further considerations being made on the structure of HEFT operators, and in particular on the expression for $\V_\mu$.} $d_p\simeq \min_i N_{\Lambda,i}^p+4$ as a sorting tool for HEFT operators, precisely to cure this issue.
Eq.~\eqref{eq.NLamp_to_Nchip} offers an alternative (but equivalent!) solution, namely to switch to $N_{\chi,i}^p$ at the cost of introducing an explicit dependence on the number of loops and external legs $n$. In principle, $N_{\chi,i}^p$ still suffers from issues of inhomogeneity over an operator's interactions: opposite to the behavior of $N_{\Lambda,i}^p$, it is not sensitive to scalar insertions, but it varies across the terms obtained expanding covariant derivatives and field strengths. However, this problem is cured by the fact that $N_{\rm HEFT}^{(s)}$ counts gauge couplings as well. 

In summary, the HEFT order of scattering amplitudes tends to depend on properties of the individual interaction terms that do not generalize to full gauge invariant operators. This is a fundamental issue that leads to challenges in the sorting of the operator series. The expansion in $N_{\rm HEFT}^s$ is a special case, in which it is possible to identify the chiral dimension $N_\chi$ as a unique classification criterion. The dependence of $N_{\rm HEFT}^{(s),\meLabel}$ on $n$ can be interpreted as a consequence of the physical expansion really being in $N_{\Lambda,\meLabel}^p$, which is sensitive to the number of field insertions.

To our knowledge, the expressions in Eqs.~\eqref{eq.NHEFT_to_Nchi},~\eqref{eq.NHEFT_to_NchiminGs} have not been proposed before. 
As our analysis borrowed several well-known ingredients from previous studies, the formulas we founds are -- predictably -- similar to previous results. However, our approach (i) gives a specific physics motivation for the HEFT expansion to count powers of $(g/4\pi), (y/4\pi), (\lambda/16\pi^2)$, and (ii) it clarifies the interplay between chiral and primary dimensions, and their role in the classification of HEFT amplitudes and operators. This translates, in particular, into the fact that Eqs.~\eqref{eq.NHEFT_to_Nchi},~\eqref{eq.NHEFT_to_NchiminGs}  
explicitly depend on the number of external legs $n$ and that~\eqref{eq.NHEFT_to_NchiminGs}
treats differently the EW and QCD perturbative expansions. These are distinctive features of the power counting proposed here.

\subsubsection{Power counting for HEFT operators}\label{sec.HEFT_counting_operators}
Having discussed the HEFT power counting for observables and Feynman diagrams, we now turn to the organization of Lagrangian interactions. 
In doing so, we will relax the assumption that HEFT operators are purely NDA normalized, and allow their definition to potentially include (positive) powers of the dimensionful couplings $g,g',g_s,y_\psi,\lambda$. We will refer to operators that do not have coupling constants in their definition as "naked" and to those containing coupling constant factors as "dressed".

This generalization is necessary because, even if we define the HEFT Lagrangian to contain only "naked" operators $\Lag_{\rm HEFT}= w_i\, C_i \cP_i+\dots$ (with $w_i$ the NDA normalization factor), corrections to the Wilson coefficients proportional to powers of the SM couplings will always arise upon renormalization. For instance,  1-loop RGE equations  generally have a form $\dot{C}_i \sim (g/4\pi)^2 C_i$, which can be seen as introducing a dressed version of the operators: $\Lag_{\rm HEFT} = w_i C_i \cP_i + w_i^\prime C_i^\prime \,(g^2\cP_i)+\dots$.
Another context in which such factors can arise  is in the matching to a UV theory, see Sec.~\ref{sec.matching}. In some cases, the presence of SM coupling constants can even be required by certain assumptions on the UV or EFT structure.
For instance, requiring the UV completion to be weakly coupled implies that some operators can only receive one-loop matching contributions, or that operators containing gauge field strengths should be accompanied by powers of the corresponding coupling constants.
In practice we can think that, for each HEFT operator $\cP_i$, there is an infinite tower of operators $\cP_i, g^2\cP_i, g^4\cP_i\dots$  (and similarly with arbitrary products of all the SM couplings) that, in general, are all present simultaneously in $\Lag_{\rm HEFT}$.\footnote{An alternative way of framing this argument is to think of only one "naked" operator being present, and interpret the $g^2$ terms as corrections to its Wilson coefficient. This choice would require the introduction of multiple versions of the coupling constant entering a given vertex $C_i + C'_i + C''_i+\dots$ (rather than multiple versions of the vertex), each assigned to different perturbative and chiral order. The logical arguments would then be developed in a completely equivalent way.} Because adding powers of the SM couplings formally increases the $N_\chi$ of the operator and of the vertices it generates, the whole tower of options needs to be taken into account in the organization of $\Lag_{\rm HEFT}$. This is of course different from the SMEFT case, in which additions of SM couplings increase the perturbative order but not the SMEFT one.

An important point is that the introduction of these constant factors does not spoil the derivation of Eqs.~\eqref{eq.NHEFT_to_Nchi} and~\eqref{eq.NHEFT_to_NchiminGs}, which can still be used to compute diagram orders, with the understanding that $N_{\chi,i}, (N_{\chi,i}-N_{g_s,i})$ account for the pre-factor's contribution as well. We will not consider arbitrary pre-factors containing $v$ or introducing new mass-dimensionful constants (such as dimensionful Wilson coefficients): a discussion of those cases is deferred to Section~\ref{sec.heft_other_normalizations}.

\vskip 1em
Eqs.~\eqref{eq.NHEFT_to_Nchi} and~\eqref{eq.NHEFT_to_NchiminGs} tell us that there are two relevant quantities that determine the order at which a certain Lagrangian interaction will contribute:  $N_{\chi,i}$ and $(N_{\chi,i}-N_{g_s,i})$. Depending on whether observables are sorted by $N_{\rm HEFT}$ or $N_{\rm HEFT}^s$, one or both of them are relevant for the organization of the HEFT operators series.
We will examine first the $N_{\rm HEFT}^s$ case, which is simpler. 

\paragraph{Counting in $N_{\rm HEFT}^s$.}
As discussed above, $N_{\chi}$ conveniently takes the same value whether it is evaluated for a gauge-invariant operator or for any of the interaction terms obtained expanding out its covariant derivatives and field strengths. By definition, it is also insensitive to the addition of scalar fields and it counts identically insertions of derivatives ($D_\mu, \de_\mu$), $\V_\mu$ structures and field strengths ($X_{\mu\nu}$).  
Table~\ref{tab.HEFT_operators_orders} shows the value of $N_\chi$ for several classes of HEFT operators up to high orders in the expansion. These results agree with the chiral dimension assignment defined in Ref.~\cite{Buchalla:2013eza}, up to a $-2$ shift in the $N_\chi$ definition. They are also consistent with Table~II of Ref.~\cite{Gavela:2016bzc}, up to the same conventions shift and to the fact that we defined the SM Yukawa term with a constant $y$ in front. 

\begin{table}[t]\centering
\renewcommand{\arraystretch}{1.1}
\begin{tabular}[t]{>{$}p{3.5cm}<{$}|*1{>{$}c<{$}}}
\hline
\textbf{operator class} & N_\chi
\\\hline
\lambda v^4 \mathcal{V}& 0
\\
X_{\mu\nu}X^{\mu\nu} \F & 0
\\
\de_\mu h \de^\mu h & 0
\\
v^2 \tr(\V_\mu \V^\mu)\F& 0
\\
\bar\psi i \slashed{D}\psi& 0
\\
yv \,\bar\psi \U\psi \F& 0

\\\hline
\end{tabular}
~~~~
\begin{tabular}[t]{>{$}p{3.5cm}<{$}|*1{>{$}c<{$}}}
\hline
\textbf{operator class} & N_\chi
\\\hline
d^2\F & 0 
\\
d \psi^2 \F& 0 
\\
\psi^4 \F& 0 
\\\hdashline
d^3 \F& 1 
\\
d^2\psi^2 \F& 1 
\\
d \psi^4 \F& 1 
\\
\psi^6 \F& 1 
\\\hdashline
d^4 \F& 2 
\\
d^3 \psi^2 \F& 2 
\\
d^2 \psi^4 \F&2
\\
d \psi^6 \F& 2
\\
\psi^8 \F& 2
\\\hdashline
d^5 \F& 3
\\
d^4 \psi^2 \F& 3
\\
d^3 \psi^4 \F& 3
\\
d^2 \psi^6 \F& 3
\\
d \psi^8 \F& 3
\\
\psi^{10} \F& 3
\\
\hline   
\end{tabular}
\caption{Chiral orders $N_\chi$  for classes of HEFT operators. The terms in the left block are the explicit SM-like interactions from Eq.~\eqref{eq.lagrangian_HEFT_LO}. The terms in the right block are characterized schematically by their content in terms of fermionic fields and $d$ insertions, where $d$ stands for one among $D_\mu,
\de_\mu,X_{\mu\nu}$ or $\V_\mu$. The functions $\F, \mathcal{V}$ are understood to have the form in~\eqref{eq.general_flare_function},~\eqref{eq.Vh} and they are only included for completeness, as their presence has no impact on $N_\chi$. 
Insertions of the $\U$ field are not indicated as the counting is transparent to them. The operators on the right are understood to be defined as "naked", \ie\ with no dimensionful coupling constants in front. Dressing them with powers of $g,g',y_\psi,\lambda$ increases their $N_\chi$.
}\label{tab.HEFT_operators_orders}
\end{table}

If HEFT observables are expanded according to the $N_{\rm HEFT}^s$ counting, the operators can be simply sorted by $N_\chi$, and Eq.~\eqref{eq.NHEFT_to_Nchi} informs directly the selection of the operators contributing to diagrams of a certain order. 
In particular, to calculate the scattering amplitude for a process with $n$ external legs up to a certain order $N_{\rm HEFT}^s \leq N_{\max}^s$, we will need to consider diagrams with  
\begin{equation}
\label{eq.NHEFTs_L_truncation}
L\leq \frac{N_{\max}^s+2-n}{2}\,,
\end{equation}
and at each loop order, we will need to include HEFT operators with 
\begin{equation}
\label{eq.Nchi_truncation_Ldependent}
  N_\chi\leq N_{\max}^s - (2L+n-2)\,.
\end{equation}
The right-hand side is maximized for $L=0$, \ie\ the highest-$N_\chi$ operators contributing to the process are those entering tree-level diagrams and they have 
\begin{equation}
\label{eq.Nchi_truncation}
    N_\chi\leq N_{\max}^s + 2 - n\,.
\end{equation}
Eqs.~\eqref{eq.NHEFTs_L_truncation},~\eqref{eq.Nchi_truncation_Ldependent} can be easily derived from Eq.~\eqref{eq.NHEFT_to_Nchi}, considering that the number of loops is maximized for $\sum N_{\chi,i}=0$, while $N_{\chi,i}$ is maximized when there is one single vertex contributing.
For instance: to calculate a $2\to 2$ process up to $N_{\rm HEFT}^{s,\meLabel}=6$, we will need to include diagrams with the number of loops and operator insertions reported in Table~\ref{tab.insertions_2to2}.
\begin{table}[t]\centering
\begin{tabular}{c|ccccc|c}
$L$&
\parbox{1.5cm}{\centering $N_\chi=0$\\[-1mm] insertions}& 
\parbox{1.5cm}{\centering $N_\chi=1$\\[-1mm] insertions}& 
\parbox{1.5cm}{\centering $N_\chi=2$\\[-1mm] insertions}& 
\parbox{1.5cm}{\centering $N_\chi=3$\\[-1mm] insertions}& 
\parbox{1.5cm}{\centering $N_\chi=4$\\[-1mm] insertions}& $N_{\rm HEFT}^{s\meLabel}$
\\[3mm]\hline
0& any& -& -& -& -& 2
\\
\rowcolor{SkyBlue!20}
0& any& 1& -& -& -& 3
\\
0& any& 2& -& -& -& 4
\\
0& any& -& 1& -& -& 4
\\
1& any& -& -& -& -& 4
\\
\rowcolor{SkyBlue!20}
0& any& -& -& 1& -& 5
\\
\rowcolor{SkyBlue!20}
0& any& 1& 1& -& -& 5
\\
\rowcolor{SkyBlue!20}
1& any& 1& -& -& -& 5
\\
0& any& -& -& -& 1& 6
\\
0& any& -& 2& -& -& 6
\\
0& any& 1& -& 1& -& 6
\\
1& any& -& 1& -& -& 6
\\
1& any& 2& -& -& -& 6
\\
2& any& -& -& -& -& 6
\\\hline
\end{tabular}   
\caption{Characterization of the diagrams required to compute a $2\to 2$ process up to order $N_{\rm HEFT}^{s,\meLabel}=6$, in terms of HEFT operator insertions and number of  loops $L$. Each diagram can only contain up to $2L+2$ vertices (see App.~\ref{app.diagrammatics}).
This table can be easily generalized to processes with different multiplicities by identifying solutions to Eq.~\eqref{eq.NHEFT_to_Nchi}.} \label{tab.insertions_2to2}
\end{table}

The counting reported in Table~\ref{tab.HEFT_operators_orders} assumes that operators are defined as ``naked", \ie\ without any dimensionful constants in front. Dressing an operator with positive powers of $g,g',g_s,y_\psi,\lambda$ will increase its $N_\chi$ and therefore push it to a higher order. This unavoidable feature of the expansion in $N_{\rm HEFT}^s$ could be seen as introducing arbitrariness in the assignment of an operator's order. 
However, if we work under the assumption that only positive powers of the dimensionful constants can be inserted in the HEFT Lagrangian, then the values in Table~\ref{tab.HEFT_operators_orders} give unambiguously \emph{the minimum order} at which each HEFT operator can appear in the expansion.
This is enough to construct a physically meaningful EFT, in the spirit of providing maximally general predictions up to a chosen truncation order: retaining the lowest-order version of each operator in a non-redundant basis ensures that all potentially relevant operators will always be included in a calculation. 
With the assumptions made throughout this work,
the operator version with the lowest $N_\chi$ is always the "naked" $\cP_i$, unless additional assumptions are made on the UV completion, that require a minimum number of couplings to be present. In that case, the series could start \eg\ at $g^2 \cP_i$.

The assumption of \emph{only positive net powers of the SM coupling constants} being allowed in $\Lag_{\rm HEFT}$ is an indispensable condition for $N_\chi$ to admit a minimum value, which is crucial for $N_{\rm HEFT}^s$ to be a consistent power counting. This assumption can be seen as a convention on the definition of HEFT operators, that can be chosen safely: as shown in App.~\ref{app.diagrammatics}, once the HEFT fields and coupling constants are defined as described in Section~\ref{sec.formalism} and all HEFT operators are normalized according to this convention, no negative powers of $g,g',g_s,y_\psi,\lambda$ can be induced from loop corrections or UV-matching to HEFT operators.
It is possible for ratios of the coupling constants (such as $g'/g$) to appear, \eg\ through EW mixing. However, because these ratios are dimensionless, they leave $N_\chi$ unaffected and are not problematic for the power counting.

\paragraph{Counting in $N_{\rm HEFT}$.}
Let us now consider the expansion in $N_{\rm HEFT}$, rather than $N_{\rm HEFT}^s$. The value of 
$(N_{\chi,i}-N_{g_s,i})$ varies among the interaction terms induced by the same operator, and it can even become negative.
The maximum and minimum values that $(N_{\chi,i}-N_{g_s,i})$ can take can be determined based on the operator structure.
Let us write a generic gauge-invariant operator as
\begin{equation}
  \mathcal{O} \sim  \kappa \,\phi^s \psi^f \,(G_{\mu\nu})^{x_G} \,(W_{\mu\nu})^{x_W} (D_\mu)^q 
\end{equation}
where, similar to Eq.~\eqref{eq.generic_interaction}, $\psi$ represents any fermion, $\phi$ any scalar, $\kappa$ is a product of dimensionful constants, and $W_{\mu\nu}$ represents any of the electroweak field strengths. 
Knowing the explicit operator structure, we can further distinguish covariant derivatives acting on colored vs colorless fields: there will be $q_G$ of the former and $q_W$ of the latter.
Then, letting the index $i$ run over the interactions obtained expanding field strengths and covariant derivatives, we will have:
\begin{align}
\label{eq.NchiNgs_max}
\max_i(N_{\chi,i}-N_{g_s,i}) &= \bar N_{\chi} 
\quad\stackrel{\eqref{eq.Nchi_amplitude}}{=} \quad x_W + x_G + q + \frac{f}{2} - 2 + N_{\chi}^{\gCoupl} - N_{g_s}\,,
\\
\min_i(N_{\chi,i}-N_{g_s,i}) &= \bar N_{\chi}- x_G- q_G 
\; = \; x_W + q_W + \frac{f}{2} - 2 + N_{\chi}^{\gCoupl} - N_{g_s} \; \geq -2\,,
\label{eq.NchiNgs_min}
\end{align}
where $N_{g_s}$ are the overall powers of $g_s$ contained in $\kappa$, and $\bar N_\chi$ is essentially the $N_\chi$ of the gauge-invariant operator defined without powers of $g_s$ in front. To find the minimum, we considered that each $G_{\mu\nu}$ or $D_\mu$ acting on a colored field carries up to 1 power of~$g_s$. Eq.~\eqref{eq.NchiNgs_min} tells us that the minimum possible value of $(N_{\chi,i}-N_{g_s,i})$ for any HEFT interaction is $-2$. 
Moreover, interactions with $(N_{\chi,i}-N_{g_s,i})= -2 \;(-1)$ must contain at least 2 powers (1 power) of $g_s$ coming from a covariant derivative or gauge field strength.

Table~\ref{tab.HEFT_operators_orders_noGs} shows the values of $\min_i(N_{\chi,i}-N_{g_s,i})$ and $N_{\chi}$ for operator classes with $N_\chi\leq 2$. As will become clear in the next paragraphs, both numbers are relevant for the characterization of HEFT operators. Compared to Table~\ref{tab.HEFT_operators_orders}, the classes definition has been refined to distinguish between QCD-like (\ie\ $G_{\mu\nu}$ or $D_\mu$ acting on colored fields) and EW-like (all remaining cases) $d$ structures. 
Only operators with at least one $d_G$ are shown, as for the remaining ones $N_{g_s,i}\equiv 0$ and the counting is the same as in Table~\ref{tab.HEFT_operators_orders}.
\\ 
As an example, the chromo-magnetic operator $ G_{\mu\nu}^A\bar Q_L \U\sigma^{\mu\nu}T^A Q_R$ belongs to the class $d_G \psi^2$ and has $N_\chi=0$. Opening the field strength, we obtain a term proportional to $\de_\mu G_\nu$, which has $(N_{\chi,i}-N_{g_s,i})=0$, and a term proportional to $g_s G_\mu G_\nu$, which has $(N_{\chi,i}-N_{g_s,i})=-1$.
An example of operator in class $d_G^2\, d^2\, \F$ is $ G_{\mu\nu} G^{\mu}_\rho\de^\nu\de^\rho \F$. It has $N_\chi=2$, and $(N_{\chi,i}-N_{g_s,i})$ is lowest (0) for the vertex $g_s^2G_\mu^2 G_\nu G_\rho\de^\nu\de^\rho h$. 
Finally, the operator $(G_{\mu\nu})^3$, belonging to class $d_G^3\F$, has $N_\chi=1$. Among its interactions, $(N_{\chi,i}-N_{g_s,i})$ is minimum and equal to $-2$ for $g_s^3 G_\mu^6$. It is equal to $-1$ for $g_s^2(\de_\mu G_\nu)G^\mu G^\nu G_\rho^2$.

\begin{table}[t]\centering
\renewcommand{\arraystretch}{1.2}
\begin{tabular}{>{$}p{3.5cm}<{$}|*2{>{$}c<{$}}}
\hline
\textbf{operator class} & \min(N_{\chi,i}-N_{g_s,i})&   N_\chi
\\\hline
G_{\mu\nu}G^{\mu\nu} \F & -2& 0
\\
\bar Q i \slashed{D}Q& -1& 0
\\\hline
d_G\, \psi^2 \F& -1& 0
\\\hdashline
d_G^3\, \F& -2& 1
\\
d_G^2\, d_W\, \F& -1& 1
\\
d_G^2\,\psi^2 \F& -1& 1
\\
d_G\, d_W\,\psi^2 \F& 0& 1
\\
d_G\, d_W^2\, \F& 0& 1
\\
d_G\, \psi^4 \F& 0&  1
\\\hdashline
d_G^4\, \F& -2& 2
\\
d_G^3\, d_W\, \F& -1& 2
\\
d_G^3\, \psi^2 \F& -1& 2
\\
d_G^2\, d_W^2\, \F& 0& 2
\\
d_G^2\, \psi^4 \F& 0& 2
\\
d_G^2\, d_W\, \psi^2 \F& 0& 2
\\
d_G\, d_W^3\, \F& 1& 2
\\
d_G\, d_W^2\, \psi^2 \F& 1& 2
\\
d_G\, d_W\, \psi^4 \F& 1& 2
\\
d_G\, \psi^6 \F& 1& 2
\\\hline
\end{tabular}
\caption{Minimum values that $(N_{\chi,i}-N_{g_s,i})$ could take for the interactions generated by a HEFT operator with the generic structure indicated. $d_G$ stands for either $G_{\mu\nu}$ or a covariant derivative acting on colored fields, while $d_W$ stands for either $\de_\mu,\V_\mu$, electroweak field strengths $W_{\mu\nu},B_{\mu\nu}$ or covariant derivatives acting on colorless fields. The operators are understood to be defined as "naked", so $\max_i (N_{\chi,i}-N_{g_s,i})=N_\chi$, whose values are in agreement with the classification given in Table~\ref{tab.HEFT_operators_orders}. We only show classes up to $N_\chi=2$ and we omit classes that do not contain $d_G$. 
}\label{tab.HEFT_operators_orders_noGs}
\end{table}

\begin{figure}[t]
\hspace*{-5mm}
\parbox{.5\textwidth}{\centering\scalebox{.9}{ 
\begin{tikzpicture}[>=stealth, scale=1.1]
    
    \filldraw[fill=blue!10, draw=blue!10]
    (1,2) -- (3.3,2) -- (3.3,3) -- (0,3) -- cycle
    (0,3) -- (3.3,3) -- (3.3,3.3) -- (0,3.3) -- cycle;
    
    \draw[blue, ultra thick] (0,3) -- (1,2) -- (3.3, 2) node[right] {\footnotesize$\displaystyle \min_i(N_{\chi,i}-N_{g_s,i}) + 1$};

    \draw[->] (0,0) -- (3.5,0) node[right] {$N_{g_s, \mathcal{M}}$};
    \draw[->] (0,0) -- (0,3.5) node[above] {$N^{\mathcal{M}}_{\mathrm{HEFT}}$};

    \foreach \x in {1,2,3}
        \draw (\x,0.05) -- (\x,-0.05) node[below] {\x};

    \foreach \y in {1,2,3}
        \draw (0.05,\y) -- (-0.05,\y) node[left] {\y};
        
    \node[below left] at (0,0) {0};

    \draw[gray, thick,dashed] (1,0) -- (0,1);
    \draw[gray, thick,dashed] (2,0) -- (0,2);
    \draw[gray, thick,dashed] (3,0) -- (0,3);
    \draw[gray, thick,dashed] (3.3,0.7) -- (0.7,3.3);
    \draw[gray, thick,dashed] (3.3,1.7) -- (1.7,3.3);
    \draw[gray, thick,dashed] (3.3,2.7) -- (2.7,3.3);

    \filldraw[red] (0,0) circle (1.5pt);
    \filldraw[red] (0,1) circle (1.5pt);
    \filldraw[red] (0,2) circle (1.5pt);
    \filldraw[red] (0,3) circle (1.5pt);
    \filldraw[red] (1,0) circle (1.5pt);
    \filldraw[red] (1,1) circle (1.5pt);
    \filldraw[red] (1,2) circle (1.5pt);
    \filldraw[red] (1,3) circle (1.5pt);
    \filldraw[red] (2,0) circle (1.5pt);
    \filldraw[red] (2,1) circle (1.5pt);
    \filldraw[red] (2,2) circle (1.5pt);
    \filldraw[red] (2,3) circle (1.5pt);
    \filldraw[red] (3,0) circle (1.5pt);
    \filldraw[red] (3,1) circle (1.5pt);
    \filldraw[red] (3,2) circle (1.5pt);
    \filldraw[red] (3,3) circle (1.5pt);
     
     \node[blue] at (-1., 2.29) {\footnotesize $N_{\chi}+\mathfrak{L}$};
     \draw[->, blue, line width=0.5mm] (-.4, 2.4) -- (-0.05,2.9);
    
\end{tikzpicture}}}
\hfill
\parbox{.5\textwidth}{\centering\scalebox{.9}{ 
\begin{tikzpicture}[>=stealth, scale=1.2]    
    \filldraw[fill=gray!40, draw=none]
    (0,0) -- (2.33,2.33) -- (0,2.33) -- cycle;

    \filldraw[fill=gray!40, draw=none]
    (0,-2) -- (3.5,-2) -- (3.5,-2.33) -- (0,-2.33)-- cycle;

    \filldraw[fill=yellow!40, draw=none]
    (0,-2) -- (2,-2) -- (2,2) -- (0,0) -- cycle;

    \filldraw[pattern color=green!60, pattern=crosshatch, draw=none]
    (0,-2) -- (1,-2) -- (1,0) -- (0,0) -- cycle;

    \draw[gray, dashed,thick] (0,0) -- (2.33,2.33);
    \draw[gray, dashed, thick] (0,-1) -- (3.33,2.33);
    \draw[gray, dashed, thick] (0,-2) -- (3.5,1.5);
    \draw[gray, dashed, thick] (1,-2) -- (3.5,0.5);
    \draw[gray, dashed, thick] (2,-2) -- (3.5,-0.5);
    \draw[gray, dashed, thick] (3,-2) -- (3.5,-1.5);
    
    \draw[->] (0,0) -- (3.5,0) node[right] {$N_{\chi}$};
    \draw[->] (0,-2.5) -- (0,2.5) node[above] {$\displaystyle \min_{i}(N_{\chi, i} - N_{g_s, i})$};

    \foreach \x in {1,3}
        \draw (\x,0.05) -- (\x,-0.05) node[below] {\x};
    \node[] at (2.2,-0.26) {2};

    \foreach \y in {-2,-1,1,2}
        \draw (0.05,\y) -- (-0.05,\y) node[left] {$\y$};
        
    \node[below left] at (0,0) {0};

    \filldraw[blue] (0,-2) circle (1.5pt);
    \filldraw[blue] (1,-2) circle (1.5pt);
    \filldraw[blue] (2,-2) circle (1.5pt);
    \filldraw[blue] (3,-2) circle (1.5pt);
    \filldraw[blue] (0,-1) circle (1.5pt);
    \filldraw[blue] (1,-1) circle (1.5pt);
    \filldraw[blue] (2,-1) circle (1.5pt);
    \filldraw[blue] (3,-1) circle (1.5pt);
    \filldraw[blue] (0,0) circle (1.5pt);
    \filldraw[blue] (1,0) circle (1.5pt);
    \filldraw[blue] (2,0) circle (1.5pt);
    \filldraw[blue] (3,0) circle (1.5pt);
    \filldraw[blue] (1,1) circle (1.5pt);
    \filldraw[blue] (2,1) circle (1.5pt);
    \filldraw[blue] (3,1) circle (1.5pt);
    \filldraw[blue] (2,2) circle (1.5pt);
    \filldraw[blue] (3,2) circle (1.5pt);

    \draw[red, ultra thick] (2,2) -- (2,-2.5);
    \node[red] at (2.2, -2.8) {\footnotesize $N^s_{\max} -\mathfrak{L}$};
    \draw[->, red, line width=0.5mm] (2,-2.2) -- (1.5,-2.2);

    \draw[green!50!black, ultra thick] (1,0) -- (1,-2.5);
    \node[green!50!black] at (.3, -2.8) {\footnotesize $N_{\max} +2 O_{\alpha_s}-\mathfrak{L}$};
    \draw[->, green!50!black, line width=0.5mm] (1,-2.2) -- (.5,-2.2);=

   \draw[green!50!black, ultra thick] (-0.7,0) -- (1,0);
   \node[green!50!black] at (-1.8, -.1) {\parbox{2.4cm}{\flushright\footnotesize $\min(\,N_{\max}-1,$\\
    $N_{\max}+2 O_{\alpha_s}-\mathfrak{L})$}};
    \draw[->, green!50!black, line width=0.5mm] (-0.4,0) -- (-0.4,-0.5);
    
\end{tikzpicture}}}
\caption{Left: the blue shaded region highlights the values of $(N_{g_s,\meLabel}, N_{\rm HEFT}^\meLabel)$ achievable by diagrams containing insertions of an operator with fixed $N_\chi, \min_i(N_{\chi,i}-N_{g_s,i})$.
We use the shorthand notation $\mathfrak{L} = 2L+n-2$. \\
Right: HEFT operators can be characterized by their $N_\chi$ and $\min_i(N_{\chi,i}-N_{g_s,i})$, which take discrete values represented with blue dots. The gray shaded regions are forbidden. The grey dashed lines are curves of constant $\max_i N_{g_s,i}$. 
The yellow region shows how a cut on $N_{\rm HEFT}^{s,\meLabel}\leq N_{\max}^s$ at the amplitudes level projects onto the operators space. The green hatched region shows the same for a double cut in $N_{\rm HEFT}^\meLabel\leq N_{\max}$ and $N_{g_s,\meLabel}\leq 2 O_{\alpha_s}$.
If the $N_{g_s,\meLabel}$ cut was removed, the green region would extend indefinitely to the right.
}
\label{fig.O_orders_scheme}
 \end{figure}
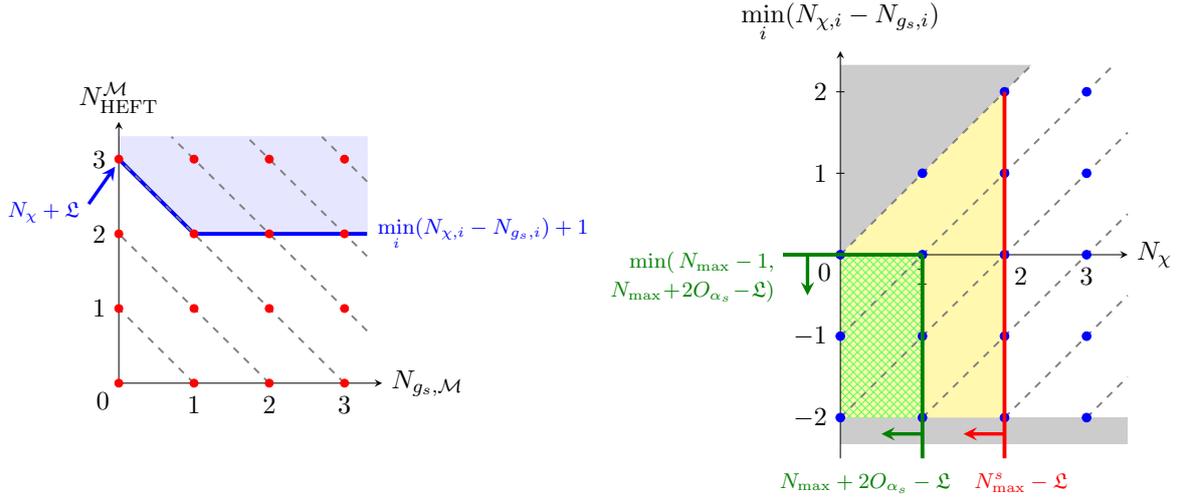

Fig.~\ref{fig.O_orders_scheme} (right) provides a graphic representation of the numerical values available for these quantities. With the conventions in which only positive, integer powers of the dimensionful coupling constants can appear in $\Lag_{\rm HEFT}$, it is possible to build operators sitting at any of the blue dots, which span the region $N_\chi\geq 0$ and $-2\leq \min_i(N_{\chi,i}-N_{g_s,i})\leq N_\chi$ at integer steps. 

\vskip 1em
Given that $(N_{\chi,i}-N_{g_s,i})$ can take negative values, we might worry that arbitrarily large negative values of $N_{\rm HEFT}^{\meLabel}$ could be reached, which would be disastrous for the HEFT power counting. Luckily, this is not the case. A hint comes from Eq.~\eqref{eq.HEFT_pc_M}, which indicates that $N_{\rm HEFT}^{\meLabel}<0$ would correspond to negative powers of $(p/\Lambda)$, $(g/4\pi)$ etc. appearing in the amplitude expression, which can hardly be the case with the assumptions made so far. In the following we derive an  
explicit proof using the expression of $N_{\rm HEFT}^{\meLabel}$ in Eq.~\eqref{eq.NHEFT_to_NchiminGs}.

From Eq.~\eqref{eq.NchiNgs_min} we see that interaction vertices with $(N_{\chi,i}-N_{g_s,i})=-2$ must have $N_{\chi}^\gCoupl=N_{g_s}$ and $x_W=q_W=f=0$, \ie\ they can only contain powers of $v, g_s$, scalars or vector bosons, but no power of $g,g',y_\psi,\lambda$, fermions or derivatives. 
If the LO HEFT Lagrangian is defined as in Eq.~\eqref{eq.lagrangian_HEFT_LO}, the only interactions satisfying these requests are the SM quartic vertex $g_s^2(G_\mu G_\nu)^2$, and higher-point interactions generated by other operators, that contain only gluons and Higgses, in arbitrary numbers $\geq 4$.
Electroweak gauge bosons cannot appear in these vertices, because their insertion is always accompanied either by a derivative (when coming from $X_{\mu\nu}$), or by a weak gauge coupling (when coming from $D_\mu, \V_\mu$ or non-abelian $gA_\mu A_\nu$ vertices in $X_{\mu\nu}$).
This leads us to the conclusion that vertices with $(N_{\chi,i}-N_{g_s,i})=-2$ must have at least 4 legs. Vertices with $(N_{\chi,i}-N_{g_s,i})=-1$, instead, can have as few as 3. SM examples are $g_s\de_\mu G_\nu G^\mu G^\nu$ and $g_s\bar\psi\gamma^\mu\psi G_\mu$.

If we want to minimize $N_{\rm HEFT}^\meLabel$ at fixed $n$ and $L$, we need to pick topologies with as many vertices as possible, and insert interactions with $(N_{\chi,i}-N_{g_s,i})=-2$ or $-1$ in all of them.
The maximum number of vertices in a diagram with $n$ external legs and $L$ loops is $V_{\max} = 2L+n-2$ (see App.~\ref{app.diagrammatics}), which is achieved when all vertices are 3-point functions. 
However, we have just established that vertices with $(N_{\chi,i}-N_{g_s,i})=-2$ have at least 4 legs. As shown in App.~\ref{app.diagrammatics}, if we insert $k$ interactions with 4 legs, then the total number of vertices in a diagram reduces to $V_{\max}-k$ at most. 
Inserting interactions with $(N_{\chi,i}-N_{g_s,i})=-1$ in all the remaining $V_{\max}-2k$ vertices yields:
\begin{align}
 N_{\rm HEFT}^\meLabel &= 2L+n-2 + \sum_{i\in{\rm vert}}(N_{\chi,i}-N_{g_s,i})   
 \\
 &\geq 2L+n-2 + (-2) \times k + (-1)\times (2L+n-2-2k) = 0\,.
\end{align}
Inserting vertices with 5 or more legs reduces the maximum number of vertices even further, and necessarily leads to diagrams with $N_{\rm HEFT}^\meLabel>0$.
This result proves that, whatever number of vertices with negative $N_{\chi,i}-N_{g_s,i}$ we insert in a diagram, in the end we will always find $N_{\rm HEFT}^\meLabel\geq0$, which is consistent with positive powers of $(p/\Lambda), (g/4\pi)$ etc. appearing in the amplitude. This result in turn ensures that in any HEFT amplitude $N_{\rm HEFT}^{s,\meLabel}\geq N_{g_s,\meLabel}$, and it justifies the representation in Fig.~\ref{fig.M_orders_scheme}.

\vskip 1em
We can now ask which values of $N_{\rm HEFT}^\meLabel, N_{g_s,\meLabel}$ a given operator can contribute to and, vice versa, how do cuts on the amplitudes orders project back onto operator orders.  In the  $N_{\rm HEFT}^s/N_\chi$ case discussed above, these questions had trivial answers. In the $N_{\rm HEFT}$ case there are more variables at play. 

Let us start from the first question: a HEFT operator is characterized by the pair of numbers $(N_\chi,\min_i(N_{\chi,i}-N_{g_s,i}))$, which can take values in the range discussed above. Inserting them into Eq.~\eqref{eq.NHEFT_to_Nchi} trivially gives 
\begin{align}
    N_{\rm HEFT}^{s,\meLabel} &\geq 2L+n-2 + N_\chi\,,
    \label{eq.NHEFTsmin_op}
\end{align}
for fixed $n, L$.
On the other hand, to obtain the minimum value of $N_{\rm HEFT}^\meLabel$ we need to consider Eq.~\eqref{eq.NHEFT_to_NchiminGs} for diagrams with exactly 1 insertion of the operator under study, and as many insertions as possible with negative values of $N_{\chi,i}-N_{g_s,i}$ in the remaining vertices, \ie\ $k$ insertions with $N_{\chi,i}-N_{g_s,i}=-2$ and $(V_{\max} -2k-1)$ with $N_{\chi,i}-N_{g_s,i}=-1$, which gives
\begin{equation}
\begin{split}
    N_{\rm HEFT}^\meLabel &\geq 2L+n-2 + \min_i(N_{\chi,i}-N_{g_s,i}) + (-2) k + (-1) (V_{\max}-2k-1)
    \\
    &= \min_i(N_{\chi,i}-N_{g_s,i}) + 1\,.
    \label{eq.NHEFTmin_op}
    \end{split}
\end{equation}
Both constraints~\eqref{eq.NHEFTsmin_op},~\eqref{eq.NHEFTmin_op} hold simultaneously and, because $N_{\rm HEFT}, N_{\rm HEFT}^s$ are not independent, they have a non-trivial intersection which is depicted in Fig.~\ref{fig.O_orders_scheme} (left). This figure reproduces the same diagram as Fig.~\ref{fig.M_orders_scheme}, superimposing a blue shaded region that highlights the values of $(N_{g_s,\meLabel}, N_{\rm HEFT}^\meLabel)$ which are accessible by a given operator. The condition~\eqref{eq.NHEFTsmin_op}, which depends only on the operator's $N_\chi$, cuts through the plane diagonally, while the condition~\eqref{eq.NHEFTmin_op}, which depends only on $\min_i(N_{\chi,i}-N_{g_s,i})$, cuts horizontally. 
Another interesting fact is that the diagram order accessible to operators with $\min_i(N_{\chi,i}-N_{g_s,i})=-2$ are exactly the same as those accessible to operators with the same $N_\chi$ and $\min_i(N_{\chi,i}-N_{g_s,i})=-1$, because for both Eq.~\eqref{eq.NHEFTmin_op} reduces to $N_{\rm HEFT}^\meLabel\geq0$.

Next, we consider how truncating the HEFT diagrams series to a certain 
\begin{align}
    N_{\rm HEFT}^\meLabel&\leq N_{\max}\,, 
    &
    N_{g_s,\meLabel}&\leq 2O_{\alpha_s} 
\end{align}
impacts the selection of HEFT operators.
First of all, it is easy to derive from Eq.~\eqref{eq.NHEFT_to_NchiminGs} that one will need to consider diagrams with 
\begin{align}
\label{eq.NHEFT_L_truncation}
    L&\leq \frac{N_{\max}+2-n}{2} + O_{\alpha_s}\,.
\end{align}
The number of loops clearly becomes unbounded if the cut in $N_{g_s}$ is removed, \ie\ $O_{\alpha_s}\to\infty$. 
At each loop order, we will need to include HEFT operators satisfying 
\begin{align}
    N_\chi &\leq N_{\max} + 2O_{\alpha_s} - (2L+n-2)\,,
    \label{eq.Nchimax_truncation}
    \\
    \min_i(N_{\chi,i}-N_{g_s,i}) &\leq N_{\max}-1\,,
    \label{eq.minNchiNgsmax_truncation}
\end{align}
where again the right-hand side of Eq.~\eqref{eq.Nchimax_truncation} is maximized for $L=0$, such that, overall, operators with  $N_\chi\leq N_{\max} + 2O_{\alpha_s} - (n-2)$ enter the process.\\
The bound in Eq.~\eqref{eq.Nchimax_truncation} can be derived from Eq.~\eqref{eq.Nchi_truncation_Ldependent}, considering that now
\begin{align}
    N_{\rm HEFT}^{s,\meLabel} &\leq N_{\max} +2O_{\alpha_s}\,.
\end{align}
The bound in Eq.~\eqref{eq.minNchiNgsmax_truncation} can be obtained reversing Eq.~\eqref{eq.NHEFTmin_op}. 

The green hatched region in Fig.~\ref{fig.O_orders_scheme} (right) shows the result in the $(N_\chi, \min_i(N_{\chi,i}-N_{g_s,i}))$ plane. Importantly, the constraint in Eq.~\eqref{eq.minNchiNgsmax_truncation} is only relevant if $(2L+n-2) < 2O_{\alpha_s}+1$. Otherwise, the strongest upper limit in the vertical direction is just given by the $\min_i(N_{\chi,i}-N_{g_s,i})\leq N_\chi$ boundary. The yellow shaded region shows for comparison the effect of a cut in $N_{\rm HEFT}^{s,\meLabel}\leq N^s_{\max}$, consistent with Eq.~\eqref{eq.Nchi_truncation_Ldependent}. 
For a correct interpretation of Fig.~\ref{fig.O_orders_scheme}, let us stress that the yellow and green regions in the right panel contain operator orders that \emph{can potentially} contribute to the relevant HEFT diagrams,  
and they represent the most restrictive operator selections one can apply at the Lagrangian level, based solely on the operator's $N_\chi$, $\min_i(N_{\chi,i}-N_{g_s,i})$. When performing a calculation for a specific process, Eqs.~\eqref{eq.NHEFT_to_Nchi} and~\eqref{eq.NHEFT_to_NchiminGs} will enforce further selections on the relevant diagrams and operators. In particular, even though the yellow and green regions might be identical, in general the final set of relevant operators will be different depending on whether amplitudes are truncated in $N_{\rm HEFT}^s$ or in $(N_{\rm HEFT}, N_{g_s})$, as a consequence of the selections shown in Fig.~\ref{fig.M_orders_scheme}. These effects are process- and operator-dependent, so they can only be evaluated on a case-by-case basis. 

\vskip 1em
Summarizing, the main result of this subsection is that, when $v$ is the only mass-dimensionful parameter present, HEFT operators can be classified according to their $N_\chi$ and $\min_i(N_{\chi,i}-N_{g_s,i})$. If observable predictions are truncated in $N_{\rm HEFT}^s$, only the former quantity is relevant for the truncation of $\Lag_{\rm HEFT}$, which follows Eq.~\eqref{eq.Nchi_truncation_Ldependent}. If predictions were truncated in $N_{\rm HEFT}$ retaining all-orders QCD corrections, then $\Lag_{\rm HEFT}$ should be ideally truncated in $\min_i(N_{\chi,i}-N_{g_s,i})$, which however contains infinite operators at each order. More realistically, a dual truncation of observable predictions in $N_{\rm HEFT}$ and $N_{g_s}\leq 2O_{\alpha_s}$ is consistent with a sequential truncation of $\Lag_{\rm HEFT}$: the operators can be first selected based on a cut in $N_\chi$ as shown in Eq.~\eqref{eq.Nchimax_truncation}, and then, if $(2L+n-1)\leq 2O_{\alpha_s}+1$, the selection can be further restricted as in Eq.~\eqref{eq.minNchiNgsmax_truncation}.
In the next section we look more systematically at algorithms for the construction of HEFT operator bases.

\subsubsection{HEFT basis reduction}\label{sec.HEFT_basis}
The construction of complete and non redundant EFT operator bases makes use of 
\begin{enumerate}[label=(\roman*)]
\item integration-by-parts (IBP) relations,
\item equation-of-motion (EOM) relations,
\item algebraic relations such as Fierz, Schouten or Bianchi identities.
\end{enumerate}
By construction, $N_\chi$ is always homogeneous across the various terms entering each of these relations, while $\min_i(N_{\chi,i}-N_{g_s,i})$ is not: for instance, IBP can relate operators of classes $d_G^2\,\F$ and $d_Gd_W \,\F$. 
If we consider "naked" operators, we find that, in fact, some relations connect operators with different $N_\chi$ as well: this happens whenever the relation contains terms that are multiplied by dimensionful coupling constants, \eg
\begin{align}
\label{eq.DD_X}
[D_\mu,D_\nu] &= i g W_{\mu\nu}^I \frac{\sigma^I}{2} + i g' B_{\mu\nu} \mathbf{h} + i g_s G_{\mu\nu}^A T^A \, ,
\end{align}
where some terms on the right-hand side might be absent depending on the charges of the field the covariant derivatives act on ($\mathbf{h}$ is the hypercharge). Another important example are EOMs, such as
\begin{align}
 i\slashed{D}Q_L &= \frac{v}{\sqrt2}\U\mathcal{Y}_Q(h) Q_R 
  \label{eq.eom_QL}
 \\
 &= \frac{v}{2\sqrt2} (\mathcal{Y}_U(h)+\mathcal{Y}_D(h))\U Q_R  + \frac{v}{2\sqrt2}(\mathcal{Y}_U(h)-\mathcal{Y}_D(h))\T\U Q_R \,,
 \label{eq.eom_QL_open}
 \\
 (D_\mu G^{\mu\nu})^A &= g_s \bar Q T^A \gamma^\nu Q\,.
 \label{eq.eom_G}
\end{align}
For instance, Eq.~\eqref{eq.eom_G} relates the operators
\begin{align}
\label{eq.eomG_ex_operators}
 \cP_1 &= (\bar Q \gamma_\nu T^A Q) (D_\mu G^{\mu\nu})^A  \F_1(h)\,,
 &
 \cP_2 &= (\bar Q \gamma_\mu T^A Q) (\bar Q \gamma^\mu T^A Q) \F_2(h)\,,
\end{align}
giving
\begin{equation}
\label{eq.eom_P1_P2}
    \cP_1 = g_s \cP_2\,,
\end{equation}
up to a re-definition of the coupling constants in $\F(h)$. Note that
\begin{align}
\big(N_\chi,\min_i(N_{\chi,i}-N_{g_s,i})\big)(\cP_1)&= (1,-1) \,,
\\
\big(N_\chi,\min_i(N_{\chi,i}-N_{g_s,i})\big)(\cP_2)&= (0,0) \,.
\end{align}

Thus, in general, the relations used in basis reduction connect operators belonging to different orders in the classification of Section~\ref{sec.HEFT_counting_operators}.
Disparities in the $N_\chi$ of the operators involved are a general feature of relations involving coupling constants, while disparities in $\min_i(N_{\chi,i}-N_{g_s,i})$ are even more widespread.
It is clear that potential inconsistencies due to these cross-order relations should be avoided in 
a HEFT basis reduction algorithm. 

Before entering the discussion in more detail, we note that the use of EOMs to reduce operator bases is not rigorous: technically, the basis reduction is actually performed via field redefinitions~\cite{Scherer:1994wi,Criado:2018sdb,Alonso:2025jvv}. We will use EOMs-like relations to illustrate the main argument of this section nonetheless, because they allow a simple visualization of the main properties we are interested in. Moreover, quantitative differences between the two methods only arise when tracing the exact relations among operators, while the mere basis definition can be safely addressed with either~\cite{Georgi:1991ch,Criado:2018sdb}.
A discussion of the role of field redefinitions is deferred to App.~\ref{app.redefinitions}, which shows, among other things, that additional (non SM-like) terms on the right-hand sides of Eqs.~\eqref{eq.eom_QL}--\eqref{eq.eom_G} should actually be accounted for.
Since these extra terms do not introduce qualitatively new features, they can be ignored for the moment.
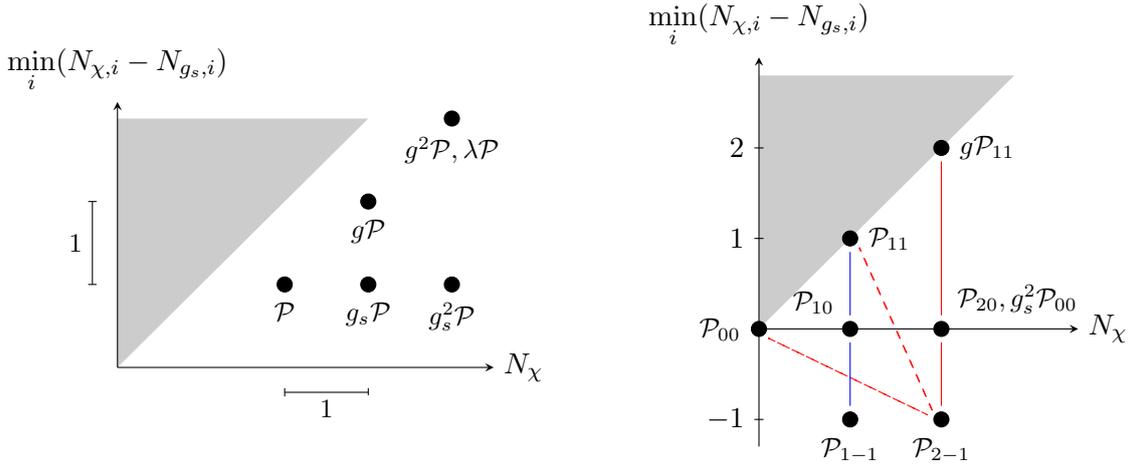
\begin{figure}[t]\centering
\parbox{.45\textwidth}{\centering
\begin{tikzpicture}[>=stealth, scale=1.1]    
    \filldraw[fill=gray!40, draw=none]
    (0,0) -- (3,3) -- (0,3) -- cycle;
    
    \draw[->] (0,0) -- (4.5,0) node[right] {$N_{\chi}$};
    \draw[->] (0,0) -- (0,3.2) node[above] {$\displaystyle \min_{i}(N_{\chi, i} - N_{g_s, i})$};

    \node[draw=black, fill=black, circle, inner sep=2pt, label=below:{\small $\cP$}] at (2,1) {};
    \node[draw=black, fill=black, circle, inner sep=2pt, label=below:{\small $g_s \cP$}] at (3,1) {};
    \node[draw=black, fill=black, circle, inner sep=2pt, label=below:{\small $g_s^2 \cP$}] at (4,1) {};
    \node[draw=black, fill=black, circle, inner sep=2pt, label=below:{\small $g \cP$}] at (3,2) {};
    \node[draw=black, fill=black, circle, inner sep=2pt, label=below:{\small $g^2 \cP, \lambda \cP$}] at (4,3) {};

    \draw (2, -0.3) -- (3, -0.3);
    \draw (2, -0.35) -- (2, -0.25);
    \draw (3, -0.335) -- (3, -0.25);
    \node[black] at (2.5, -0.5) {\small $1$};

    \draw (-0.3, 1) -- (-0.3, 2);
    \draw (-0.35, 1) -- (-0.25, 1);
    \draw (-0.35, 2) -- (-0.25, 2);
    \node[black] at (-0.5,1.5) {\small $1$};

\end{tikzpicture}}
\hfill
\parbox{.45\textwidth}{\centering
\begin{tikzpicture}[>=stealth, scale=1.2]    

    \filldraw[fill=gray!40, draw=none]
    (0,0) -- (2.8,2.8) -- (0,2.8) -- cycle;

    \foreach \y in {-1,1,2}
        \draw (0.05,\y) -- (-0.05,\y) node[left] {$\y$};

    \draw[->] (0,0) -- (3.5,0) node[right] {$N_{\chi}$};
    \draw[->] (0,-1.3) -- (0,3) node[above] {$\displaystyle \min_{i}(N_{\chi, i} - N_{g_s, i})$};

    \node[draw=black, fill=black, circle, inner sep=2pt, label=left:{\small $\cP_{00}$}] at (0,0) {};
    \node[draw=black, fill=black, circle, inner sep=2pt, label=below:{\small $\cP_{2-1}$}] at (2, -1) {};
    \node[draw=black, fill=black, circle, inner sep=2pt, label=below:{\small $\cP_{1-1}$}] at (1, -1) {};
    \node[draw=black, fill=black, circle, inner sep=2pt, label=right:{\small $g \cP_{11}$}] at (2, 2) {};
    \node[draw=black, fill=black, circle, inner sep=2pt, label=right:{\small $\cP_{11}$}] at (1, 1) {};
    \node[draw=black, fill=black, circle, inner sep=2pt, label=above left:{\small $\cP_{10}$}] at (1, 0) {};
    \node[draw=black, fill=black, circle, inner sep=2pt, label=above right:{\small $\cP_{20},g_s^2 \cP_{00}$}] at (2,0) {};

    \draw[red,dashed] (0.1, -0.1) -- (1.85, -0.95) -- cycle;
    \draw[red,dashed] (1.9, -0.9) -- (1.1, 0.9) -- cycle;

    \draw[blue] (1, -0.15) -- (1, -0.85) -- cycle;
    \draw[blue] (1, 0.85) -- (1, 0.15) -- cycle;
    
    \draw[red] (2, -0.15) -- (2, -0.85) -- cycle;
    \draw[red] (2, 0.15) -- (2, 1.85) -- cycle;
    
\end{tikzpicture}}
\caption{Left: visualization of how "dressing" an operator $\cP$ increases its $N_\chi$ and $\min_i(N_{\chi,i}-N_{g_s,i})$ orders. 
Right: graphic representation of the generic relations in Eqs.~\eqref{eq.generic_relation_1} (blue) and~\eqref{eq.generic_relation_2} (red). The solid lines trace the true relations, which involve dressed operators, while the dashed lines trace relations among naked operators.
}\label{fig.Morders_basis_reduction_examples}
\end{figure}

As derived in Section~\ref{sec.HEFT_counting_operators}, HEFT operators can be classified according to their $N_\chi$ and $\min_i(N_{\chi,i}-N_{g_s,i})$, and the prescriptions for $\Lag_{\rm HEFT}$ truncation act as vertical and/or horizontal upper cuts in the plane representing these two quantities, see the right panel of Fig.~\ref{fig.O_orders_scheme}. This suggests that a HEFT operator basis can be built proceeding order by order in $N_\chi$:  start by collecting all the possible naked invariants with $N_\chi=0$, reduce them to a minimal, non-redundant set, then move on to $N_\chi=1$, do the same while keeping the $N_\chi=0$ basis fixed, and so on.  During the reduction process at a certain $N_\chi$, one will generally meet
\begin{itemize}
    \item[(i)] relations among naked operators with same $N_\chi$ but diverse $\min_i(N_{\chi,i}-N_{g_s,i})$.\\[2mm]
    If $\cP_{ij}$ is a generic operator with $(N_\chi, \min_i(N_{\chi,i}-N_{g_s,i}))=(i,j)$, then such a relation can be of the form
    \begin{equation}
    \label{eq.generic_relation_1}
      \cP_{1-1} + \cP_{10} + \cP_{11} = 0\,.  
    \end{equation}
    \item[(ii)] relations connecting operators of the $N_\chi$ of interest to operators with lower $N_\chi$. The values of $\min_i(N_{\chi,i}-N_{g_s,i})$ can again be diverse across all the operators involved in the relation. For instance, at $N_\chi=2$ one can have a generic:
    \begin{equation}
    \label{eq.generic_relation_2}
     g \cP_{11} + g_s^2 \cP_{00}  + \cP_{2-1} + \cP_{20} =0\,.
    \end{equation}
\end{itemize}
Fig.~\ref{fig.Morders_basis_reduction_examples} (right) shows a graphical representation of relations~\eqref{eq.generic_relation_1} (blue lines) and~\eqref{eq.generic_relation_2} (red lines): $N_\chi$ is always homogeneous across all the terms in a relation, so, when allowing for dressed operators (\eg\ $g_s^2\cP_{00}, g \cP_{11}$), both types (i) and (ii) relate vertical nodes in the graph. However, at the level of naked operators, relations of type (ii)  relate nodes in different columns.
Fig.~\ref{fig.Morders_basis_reduction_examples}
(left) shows how "dressing" an operator with powers of $g_s$ increases its $N_\chi$ leaving its $\min_i(N_{\chi,i}-N_{g_s,i})$ unchanged, while dressing it with powers of an EW coupling increases both orders.

\vskip 1em
We argue that a complete, non redundant basis of naked HEFT operators compatible with both $(N_{\rm HEFT}, N_{g_s})$ and $N_{\rm HEFT}^s$ expansions can actually be obtained simply by building non-redundant bases \emph{order by order in $N_\chi$} with standard algorithms available in the literature. In particular:
\begin{itemize}
\item the counting $\min_i(N_{\chi,i}-N_{g_s,i})$ can be ignored in the basis construction. \\[2mm]
Defining a "basis" at each 
($N_\chi$, $\min_i(N_{\chi,i}-N_{g_s,i}))$ order pair would be a cumbersome exercise with limited physical applicability, as even IBP relations (\ie\ momentum conservation) equate operators with different $\min_i(N_{\chi,i}-N_{g_s,i})$.  Fortunately, this is unnecessary to obtain complete bases.

\item consistently with the first point, relations of type (i) can be employed to remove \emph{any of the operators} appearing in them. 
\item relations of type (ii) should be employed to remove \emph{one of the operators with the largest $N_\chi$}, \ie\ one of those that appear as "naked" in the relation.\\[2mm]
In the example of Eq.~\eqref{eq.generic_relation_2}, one can remove either $\cP_{20}$ or $\cP_{2-1}$.
This rule is compatible with the usual conventions for the application of EOMs, that favor the removal of operators containing structures descending from the kinetic terms of the fields ($i\slashed{D}\psi,D_\mu X^{\mu\nu},\square h,D_\mu\V^\mu$) and, more in general with the removal of operators with higher numbers of derivatives~\cite{Criado:2018sdb,Grzadkowski:2010es}.
\end{itemize}

These rules guarantee that the same operator bases can be adopted irrespectively of whether amplitudes are sorted by $N_{\rm HEFT}$ or $N_{\rm HEFT}^s$. Moreover, the bases available in the literature already satisfy them~\cite{Buchalla:2013rka,Brivio:2016fzo,Sun:2022ssa,Sun:2022snw} 

\vskip 1em
To motivate our ansatz, let us take a step back and consider a generic EFT, with some valid, unspecified power counting. Whenever a relation exists among a certain set of operators, even belonging to different power counting orders, \emph{in principle it can always be used to remove any of them from the operator basis, 
even if the diagram series is truncated within the range of values spanned by the relation.} What we mean by this statement is that, whatever operator is removed, \emph{the EFT predictions will remain complete at each order in the power counting}, independently of the operators orders and diagrams truncations.
However, in many cases, removing lower-order operators participating in the relation is not recommendable because it introduces negative powers of the expansion parameters in the Lagrangian, and it creates an unnatural algorithm for the basis construction. This happens in several well-known EFTs. 
For instance, consider the SM Higgs EOM
\begin{align}
 \square H_i = -\frac{m_h^2}{2} H_i +2 \lambda (H^\dag H) H_i - \sum_\psi \bar \psi_{Li} Y_\psi \psi_R \,,
\end{align}
where the last term is understood to sum over the three Yukawa terms, 
which defines relations among dimension-6 and -4 SMEFT operators, such as
\begin{align}
  (H^\dag H)(H^\dag \square H) +{\rm h.c.} &=  - m_h^2 (H^\dag H)^2 +4 \lambda (H^\dag H)^3
  \nonumber\\
  &  -\sum_\psi (H^\dag H)\left(\bar \psi_L H Y_\psi \psi_R + {\rm h.c.}\right)
\end{align}
In principle, this relation could be used to remove the term $(H^\dag H)^2$, trading it for 
\begin{align}
-\frac{\Lambda^2}{m_h^2}\left[
 \left(\frac{(H^\dag H)(H^\dag \square H)}{\Lambda^2} 
 + \sum_\psi\frac{(H^\dag H)}{\Lambda^2}(\bar \psi_L H Y_\psi \psi_R)+
  +{\rm h.c.}\right)
  -4\lambda \frac{(H^\dag H)^3}{\Lambda^2}\right]. 
\end{align}
From the point of view of ensuring complete predictions order by order in $\Lambda$ up to $\mathcal{O}(\Lambda^{-2})$, a Lagrangian containing these interactions would yield perfectly correct results. However, the appearance of negative powers of $(m_h/\Lambda)\sim (v/\Lambda)$ makes the SMEFT expansion cumbersome and it significantly complicates both the basis construction and the organization of the SMEFT diagrams series.

In the case of HEFT with $N_{\rm HEFT}^{(s)}$ power counting, relations of type (ii) behave similarly to this example: although removing an operator with lower $N_\chi$ would yield correct, complete predictions, it would be inconvenient, because it would introduce negative powers of the coupling constants $g,g',y_\psi,\lambda$, violating the power counting conventions established above. For instance, using Eq.~\eqref{eq.generic_relation_2} to remove $\cP_{11}$ would introduce $-(g_s^2/g) \cP_{00} - (\cP_{2-1}+\cP_{20})/g$. In conclusion, relations of type (ii) should be used to remove one of the highest $N_\chi$ operators appearing in them, for the same reason why in SMEFT they are used to remove one of the highest-dimensional operators.

Relations of type (i), on the other hand, are not subject to this power counting argument, so they can be used freely to remove \emph{any} of the operators appearing in them.

\vskip 1em
It is worth illustrating in more detail the argument that using redundancy relations to remove a low-order operator from an EFT Lagrangian does not affect the completeness of the results, which might appear counterintuitive. For instance, one could worry that, in combination with the truncation of the diagrams series, this operation could artificially remove  some physical contributions to the EFT prediction. However, this is not the case. 

Let us start by noting that the completeness of a HEFT basis at a certain $N_\chi$ order can be evaluated considering only tree-level ($L=0$) diagrams with $N_{\rm HEFT}^s=N_\chi+n-2$, for fixed $n\geq 2$. That is: a basis is complete \emph{iff} it yields complete expressions for tree-level amplitudes at that specific $N_{\rm HEFT}^s$ order. 
This is analogous to observing that the completeness of a dimension-$k$ SMEFT basis can be evaluated based only on tree-level diagrams of order $\Lambda^{4-k}$, and it naturally stems from  a recursive procedure: the $N_\chi=0$ operators set is complete \emph{iff} it gives complete predictions at order $N_{\rm HEFT}^s = n-2$. Once this is established, the contributions from $N_\chi=0$ operators to all higher order amplitudes must be complete as well, therefore predictions at order $N_{\rm HEFT}^s=n-1$ are complete \emph{iff} the contributions from $N_\chi=1$ operators are complete, etc.
This observation guarantees, in particular, that higher-order diagrams do not need to be considered when evaluating the completeness of a basis, which is why operator relations that, in principle, would contain infinite terms (\eg\ from field redefinitions), can be safely truncated.

Let us now focus on the example in Eq.~\eqref{eq.generic_relation_1}. We would meet this relation when reducing the $N_\chi=1$ basis, so we only need to consider tree level diagrams with $N_{\rm HEFT}^s=n-1$ for all $n\geq 2$.
In terms of $N_{\rm HEFT}$ orders, the relevant diagrams to consider for each operator have
\begin{align}
    \cP_{1-1}& \qquad\to\qquad  n-1\geq N_{\rm HEFT}\geq n-3\,,
    \\
    \cP_{10}&  \qquad\to\qquad  n-1\geq N_{\rm HEFT}\geq n-2\,,
    \\
    \cP_{11}&  \qquad\to\hspace{2.2cm} N_{\rm HEFT}= n-1\,,
\end{align}
where we have accounted for the fact that $N_{\rm HEFT}\leq N_{\rm HEFT}^s$.

Let us now imagine to use Eq.~\eqref{eq.generic_relation_1} to remove $\cP_{1-1}$ from the basis. The worry is that, doing so, we could miss potentially relevant contributions in $(N_{\rm HEFT}^s,N_{\rm HEFT})  =(n-1,n-3)$ diagrams. However, the equality in Eq.~\eqref{eq.generic_relation_1} implies that any (on-shell) scattering amplitude computed with $\cP_{1-1}$ insertions must be \emph{identical}  to the same amplitude computed with insertions of the combination $-(\cP_{10}+\cP_{11})$. In particular, the results obtained using one or the other must be equal order by order in $N_{\rm HEFT}\leq n-1$.\footnote{
In diagrams with $N_{\rm HEFT}>n-1$, differences could arise, if operators with higher $\min_i(N_{\chi,i}-N_{g_s,i})$  enter the equivalence relation. However, this will only affect the basis construction at higher orders, and it can be ignored here.}  In this case, this means that $\cP_{1-1}$ contributions to $N_{\rm HEFT}=n-3$ amplitudes must actually be absent, because it is not possible to obtain them with $\cP_{10}, \cP_{11}$ insertions.
Similarly, if we use~\eqref{eq.generic_relation_1} to remove $\cP_{10}$, we will find that, necessarily, contributions of this operator to $N_{\rm HEFT}=n-2$ amplitudes are either absent or equivalently accounted for by $\cP_{1-1}$ insertions. 

We conclude that no inconsistencies can arise even when the diagrams series is truncated, \emph{provided that the diagrams and operators series are truncated consistently}, as derived in Section~\ref{sec.HEFT_counting_operators}. This crucially ensures that a given operator is removed by the $\Lag$ truncation \emph{iff} the diagrams to which it contributes are removed by the amplitudes truncation.

\vskip1em

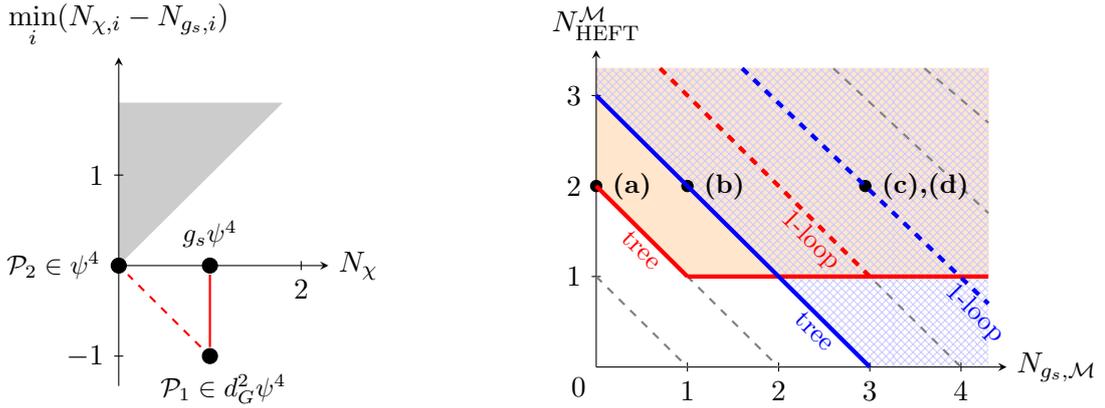
\begin{figure}[t]\centering
\parbox{.4\textwidth}{\centering
\begin{tikzpicture}[>=stealth, scale=1.2]    

    \filldraw[fill=gray!40, draw=none]
    (0,0) -- (1.8,1.8) -- (0,1.8) -- cycle;

    \foreach \y in {-1,1}
        \draw (0.05,\y) -- (-0.05,\y) node[left] {$\y$};

    \foreach \x in {2}
        \draw (\x,0.05) -- (\x,-0.05) node[below] {$\x$};
        
    \draw[->] (0,0) -- (2.3,0) node[right] {$N_{\chi}$};
    \draw[->] (0,-1.33) -- (0,2.3) node[above] {$\displaystyle \min_{i}(N_{\chi, i} - N_{g_s, i})$};

    \node[draw=black, fill=black, circle, inner sep=2pt, label=left:{\small $\cP_2\in \psi^4$}] at (0,0) {};
    \node[draw=black, fill=black, circle, inner sep=2pt, label=below:{\small $\quad\cP_1\in d_G^2\psi^4$}] at (1,-1) {};
    \node[draw=black, fill=black, circle, inner sep=2pt, label=above:{\small $g_s\psi^4$}] at (1,0) {};

    \draw[red, thick] (1,-0.9) -- (1,-0.1);
    \draw[red,dashed, thick] (0.9,-0.9) -- (0.1,-0.1);
    
\end{tikzpicture}}
\hfill
\parbox{.5\textwidth}{\centering
\begin{tikzpicture}[>=stealth, scale=1.2]    

     \filldraw[fill=orange!20, draw=orange!20]
    (1,1) -- (4.3,1) -- (4.3,3.3) -- (0,3.3) -- (0,2) -- cycle;

     \filldraw[pattern color=blue!20, pattern=crosshatch, draw=none]
    (3,0) -- (4.3,0) -- (4.3,3.3) -- (0,3.3) -- (0,3) -- cycle;

    \foreach \y in {1,2,3}
        \draw (0.05,\y) -- (-0.05,\y) node[left] {$\y$};

    \foreach \x in {1,2,3,4}
        \draw (\x,0.05) -- (\x,-0.05) node[below] {$\x$};
        
    \draw[->] (0,0) -- (4.5,0) node[right] 
     {$N_{g_s,\mathcal{M}}$};
    \draw[->] (0,0) -- (0,3.5) node[above]{$N_{\mathrm{HEFT}}^{\mathcal{M}}$};

    \draw[gray, dashed, thick] (0,1) -- (1,0);
    \draw[gray, dashed, thick] (0,2) -- (2,0);
    \draw[gray, dashed, thick] (0,3) -- (3,0);
    \draw[gray, dashed, thick] (0.7,3.3) -- (4,0);
    \draw[gray, dashed, thick] (1.6,3.3) -- (4.3,0.7);
    \draw[gray, dashed, thick] (2.6,3.3) -- (4.3,1.7);
    \draw[gray, dashed, thick] (3.6,3.3) -- (4.3,2.7);

    {\footnotesize
    }
    
    \node[below left] at (0,0) {0};

    \node[draw=black, fill=black, circle, inner sep=1.5pt, label=right:{\small \bf (a)}] at (0,2) {};
    \node[draw=black, fill=black, circle, inner sep=1.5pt, label=right:{\small \bf (b)}] at (1,2) {};
    \node[draw=black, fill=black, circle, inner sep=1.5pt, label=right:{\small \bf (c),(d)}] at (2.95,2) {};

    \draw[red, ultra thick] (0,2) -- (1,1);
    \draw[red, ultra thick] (1,1) -- (4.3,1);
    \draw[blue, ultra thick] (0,3) -- (3,0);
    \draw[red, ultra thick, dashed] (0.7,3.3) -- (3,1);
    \draw[blue, ultra thick, dashed] (1.6,3.3) -- (4.3,0.7);

    \node[red, rotate = -45] at (.5, 1.3) {\small tree};
    \node[blue, rotate = -45] at (2.4, 0.4) {\small tree};
    \node[red, rotate = -45] at (2.35, 1.4) {\small 1-loop};
    \node[blue, rotate = -45] at (4.15, .6) {\small 1-loop};
    
\end{tikzpicture}}
\caption{Left: graphical representation of the relation among operators $\cP_1,\cP_2$ defined in Eq.~\eqref{eq.eomG_ex_operators}.
Right: representation of the amplitude orders at which the two naked operators can contribute, for tree-level (solid) and 1-loop (dashed) diagrams. The dots labeled with (a)--(d) indicate the orders of the diagrams in Fig.~\ref{fig.diagrams_qqqq}. 
}\label{fig.P1P2_eom_example}
\end{figure}

To conclude this section, let us discuss a concrete example using the operators $\cP_1,\cP_2$ defined in Eq.~\eqref{eq.eomG_ex_operators}, that are represented in Fig.~\ref{fig.P1P2_eom_example} (left): $\cP_1$ belongs to class $d_G^2\psi^2\F$, of orders  $(1,-1)$, $\cP_2$ belongs to class $\psi^4\F$, of orders $(0,0)$, and therefore its $g_s$-dressed version $g_s \cP_2$ sits at orders $(1,0)$. The EOM~\eqref{eq.eom_P1_P2} (solid red line) equates the effects of $\cP_1$ and $g_s \cP_2$, which share the same $N_\chi$ but have different $\min_i(N_{\chi,i}-N_{g_s,i})$.

\begin{figure}[t] \centering
    \begin{subfigure}[t]{0.15\textwidth}
        \centering
        \raisebox{0.55em}{\includegraphics[width=\textwidth]{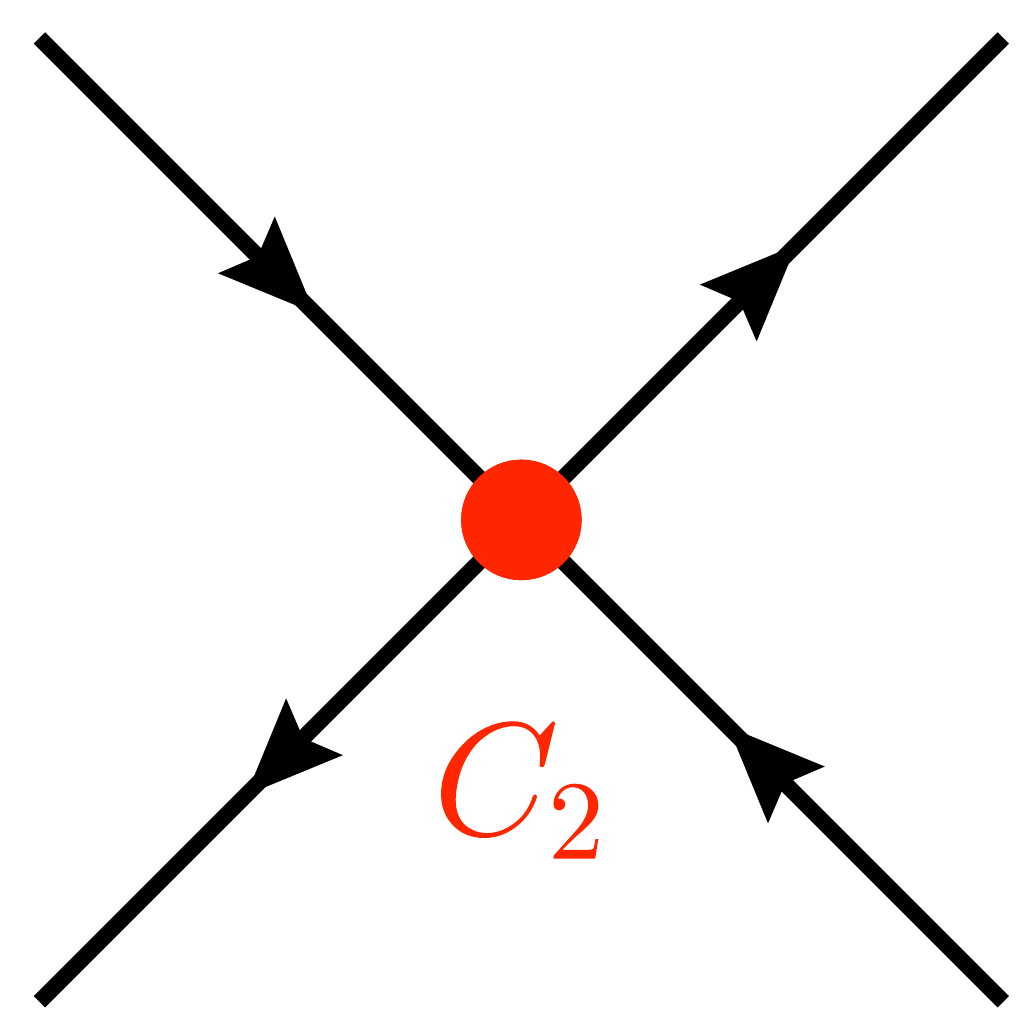}}
        \caption{}
    \end{subfigure}
    \hspace{1em}
    \begin{subfigure}[t]{0.2\textwidth}
        \centering
        \raisebox{0.85em}{\includegraphics[width=\textwidth]{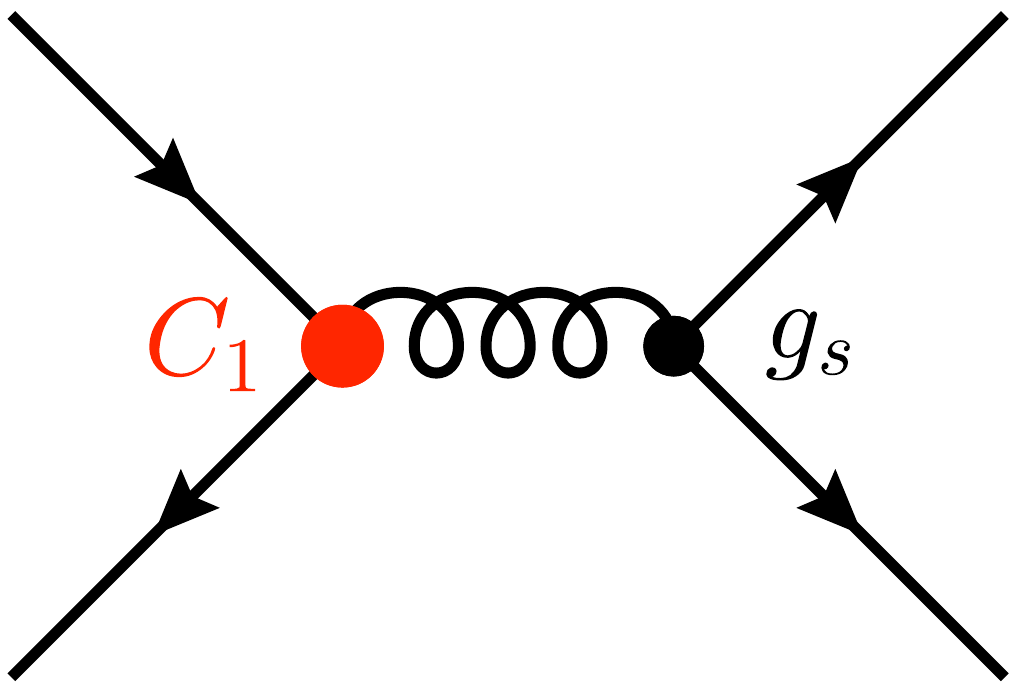}}
        \caption{}
    \end{subfigure}
    \hspace{1em}
    \begin{subfigure}[t]{0.22\textwidth}
        \centering
        \raisebox{0.85em}{\includegraphics[width=\textwidth]{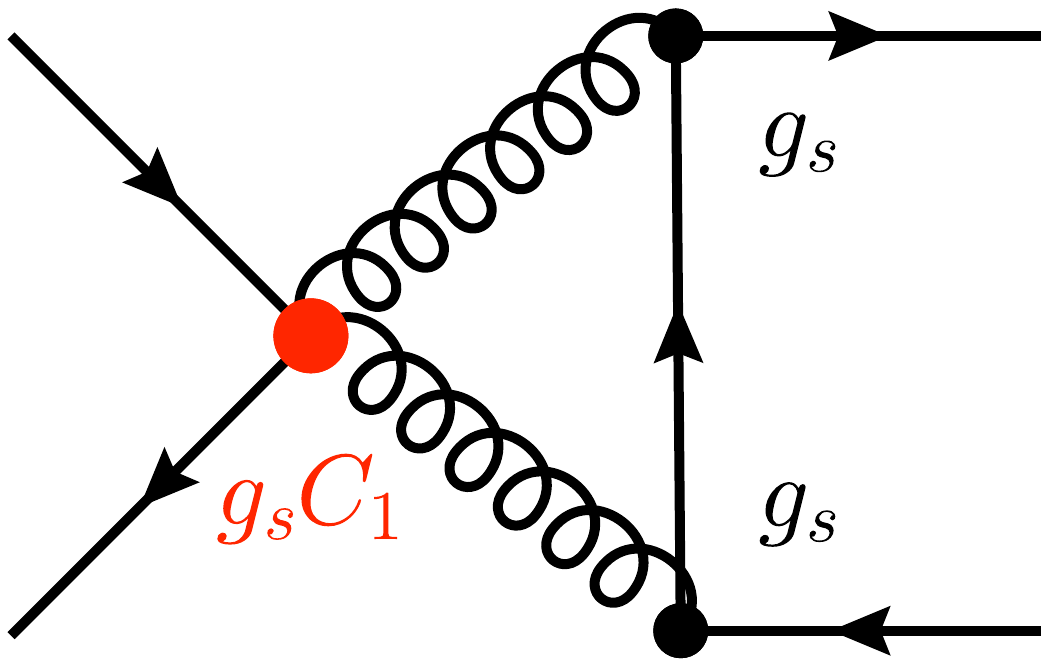}}
        \caption{}
    \end{subfigure}
    \hspace{1em}
    \begin{subfigure}[t]{0.25\textwidth}
        \centering
        \raisebox{0.75em}{\includegraphics[width=\textwidth]{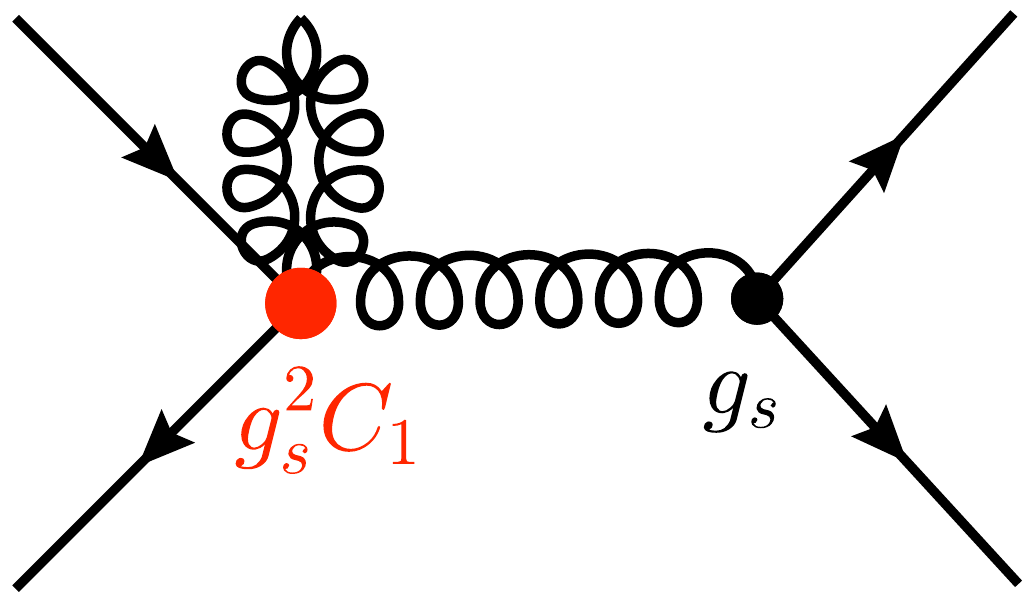}}
        \caption{}
    \end{subfigure}

    \caption{Example diagrams for the process $\bar QQ\to \bar QQ$, with insertions of the operators $\cP_1$, $\cP_2$ defined in Eq.~\eqref{eq.eomG_ex_operators}, that are linked by the EOM relation~\eqref{eq.eom_P1_P2}.}
    \label{fig.diagrams_qqqq}
\end{figure}

Let us examine their impact on the process $\bar QQ\to\bar QQ$: the naked operator $\cP_2$ can induce $2\to 2$ scattering amplitudes of the orders highlighted by the orange region in Fig.~\ref{fig.P1P2_eom_example} (right), while $\cP_1$ can induce amplitudes of the orders highlighted in the blue region.
Example diagrams are shown in Fig.~\ref{fig.diagrams_qqqq}, and their orders are represented by dots in Fig.~\ref{fig.P1P2_eom_example} (right): diagram (a) contains the naked $\cP_2$, and it has orders $(N_{g_s},N_{\rm HEFT}) = (0,2)$.  Diagrams (b), (c), (d) contain $\cP_1$, and they have orders $(N_{g_s},N_{\rm HEFT}) = (1,2), (3,2), (3,2)$. The EOM equivalence $\cP_1=g_s \cP_2$ is verified by the fact that $g_s \times $(a) is of the same order as~(b). 
Diagram (a) is an example of low-order $\cP_2$ contribution that would be missed if one chose to retain $\cP_1$ while simultaneously forbidding $\cP_1/g_s$ insertions. By contrast, retaining $\cP_2$ in the basis ensures that all $\cP_1$ contributions are covered.

We have argued that, since $\cP_1=g_s \cP_2$, it is impossible for $\cP_1$ diagrams to reach lower orders in $N_{\rm HEFT}$ compared to $P_2$ diagrams, despite $\cP_1$ having a lower $\min_i(N_{\chi,i}-N_{g_s,i})$. Graphically, this can be seen in the right panel of Fig.~\ref{fig.P1P2_eom_example} as the fact that, shifting the orange area to the right by one unit (corresponding to multiplying $\cP_2$ by $g_s$), the overlap with the blue hatched region remains partial: the lower end of the blue region, which remains non-overlapping with the orange, encloses the unreachable orders in $N_{\rm HEFT}^\meLabel$.
For an explicit test on the $\bar QQ\to\bar QQ$ process, consider that the $\cP_{1}$ interaction that minimizes $(N_{\chi,i}-N_{g_s,i})$ is $g_s^2 Q^2 G^3$ and the lowest-order diagram in which it can be inserted is precisely diagram~(d), which has still $N_{\rm HEFT}^\meLabel=2$. In practice, to have lower $(N_{\chi,i}-N_{g_s,i})$ one needs to have higher powers of $g_s$ and therefore more gluon insertions, which inevitably need to be tied up in loops. As a result the smaller $(N_{\chi,i}-N_{g_s,i})$  contribution to $N_{\rm HEFT}$ is compensated by the higher perturbative order of the diagram.

\subsubsection{Alternative normalizations of the HEFT operators}\label{sec.heft_other_normalizations}
As done for the SMEFT case in Section~\ref{sec.smeft_other_normalizations},  we discuss the impact of choosing alternative normalizations for HEFT operators onto the conclusions derived in this section.

Consider for instance the $N_\chi=4$ operator $\cP = B_{\mu\nu}B^{\mu\nu}\tr[\V^\rho\V_\rho]$. We can include it in the Lagrangian in various ways, \eg
\begin{align}
 \Lag &\supset  \frac{C_1}{\Lambda^2}\, \cP\,, 
 &
 \Lag &\supset C_{2}\, \cP  \,,
 &
\Lag &\supset \frac{C_{3}}{v^2}\, \cP  \,,
 &
 \Lag &\supset \frac{C_{4}}{\Lambda^2}\,  (4\pi)^{2x}\left(\frac{g'}{4\pi}\right)^2  \cP \,.
\end{align}
The first version is the NDA-normalized one considered so far, so the Wilson coefficient $C_1$ is dimensionless. $C_2$ has dimensions $M^{-2}$, while $C_3$ has dimensions $\hbar$. Finally, $C_4$ has dimensions $\hbar^{x}$; we leave $x$ as a free power. Clearly, modifying the normalization of an effective operator only affects the $N^\gCoupl,\alpha^\gCoupl$ countings introduced in Section~\ref{sec.pc}. In this case:
\begin{center}
\renewcommand{\arraystretch}{1.2}
\begin{tabular}{>{$}c<{$}|*3{>{$}c<{$}}}
\hline
& N_\Lambda^\gCoupl & N_{4\pi}^\gCoupl& N_\chi^\gCoupl
\\\hline
C_1& 0& 0& 0\\
C_2& -2& 0& -2\\
C_3& 0& -2& -2\\
C_4& 0& -2x& -2x
\\\hline
\end{tabular}
\end{center}
such that the chiral order of an amplitude would be given by 
\begin{align}
 N_{\chi,\meLabel} &= N_g + N_{g'}+N_{g_s} + N_{y_\psi} + 2N_\lambda - 2 N_{C_2} - 2 N_{C_3}- 2x N_{C_4}\,.    
\end{align}
The HEFT order of an amplitude $N_{\rm HEFT}^\meLabel$, defined in Eq.~\eqref{eq.HEFT_pc}, can then be written as
\begin{align}
N_{\rm HEFT}^\meLabel &= N_{\chi,\meLabel} - N_{g_s,\meLabel} + 2 N_{C_2,\meLabel} + 2 N_{C_3,\meLabel} +2x N_{C_4,\meLabel}+ (n-2)\,,
\end{align}
which corrects Eq.~\eqref{eq.NHEFT_to Nchi_me} to account for the dimensionful coefficients $C_2,C_3,C_4$.
Correspondingly, Eqs.~\eqref{eq.NHEFT_to_Nchi},~\eqref{eq.NHEFT_to_NchiminGs} would need to be modified into
\begin{align}
N_{\rm HEFT}^{s,\meLabel} &= n-2+2L + \sum_{i\in\text{vert.}} (N_{\chi,i}+2N_{C_2,i}+2N_{C_3,i}+2x N_{C_4,i})\,,
\\
N_{\rm HEFT}^{\meLabel} &= n-2+2L + \sum_{i\in\text{vert.}} (N_{\chi,i}-N_{g_s,i}+2N_{C_2,i}+2N_{C_3,i}+2xN_{C_4,i})\,.
\end{align}
These two equations simply indicate that, in the presence of mass- or $\hbar$-dimensionful Wilson coefficients, their chiral dimension should be neglected in the computation of $N_{\rm HEFT}^{(s)}$.
An equivalent but simpler way to formulate this generalization is that, {\it in Eqs.~\eqref{eq.NHEFT_to_Nchi},~\eqref{eq.NHEFT_to_NchiminGs}, $N_{\chi,i}$ should be understood as counting the chiral dimensions carried by the operator and by any factors of $g,g',g_s,y_\psi,\lambda$ in its prefactor, while ignoring chiral dimensions carried by other constants.} All the conclusions derived in this section can be directly generalized with this prescription. However, the use of NDA normalization is clearly useful in this case, as it gives directly the relevant value of $N_{\chi,i}$ without requiring subtractions.

To visualize the impact of alternative normalizations on the parameterization of scattering amplitudes, we can consider the operator $\cP$ defined above. Extracting for instance a $BB\pi\pi$ interaction of the form
\begin{equation}
  \cP\supset (\de_\mu B_\nu)^2 \frac{\de_\rho\pi^I\de^\rho\pi^I}{v^2} + \dots
\end{equation}
we can evaluate the corresponding contact diagram for $BB\to\pi\pi$ scattering  for each of the four normalization options. They scale respectively as: 
\begin{align}
 \mathcal{M}_1 &\sim (4\pi)^2 
 \left(\frac{p}{\Lambda}\right)^{4} \left(\frac{\Lambda}{4\pi v}\right)^{2}
 C_1\,,
\\
 \mathcal{M}_2 &\sim (4\pi)^2 
 \left(\frac{p}{\Lambda}\right)^{4}
\left(\frac{\Lambda}{4\pi v}\right)^{2}
 \left(\Lambda^2\, C_2\right)\,,
 \\
 \mathcal{M}_3 &\sim (4\pi)^2 
 \left(\frac{p}{\Lambda}\right)^{4}
 \left(\frac{\Lambda}{4\pi v}\right)^{4}
 \left((4\pi)^2
 C_3 \right)\,,
 \\
 \mathcal{M}_4 &\sim (4\pi)^2 
 \left(\frac{p}{\Lambda}\right)^{4}
 \left(\frac{\Lambda}{4\pi v}\right)^{2}
 \left(\frac{g'}{4\pi }\right)^{2}
 \, \left((4\pi)^{2x}C_4\right)\,.
\end{align}
We can note that: 
\begin{itemize}
\item as in SMEFT, introducing a mass-dimensionful Wilson coefficient ($C_2$)  modifies the powers of $1/\Lambda$ manifestly appearing in the expressions, making them different from the powers of $p$, or even absent. 
However, this has no impact on the power counting, in the sense that the use of $C_2$ instead of $C_1$ obviously does not increase or lower the $N_{\rm HEFT}^{(s)}$ order of the diagram. It can just be seen as making the suppression orders less transparent, in the same way as in SMEFT.

\item introducing Wilson coefficients with non-zero chiral dimensions ($C_2,C_3,C_4$) modifies the powers of $\Lambda$ and/or $(4\pi)$ manifestly appearing in the expressions, but, again, has no impact on the power counting. As per the prescription above, this choice does not alter the $N_{\rm HEFT}^{(s)}$ order of HEFT diagrams and, correspondingly, the chiral dimensions carried by $C_i$ should not be counted in the $N_\chi$ employed in the classification of HEFT operators,

\item normalizing operators by $1/v$ rather than $1/\Lambda$ (as with $C_3$) has the dual effect of introducing Wilson coefficients with non-vanishing chiral dimension and of introducing factors $(\Lambda/4\pi v)$ in the amplitude expansion. The latter can have numerical relevance, but, as they do not increase $N_{\rm HEFT}^{(s)}$, there is no overall impact on the power counting.

\item adding positive powers of the SM couplings constants $g,g',g_s,y_\psi,\lambda$ (as with $C_4$) increases the $N_\chi,\min_i(N_{\chi,i}-N_{g_s,i})$ orders of the operator and increases the perturbative, $N_{\rm HEFT}$ and $N_{\rm HEFT}^s$ orders at which the operator will contribute in diagrams.

As discussed above, this is consistent with loop corrections to a given operator scaling with $(g/4\pi)^2$, $(\lambda/16\pi^2)$ etc. 

\end{itemize}
Summarizing, normalization choices for HEFT operators are of course arbitrary and left to the user to make. Adopting NDA-normalized, "naked" operators gives a conservative result, that assigns HEFT operators to the lowest possible order in the Lagrangian expansion, and yields final expressions for amplitudes and observables that manifestly match the expansion in the master formula in Eq.~\eqref{eq.xs_MF_SMEFT}.

Dressing the operator with powers of  $g,g',g_s,y_\psi,\lambda$ can be motivated \eg\ by the assumption of a weakly interacting UV completion (see \eg~\cite{Arzt:1994gp,Giudice:2007fh,Buchalla:2022vjp}) or by imposing flavor symmetries (such as Minimal Flavor Violation~\cite{Chivukula:1987py,Hall:1990ac,DAmbrosio:2002vsn}) on the HEFT Lagrangian.
This normalization choice is of course allowed, as long as the powers are positive, as per the conventions motivated in Section~\ref{sec.HEFT_counting_operators}. It increases the chiral order of the operator, as well as the perturbative, $N_{\rm HEFT}$ and $N_{\rm HEFT}^s$ orders of the diagrams in which the operator is inserted. This is an important difference from the SMEFT case.

Any other normalization choices, including scalings by powers of $v$, generally yield dimensionful Wilson coefficients, which however has no impact on the power counting:  the $N_{\rm HEFT}^{(s)}$ order of scattering amplitudes can be computed from Eqs.~\eqref{eq.NHEFT_to_Nchi},~\eqref{eq.NHEFT_to_NchiminGs}, with the understanding that $N_{\chi,i}$ counts the chiral dimensions carried by the effective operator and by powers of $g,g',g_s,y_\psi,\lambda$ in its prefactor, but not by other dimensionful constants. The same $N_{\chi,i}$ should be used in the classification of HEFT operators into $(N_{\chi,i},\min_i(N_{\chi,i}-N_{g_s,i}))$ orders.

We stress that these arguments hold for the HEFT Lagrangian without reference to specific UV completions. In the spirit of a "bottom-up" EFT, they ensure a conservative organization of the EFT series expansion, in which effective operators and Feynman diagrams are assigned to the lowest possible order they could contribute to. The matching to specific UV models can give expressions for the Wilson coefficients that contain powers of the SM coupling constants, thereby increasing the chiral order of the corresponding operators. Crucially, because the matching must reproduce an expansion of the UV theory in powers of $\delta_{\rm HEFT}$, the matching expressions will only contain \emph{positive} powers of those couplings, and therefore it can only \emph{increase} the order of certain contributions compared to the general EFT calculation.
 
\subsection{Option (b): two independent scales $f>v$}\label{sec.xi}
In this section we explore an alternative approach to the power counting of HEFT, corresponding to option~\ref{option.xi} presented above. We consider a less minimal version of HEFT, where a new independent scale $f$ is adopted as a suppression for scalar insertions, replacing $v$, which is defined as the scale of SM masses. This means that we will now parameterize
\begin{align}
\label{eq.U_F_f}
    \U &= \exp\left(\frac{i\pi_I\sigma^I}{f}\right)\,,
    &
    \F_i(h)& = \sum_{n=0}^\infty a_{i,n} \left(\frac{h}{f}\right)^n\,.
\end{align}
In this way, the new scale $f$ inherits the constraint
\begin{equation}
    4\pi f \geq \Lambda\,,
\end{equation}
leading to the impossibility of expanding in powers of $(4\pi f/\Lambda)$, while the EW scale $v$ remains independent. As discussed below, $v$ will be introduced as a free parameter inserted in numerators so, in principle, $4\pi v/\Lambda\leq 1$ is required for perturbativity, as in SMEFT.  If we introduce the parameter $\xi\geq 0$ defined in Eq.~\eqref{eq.xi_def}, we can further specify that: 
\begin{align}
\label{eq.vorder_xi}
  1\geq   \frac{4\pi v}{\Lambda} = \sqrt{\xi}\, \frac{4\pi f}{\Lambda}\geq \sqrt{\xi}\,.
\end{align}
Thus,  if $\xi$ is sufficiently small, we can have $\Lambda\gg v$. In this sense, introducing $f\neq v$ extends the validity range of HEFT. 

Let us now examine the physical HEFT expansion defined by Eq.~\eqref{eq.delta_HEFT}. At the beginning of Section~\ref{sec.HEFT_PC} we argued that, in order to ensure that SM-mass insertions are treated as momentum insertions, the power counting should count suppression due to either the coupling constants $g,g',y_\psi,\lambda$ or the scale $v$. In the scenario considered in Section~\ref{sec.HEFT_PC_a}, the former option was the only viable one. In the scenario considered here, instead, we can choose to sort observable contributions in HEFT by
\begin{align}
\label{eq.NHEFT_xi}
N_{\rm HEFT}^\xi &= \xsLamP + N_v = \xsLamP + 2 N_\xi\,,
\end{align}
where $N_\xi$ counts $\xi$ insertions. In the last step we assumed that $4\pi f/ \Lambda\simeq 1$ (or that in any case we do not expand in this quantity), such that $(4\pi v/\Lambda)\sim \sqrt{\xi}$. 

Mimicking the procedure in Eqs.~\eqref{eq.NHEFT_to Nchi_xs} and~\eqref{eq.NHEFT_to Nchi_me}, we can also write
\begin{align}
N_{\rm HEFT}^\xi &= \xsLamP + N_v = \alpha_\chi^p + 2(n-2) + N_v \,,
\\
N_{\rm HEFT}^{\xi,\meLabel} &= N_{\Lambda,\meLabel}^p + N_{v,\meLabel} = N_{\chi,\meLabel}^p + (n-2) +  N_{v,\meLabel}\,,
\end{align}
for observables and amplitudes respectively. The latter can be further specified as 
\begin{align}
\label{eq.NHEFTxi_to_Nchi}
\framebox{$\displaystyle
N_{\rm HEFT}^{\xi,\meLabel} = \sum_{i\in\text{vert}}(N_{\Lambda,i}^p + N_{v,i}) = 
 n - 2 + 2L  + \sum_{i\in\text{vert}} (N_{\chi,i}^p + N_{v,i}) 
 $}
\end{align}
where $N_{v,i}$ counts $v$ insertions and the sum is over the vertices in the diagram. 

In Eq.~\eqref{eq.NHEFTxi_to_Nchi} we kept both the form as a function of $N_{\Lambda,i}^p$ and the form as a function of $N_{\chi,i}^p$. As discussed at the end of Section~\ref{sec.HEFT_pc_a_amplitudes}, both quantities are inhomogeneous over the interactions generated by each operator: $N_{\Lambda,i}^p$ depends on the number of scalar field insertions, which, in the expression on the right-hand side, is accounted for by $n$. On the other hand, $N_{\chi,i}^p$ varies over the components of covariant derivatives and gauge field strengths. When expanding in $N_{\rm HEFT}^{s}$, this issue is removed by the fact that the gauge coupling constants are also counted as HEFT expansion parameters. In Eq.~\eqref{eq.NHEFTxi_to_Nchi}, however, this is not the case. In this sense, the convenience of using $N_\chi^p$ instead of $N_\Lambda^p$ in the expression for $N_{\rm HEFT}^\xi$ is less obvious. 

\subsubsection{Power counting for HEFT operators}
Having defined an order $N_{\rm HEFT}^{\xi,\meLabel}$ for scattering amplitudes, we would now like to identify the corresponding rules for the organization of $\Lag_{\rm HEFT}$. 
For such rules to be fully defined, we need to identify a prescription for how $v$ enters the Lagrangian. However, in the construction of the EFT "from the bottom up", there are only a few physical conditions one can impose, that are not enough to fix $v$ insertions for all the effective interactions. Requiring $v$ to be the scale controlling the masses of the SM fermions and Higgs boson implies that the Yukawa terms and scalar potential take the form
\begin{align}
\label{eq.YV_xi_new}
  \Lag_{\rm HEFT }\supset\,&
  -\frac{1}{\sqrt2} \left(\bar d_L Y_d d_R + \text{h.c.}\right) \left( {\color{blue}v} +  a_{d,1} h +  a_{d,2}\frac{h^2}{f} + \dots\right) 
  \nonumber\\
  &
  - \lambda\left({\color{blue} v^2 h^2} + f\, a_{V,3} h^3 + \frac{1}{4} a_{V,4} h^4 + a_{V,5} \frac{h^5}{f} + \dots\right)\,,
\end{align}
and analogously for up-type quarks and charged leptons. The condition imposed only regulates the mass terms, highlighted in blue. The remaining $h$ interactions are left free, and one could insert different powers of $\xi$ in each of them. This is true for all the $\F(h)$ functions appearing in $\Lag_{\rm HEFT}$, and it represents an unavoidable ambiguity in the formulation of the "non-minimal" HEFT with $v$ and $f$. A possible way to reduce this ambiguity is by resorting to UV-motivated arguments. For instance, taking the scalar potential parameterization for Composite Higgs models given in Ref.~\cite{Panico:2015jxa} and taking the leading term in $\xi\ll 1$, one has that couplings with odd $h$ insertions scale with $\lambda \sqrt{\xi}/f^{n-4}$, while couplings with even $h$ insertions scale with $\lambda/f^{n-4}$. These arguments are of course model-dependent, so alternative UV scenarios might lead to different rationales.
In the following, we will make the minimal assumption that $v$ enters with positive powers in all HEFT interactions, leaving the exact $v$-dependence of $\F(h)$ functions otherwise unspecified. 

Turning our attention to gauge masses:  in HEFT they arise from the term $\tr[\V_\mu\V^\mu]$, which also yields the kinetic term for the Goldstone bosons. In order to achieve the correct normalizations
\begin{align}
-\frac{f^2}{4}\tr[\V_\mu\V^\mu] = 
 \frac{\de_\mu\pi^I\de^\mu\pi^I}{2} + \frac{v^2 g^2}{4} W_\mu^+W^{-\mu} + \frac{v^2 (g^2+g^{\prime 2})}{4} Z_\mu Z^{\mu}\,,
\end{align}
one is then forced to introduce $v$ into the covariant derivative acting on the $\U$ field~\cite{Gavela:2016bzc}:
\begin{align}
\label{eq.DU_xi}
 \V_\mu &= D_\mu \U \U^\dag = \de_\mu \U \U^\dag +\sqrt{\xi}\,  \frac{ig}{2}W_\mu^I \sigma^I -\sqrt{\xi} \, \frac{ig'}{2}B_\mu \T \,.
\end{align}
In order to re-obtain the usual SM Lagrangian, the remaining covariant derivatives and field strengths maintain instead the form without $\sqrt{\xi}$. This condition, albeit unusual, can be justified by arguing that Eq.~\eqref{eq.DU_xi} makes a statement about the longitudinal components of the $W,Z$ bosons, while the remaining $D_\mu, X_{\mu\nu}$ only contain transverse components.
With this parameterization, all interactions contained in $\V_\mu$ have the same $N_{\chi,i}^p + N_{v,i}=1$:  $\de_\mu \U \U^\dag$ has $N_{\chi,i}^p=1, N_{v,i}=0$, while the countings are flipped for the $\sqrt{\xi}\, g A_\mu $ terms.

We are now ready to go back to organization of the HEFT operator series. As anticipated, neither $N_{\Lambda,i}^p$ nor $N_{\chi,i}^p$ are homogeneous over all the interactions stemming from an operator, which means that a given operator can produce diagrams of different $N_{\rm HEFT}^{\xi,\meLabel}$ depending on the contributing vertex. 
However, we can note that the quantity\footnote{This equation assumes that the only dimensionful constants present are those listed. If additional dimensionful constants are present, they would need to be subtracted as well from the operator's $N_{\chi,i}$.}
\begin{align}
\label{eq.Nchip_to_NchiNg}
    N_{\chi,i}^{p} = N_{\chi,i} - N_{g,i} - N_{g',i} - N_{g_s,i} - N_{y_\psi,i} - 2 N_{\lambda,i}\,,
\end{align}
is formally very similar to the $(N_{\chi,i}-N_{g_s,i})$ employed in the classification of HEFT operators with $N_{\rm HEFT}$, presented in Section~\ref{sec.HEFT_counting_operators}.
We can leverage this analogy to reach a number of important conclusions: first of all, even if $N_{\Lambda,i}^p,N_{\chi,i}^p$ are inhomogeneous over the interaction terms from an operator, the power counting in $N_{\rm HEFT}^{\xi,\meLabel}$ is consistent with gauge invariance, as the latter is manifestly preserved by $N_{\chi}$ and also order-by-order in the couplings at the level of scattering amplitudes. Moreover, the algorithms for the organization of $\Lag_{\rm HEFT}$ and the reduction of HEFT operator bases derived in Sections~\ref{sec.HEFT_counting_operators} and~\ref{sec.HEFT_basis} remain valid in this scenario: all the formulas derived there can be applied here by replacing $(N_{\chi,i}-N_{g_s,i})\mapsto (N_{\chi,i}^p + N_{v,i})$ and $O_{\alpha_s}\mapsto O_{\alpha_s} + 2(O_g + O_{g'}+O_{y_\psi} + 2O_\lambda)$ where $O_g$ etc. are the orders at which the perturbative series in each of the EW couplings is truncated. 

The derivation of those results proceeds as described in Section~\ref{sec.HEFT_counting_operators}: in particular, it remains true that $(N_{\chi,i}^p + N_{v,i})\geq -2$, with the minimum being achieved by $(k\geq 4)$-point interactions, that in this case can contain indifferently gluons, Higgs bosons, Goldstone bosons or weak gauge bosons. The only special case is represented by the $h^3$ interaction, that has $N_{\Lambda,i}^p=-2$ but only 3 legs. However, assuming that the coupling will receive at least one $v$ insertion -- which is quite reasonable -- restores $(N_{\chi,i}^p + N_{v,i})\geq -1$. Then, the proof that $N_{\rm HEFT}^{\meLabel}\geq 0$ applies identically, giving $N_{\rm HEFT}^{\xi,\meLabel}\geq 0$.

The maximum and minimum values that $(N_{\chi,i}^p+N_{v,i})$ can take among the interactions stemming from an operator can be derived as follows: 
\begin{align}
\max_i (N_{\chi,i}^p + N_{v,i})& = q_D + q_\de + x + \mathbf{v} + \frac{f}{2} - 2 + N^\kappa_{v}
= \bar N_\chi + N^\kappa_{v}\,,
\\
  \min_i (N_{\chi,i}^p + N_{v,i})& = q_\de + \mathbf{v} + \frac{f}{2} - 2 + N_{v}^\kappa\geq -2\,.  
\end{align}
where $q_D, q_\de, x$ are respectively the numbers of $D_\mu,\de_\mu$ and $ X_{\mu\nu}$ insertions in the operator's definition, $\mathbf{v}$ is the number of $\V_\mu$ insertions and $f$ the number of fermions. $\bar N_\chi$ is the chiral dimension of the naked operator and $N_v^\kappa$ are the powers of $v$ in the operator's prefactor. 
The minimum value of $(N_{\chi,i}^p+N_{v,i})$ is reached when taking the terms $\sim g A_\mu$ from all the covariant derivatives and field strengths. 

Table~\ref{tab.HEFTxi_operators_orders} reports the maxima ($=\bar N_\chi$) and minima of $(N_{\chi,i}^p+N_{v,i})$ for a number of operator classes. 
These numbers are of course dependent on the prescription for $v$ insertions. Here we assume that $\F(h)$ does not contain any $v$, such that \eg\ the $y \bar\psi\psi h$ interaction has $(N_{\chi,i}^p+N_{v,i})=-1$, $\lambda h^4$ has $-2$ etc. As long as only positive powers of $v$ can be inserted by further considerations, the orders reported represent the minimum ones available.

\begin{table}[t]\centering
\renewcommand{\arraystretch}{1.1}
\begin{tabular}[t]{>{$}p{3.5cm}<{$}|*4{>{$}c<{$}}}
\hline
\textbf{operator class} & \bar N_\chi& 
\min_i (N_{\chi,i}^p+N_{v,i}) 
& \min_i (N_{\Lambda,i}^p+N_{v,i}) 
& d_p 
\\\hline
\lambda f^4 (\mathcal{V}-\xi h^2/f^2)& -2&  
-2 & 0&  0
\\
X_{\mu\nu}X^{\mu\nu} \F & 0& 
-2& 0& 4
\\
\de_\mu h \de^\mu h & 0& 
0& 0& 4
\\
f^2 \tr(\V_\mu \V^\mu)\F& 0& 
0& 0& 4
\\
\bar\psi i \slashed{D}\psi& 0& 
-1& 0& 4
\\
yf\,\bar\psi \U\psi (\F-1)& -1& 
-1& 0& 3
\\\hline
\V^2\F& 0& 
0 & 0& 4
\\
X \psi^2 \F& 0 & 
-1& 1& 5
\\
\V \psi^2 \F& 0 & 
0& 1& 5
\\
\psi^4 \F& 0 &
0&  2& 6
\\\hdashline
X^3 \F& 1 &  
-2& 2& 6
\\
X\V (\de\F)& 1&
0 & 2& 6
\\
X\V^2\F& 1&  
0 & 2& 6
\\
\psi^2 (\de^2\F)& 1 & 
1& 3& 7
\\
\V \psi^2 (\de\F)& 1 & 
1& 3& 7
\\
\V^2 \psi^2 \F& 1 &  
1& 3& 7
\\
\psi^4 (\de\F)& 1 & 
1& 4& 8
\\
X \psi^4 \F& 1 & 
0&  4& 8
\\
\V \psi^4 \F& 1 & 
1& 4& 8
\\
\psi^6 \F& 1 &  
1& 5& 9
\\\hdashline
X^2 \V^2 \F & 2& 
0& 4& 8
\\
\V^2 (\de^2\F)& 2 & 
2 & 4& 8
\\
\V^4 \F& 2 & 
2& 4& 8
\\
X^3 \psi^2 \F& 2 &
-1 & 5& 9
\\
\V^3 \psi^2 \F& 2 &
2 & 5& 9
\\
\hline   
\end{tabular}
\caption{
Values of $\bar N_\chi = \max_i (N_{\chi,i})$, $\min_i(N_{\chi,i}^p + N_{v,i})$ and $\min_i(N_{\Lambda,i}^p + N_{v,i})$  for some classes of HEFT operators. The terms in the upper block are the explicit SM-like interactions from Eq.~\eqref{eq.lagrangian_HEFT_LO}. The lower block contains classes of "naked" HEFT operators, characterized by their content in terms of fermionic fields $\psi$, field strengths $X$, $\V_\mu$ and derivatives $D$. The Lagrangian is understood to be formulated with two scales $f>v$, assuming the normalization in~\eqref{eq.DU_xi} and that $\F(h)$ does not contain $v$ insertions.  The last column reports the primary dimension $d_p$ defined in~\cite{Gavela:2016bzc}. 
}\label{tab.HEFTxi_operators_orders}
\end{table}

Table~\ref{tab.HEFTxi_operators_orders} also reports, for comparison, the values of $\min_i(N_{\Lambda,i}^p+N_{v,i})$ and of the primary dimension $d_p$ defined in Ref.~\cite{Gavela:2016bzc}. With our normalization assumptions, we find that 
\begin{equation}
\min_i(N_{\Lambda,i}^p+N_{v,i}) = d_p-4\,.
\end{equation}
The large numerical differences between the two columns are entirely due to the fact that $N_{\Lambda,i}^p$ is proportional to the number of legs in the vertex, while $N_{\chi,i}^p$ is transparent to that. In fact, one can easily check that
\begin{align}
  \min_i(N_{\Lambda,i}^p+N_{v,i}) &- \min_i(N_{\chi,i}^p+N_{v,i}) = -\min_i N_{4\pi}^p 
  \\
  &= \max_i \left(s_i+f_i-2\right)
  = f + \mathbf{v} + 2x + q_D +q_{\de\F} -2\,.
\end{align}
In the calculation of $N_{\rm HEFT}^{\xi,\meLabel}$ this difference is compensated by the presence of $n$ in the formula with $N_{\chi}^p$. 

In practice, the analogy with the discussion in Section~\ref{sec.HEFT_counting_operators} suggests that, even in the non-minimal setup with $f>v$, the HEFT Lagrangian can be organized by $N_\chi$, ignoring the values of $\min_i(N_{\chi,i}^p+N_{v,i})$ in the definition of operator bases, and that a truncation of the operator series can be performed as in Fig.~\ref{fig.O_orders_scheme} (right), with the appropriate re-assignment of the relevant variables. However, Table~\ref{tab.HEFTxi_operators_orders} indicates that, 
although $N_\chi$ still provides a perfectly consistent counting parameter, 
in this case, it does not constitute
a particularly good indicator for the potential relevance of a given operator: as the insertion of gauge couplings does not count towards increasing the HEFT suppression order, the relevant quantity $(N_{\chi,i}^p+N_{v,i})$ can always take values in a very broad range, and in fact it is often quite different from $N_\chi$. Indeed, the reported values of $\min_i(N_{\chi,i}^p+N_{v,i})$ are all quite low and similar among classes, exhibiting very low discriminating power. 
This can translate into the fact that the rules derived generalizing the prescriptions in Section~\ref{sec.HEFT_counting_operators} might not be particularly restrictive or informative. The information contained in $\min_i(N_{\Lambda,i}^p+N_{v,i})$ or, equivalently, in $d_p$, can then provide a valuable handle. Either way, one is faced with the difficulty of identifying a unique feature that controls the power counting in $N_{\rm HEFT}^\xi$ at the operator level. 

In summary, if we wish to avoid a HEFT power counting that expands in the SM coupling constants, we necessarily need to introduce a new expansion parameter to weigh down SM mass insertions: the most natural choice is a ratio of scales $v/f=\sqrt{\xi}$. Expanding in $(p/\Lambda)\sim\sqrt\xi$ leads to a different classification of HEFT Feynman diagrams, that depends on properties of the interaction vertices that are not uniquely determined by the form of the gauge-invariant operators. This is a fundamental challenge, that can be only partially addressed by resorting to $N_\chi$, in analogy with the treatment for the $N_{\rm HEFT}$ expansion, and/or to $d_p$.

\subsection{Power counting and renormalization}\label{sec.renormalization}
In this subsection we examine more closely some  aspects that emerge when computing HEFT amplitudes at one or more loops.
We assume that UV divergences are always cured by working in dimensional regularization and in the $\overline{\text{MS}}$ renormalization scheme. The use of a mass-independent regularization scheme is important to preserve the NDA scaling  at loop level, as it prevents loop integrals from producing spurious powers of $\Lambda$ or other mass-dimensionful regulators.

An important requirement for the consistency of an EFT is that all the counterterms needed to renormalize UV divergences are retained by the Lagrangian truncation. This condition is automatically respected as long as the Lagrangian truncation consistently retains all the effective operators that can contribute to EFT amplitudes up to the desired order. 
In the case of SMEFT, the divergence of a diagram of order $N_{\rm SMEFT}$ must be canceled by a counterterm which is a dimension $(N_{\rm SMEFT}+4)$ operator, which should of course be retained in the Lagrangian truncation. This is the well-known order-by-order renormalizability of the SMEFT. Since the SMEFT expansion is orthogonal to the loop one (\ie\ $N_{\rm SMEFT}$ does not depend on the number of loops $L$), this is true at any $L$.

In the case of HEFT, this condition is also respected when truncating in $N_\chi$, because the counterterms required to cancel divergences are essentially operators giving tree-level, contact diagrams with exactly the same $N_{\rm HEFT}^{s}$ as the  divergent amplitude. As long as that operator class is retained in the truncation, all counterterms will be present.

Specifically, the divergence generated by a $L$-loop diagram with operator insertions yielding a certain $\sum_i N_{\chi,i}$, must be reabsorbed by a counterterm of order
\begin{align}
\label{Eq.Nchi_CT}
N_{\chi,{\rm CT}} = 2L + \sum_{i\in\text{vert.}} N_{\chi,i} = N_{\rm HEFT}^s - (n-2)\,,
\end{align}
where, as specified above, $N_{\chi}$ counts the chiral dimensions carried by fields and derivatives and by any SM coupling constant appearing in the normalization factor.
This relation can be easily found equating the expressions for $N_{\rm HEFT}^s$ for the counterterm and loop diagrams. 
Eq.~\eqref{Eq.Nchi_CT} formalizes the common lore that in chiral theories "each order is renormalized by the next one" 
and it indicates that \emph{HEFT is in fact renormalizable order-by-order in the power counting defined in Section~\ref{sec.HEFT_PC_a}}, in the sense that all the UV divergences from loop diagrams up to a certain order in the  $N_{\rm HEFT}^s$ expansion are reabsorbed by a finite set of counterterms which are contained in the HEFT Lagrangian truncated as explained in Section~\ref{sec.HEFT_counting_operators}. In other words, the truncated Lagrangian is renormalizable, as long as only loop diagrams within the corresponding amplitude truncation are considered.

It is worth highlighting that, in general, the set of counterterms with a certain $N_{\chi,\text{CT}}$ will include "naked" HEFT interactions of chiral order $N_\chi$, as well as interactions of lower $N_\chi$, "dressed" with positive powers of the SM coupling constants. For instance,  all 1-loop diagrams containing only $N_\chi=0$ vertices require counterterms with $N_{\chi,\text{CT}}=2$. Comparing for instance to the results in Refs.~\cite{Buchalla:2017jlu,Alonso:2017tdy}, it is easy to verify that this is indeed the case, and that some counterterms correspond to naked $N_\chi=2$ operators (\eg\ $\tr(\V_\mu\V^\mu)^2$), while others correspond to dressed $N_\chi=1$ (\eg\ $ig' B_{\mu\nu}\tr(\T[\V^\mu,\V^\nu])$) and $N_\chi=0$ (\eg\ $g^2 \tr(\V_\mu\V^\mu)$) operators.  
This pattern extends to the Renormalization Group equations~\cite{Yan:2004xi,Gavela:2014uta,Buchalla:2017jlu,Buchalla:2020kdh,Morales:2025jyu}: the $\beta$-function of a certain HEFT operator generally receives contributions from operators with the same $N_\chi$, as well as from operators with lower $N_\chi$, multiplied by SM couplings.

\subsection{Power counting and matching}\label{sec.matching}
If a UV theory is matched to the HEFT, its power counting must remain consistent. This consistency is guaranteed when the \textit{minimal} suppression factors are adopted, ensuring that the HEFT reflects the least-suppressed realization of any given interaction. The matching to a specific UV model can then only introduce additional layers of suppression. In practice, these may arise from loop factors or flavor structures that modify the naive EFT scaling, effectively promoting certain operators to higher order in the HEFT expansion.
\par
A simple example is provided by four-fermion operators. From general EFT considerations, such terms can appear already in the LO HEFT Lagrangian. However, as shown in~\cite{deBlas:2017xtg}, when matching simplified UV models, operators that involve both left- and right-handed fields typically acquire a Yukawa suppression $y^2$, which shifts them effectively to NLO level in HEFT. In contrast, operators of the same chirality arise with two powers of new-physics couplings. Adopting the minimal assumption that four-fermion operators belong to the LO Lagrangian thus provides a conservative and consistent starting point: it includes the lowest possible order at which such interactions can appear, while allowing the matching to the UV model to generate additional suppression factors.

Such suppression factors are sometimes encoded directly at the EFT level when specific UV dynamics are envisaged. A prime example is the EFT of the strongly-interacting light Higgs~\cite{Giudice:2007fh}, a SMEFT-like construction that incorporates the extra suppression factors characteristic of composite or strongly-coupled Higgs scenarios~\cite{Contino:2006qr, Agashe:2004rs}. Analogous reasoning applies to loop-suppression factors in the SMEFT arising from weakly coupled UV completions~\cite{Arzt:1994gp, Buchalla:2022vjp}, or to flavor suppressions implemented through minimal flavor violation~\cite{DAmbrosio:2002vsn}, both of which effectively reduce the number of independent operators.
\par
Explicit examples of matching to the HEFT in concrete UV contexts have been provided in literature, for instance the case of the  scalar singlet extension of the Standard Model is discussed in Refs.~\cite{Buchalla:2016bse, Dawson:2023oce, Dittmaier:2021fls} and the scalar triplet extension in \cite{Song:2024kos, Song:2025kjp}. 
Depending on the assumed hierarchy of the UV parameters, the resulting power counting is the expansion in chiral dimension or a SMEFT-like expansion~\cite{Buchalla:2016bse}. Moreover, as emphasized in~\cite{Dawson:2023oce}, the interpretation of which parameters are regarded as physical—e.g. when rewriting the Lagrangian in terms of masses and couplings—can also lead to distinct matching outcomes.
A systematic study of the interplay between UV matching and power counting, including these ambiguities, is left for future work.

\section{Examples} \label{sec.examples}
In this section we discuss some concrete examples to put the power counting rules into practice. A more detailed study, including the phenomenological implications of our power counting, will be presented in a companion paper focusing on $gg\to hh$ production at the LHC~\cite{Brivio:2025sib}.

\subsection{Example one: $\bar{q}q \to WW$}
\begin{figure}[t] \centering
    \begin{subfigure}[t]{0.3\textwidth}
        \centering
        \raisebox{0.7em}{\includegraphics[width=\textwidth]{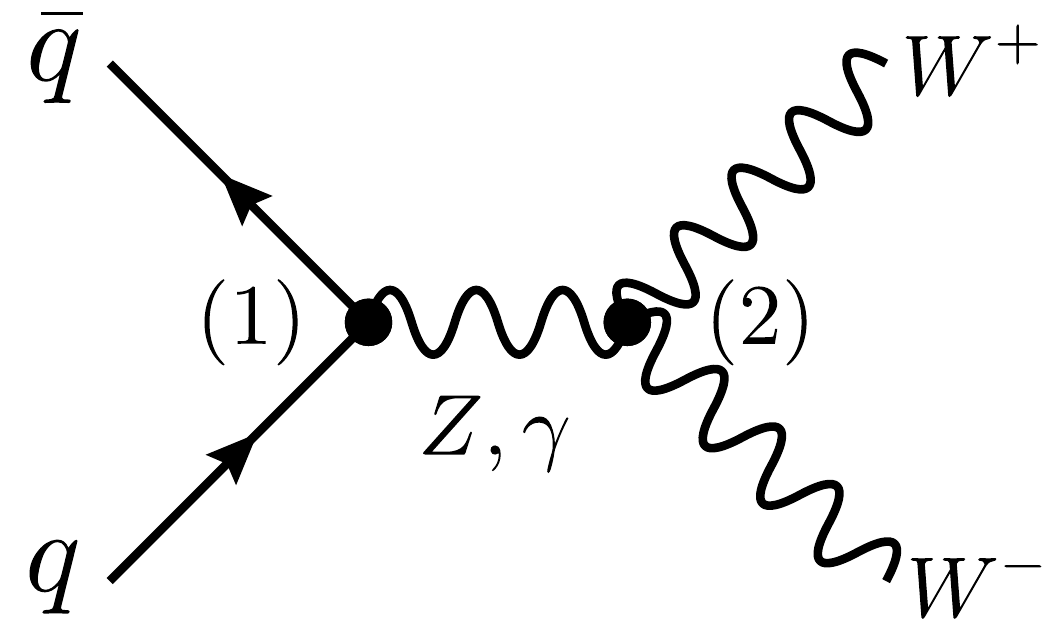}}
        \caption{$N_{\chi, 1} = N_{\chi, 2} = 0$}
    \end{subfigure}\hfill
    \begin{subfigure}[t]{0.3\textwidth}
        \centering
        \raisebox{0.58em}{\includegraphics[width=\textwidth]{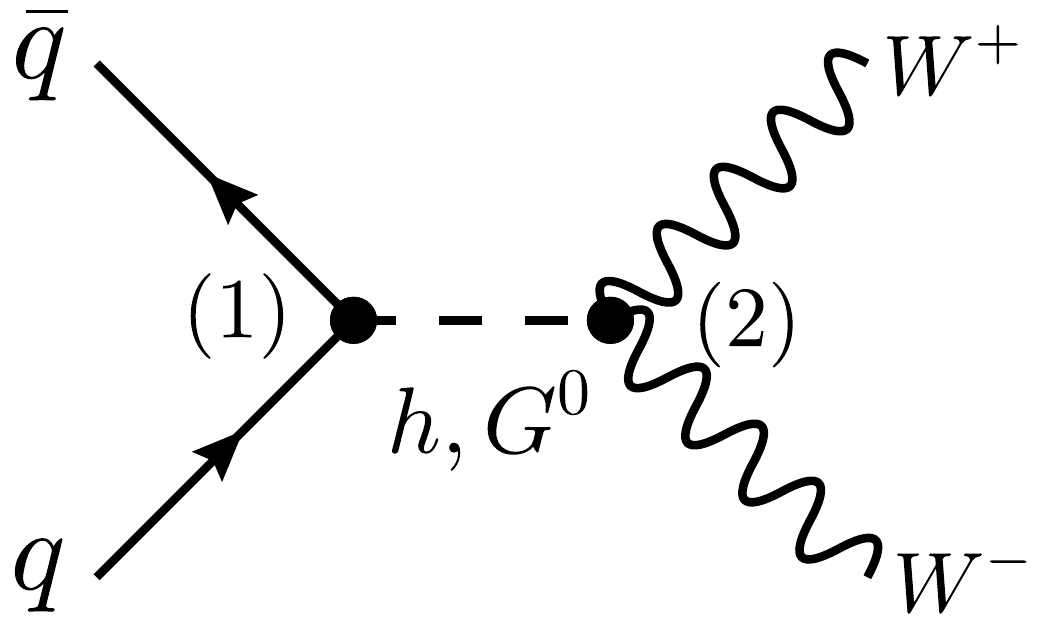}}
        \caption{$N_{\chi, 1} = N_{\chi, 2} = 0$}
    \end{subfigure}\hfill
    \begin{subfigure}[t]{0.2\textwidth}
        \centering
        \includegraphics[width=\textwidth]{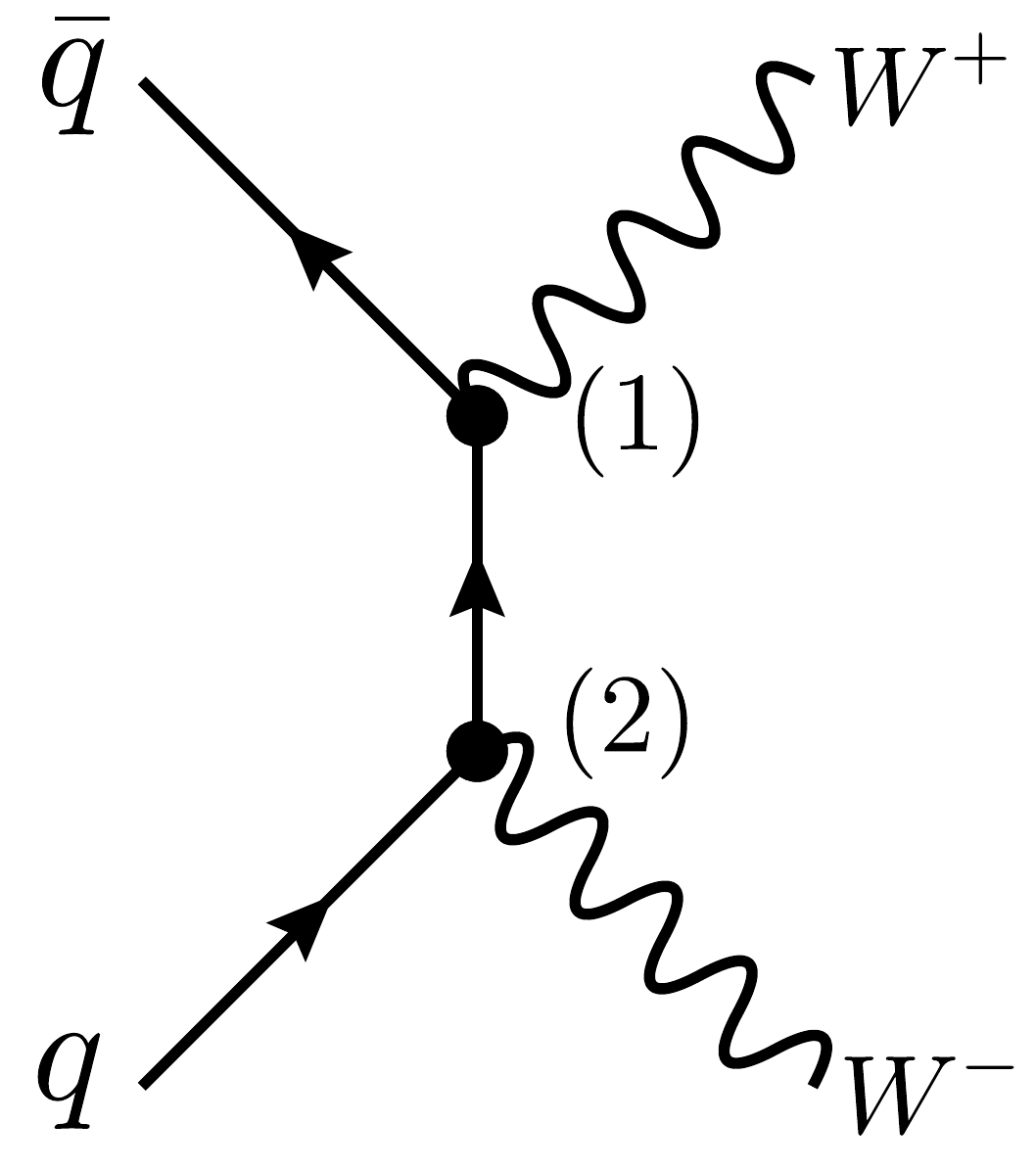}
        \caption{$N_{\chi, 1} = N_{\chi, 2} = 0$}
    \end{subfigure}
    \caption{LO Feynman diagrams for $q\bar{q}\to W^+ W^-$. The chiral dimension $N_{\chi, i}$ refers to the vertex labeled as (i).}
    \label{fig:FDqqWW}
\end{figure}

Fig.~\ref{fig:FDqqWW} shows the tree-level Feynman diagrams for $q\bar{q}\to WW$ production with insertions of operators from the LO HEFT Lagrangian given in Eq.~\eqref{eq.lagrangian_HEFT_LO}.
Using Eqs.~\eqref{eq.NHEFT_to_Nchi}, \eqref{eq.NHEFT_to_NchiminGs} and the fact that all LO interactions have $N_\chi=0$, one finds that all diagrams scale in the same way, with $N_{\rm HEFT}^{\mathcal{M}}=N_{\rm HEFT}^{s,\mathcal{M}}=2$ and $N_{\rm HEFT}^{\xi,\meLabel}=0$, which is consistent with gauge invariance.\footnote{As can be verified, the same scaling is obtained for the diagram arising from the Goldstone equivalence theorem: the Higgs–fermion vertex shown in Fig.~\ref{fig:FDqqWW}~(b) is already present, and the Higgs–Goldstone–Goldstone vertex also carries $N_{\chi,i}=0$.}
These values are reported in Table~\ref{tab.qqWW_diagram_counts}, that also shows a breakdown into $N_{\Lambda,\meLabel}^p$ and the individual coupling counts $N_{g,\meLabel}$ etc. 
For reference, the $N_{\chi,i}$ of the vertices entering the diagrams can be read off from Table~\ref{tab.operator_counting_examples}. 

To illustrate the physical meaning of the countings, let us consider first the diagram in Fig.~\ref{fig:FDqqWW}~(a), which scales as 
\begin{equation}
\mathcal{M}_{\ref{fig:FDqqWW}a} \sim (4 \pi)^2 \left( \frac{g}{4\pi}\right)^2\,,
\end{equation}
where, consistent with Eq.~\eqref{matrix_element}, the first factor $(4\pi)^2$ is associated to the external legs and will be canceled by phase space factors when computing the cross section. 
The absence of $(p/\Lambda)$ suppressions, corresponding to $N_{\Lambda,\meLabel}^p=0$, can be understood from the cancellation among $p$ factors in the propagator ($1/p^2$), the trilinear gauge coupling $\sim p$ and the two external spinors $\bar u v \sim p$. This result was of course expected as the diagram considered is a SM one.
Thus, in this case, the HEFT expansion counts $N_{\rm HEFT}^\meLabel = N_{\rm HEFT}^{s,\meLabel}= N_{g,\meLabel}$.

The diagram in Fig.~\ref{fig:FDqqWW}~(c) scales exactly in the same way, with the $(1/p)$ factor from the fermion propagator canceling against the $p$ from the spinors.

The diagram  Fig.~\ref{fig:FDqqWW}~(b) instead scales as
\begin{equation}
\mathcal{M}_{\ref{fig:FDqqWW}b} \sim (4 \pi)^2\,  \left( \frac{p}{\Lambda} \right)^{-1} \left( \frac{y_q}{4\pi}\right) \left( \frac{g}{4\pi}\right)^2 \left( \frac{4\pi v}{\Lambda} \right)\,.
\label{eq.M7b}
\end{equation}
The presence of the super-renormalizable $hWW$ coupling, that contains a power of $v$, essentially leaves a factor $(4\pi v/p)$. While in the SM(EFT) this would be considered as an order 1 quantity (consistent with the fact that $\Lambda$ cancels in the ratio), in HEFT it actually represents an enhancement.  The overall HEFT order of the diagram remains positive, thanks to the 3 weak coupling insertions. This is of course not an accident, and it is related to all SM interactions having overall $N_{\chi,i}=0$.

Note that the scaling in Eq.~\eqref{eq.M7b} could be rearranged, for instance, into
\begin{equation}
\mathcal{M}_{\ref{fig:FDqqWW}b} \sim (4 \pi)^2\, 
\left(\frac{p}{\Lambda}\right)\left( \frac{y_q}{4\pi}\right) \left( \frac{4\pi v}{\Lambda} \right)^{-1}\,,
\end{equation}
by parameterizing the $hWW$ coupling as $(m_W^2/v) \sim (p^2/v)$ rather than as $g^2v$. This expression does not match the conventions adopted in the derivation of Eq.~\eqref{matrix_element}, but it still gives the correct values of $N_{\rm HEFT}^{(s),\meLabel}=2$ and $N_{\rm HEFT}^{\xi,\meLabel}=0$, counting respectively the $(p/\Lambda)+(y_q/4\pi)$ and the $(p/\Lambda)-(4\pi v/\Lambda)$ factors. 
This property follows from the fact that all three power counting rules were defined by requiring that $p\sim m\sim gv$. This crucially removes all possible ambiguities associated to the interpretation of the suppression factors.

\begin{figure}[t] \centering
\includegraphics[width=\textwidth]{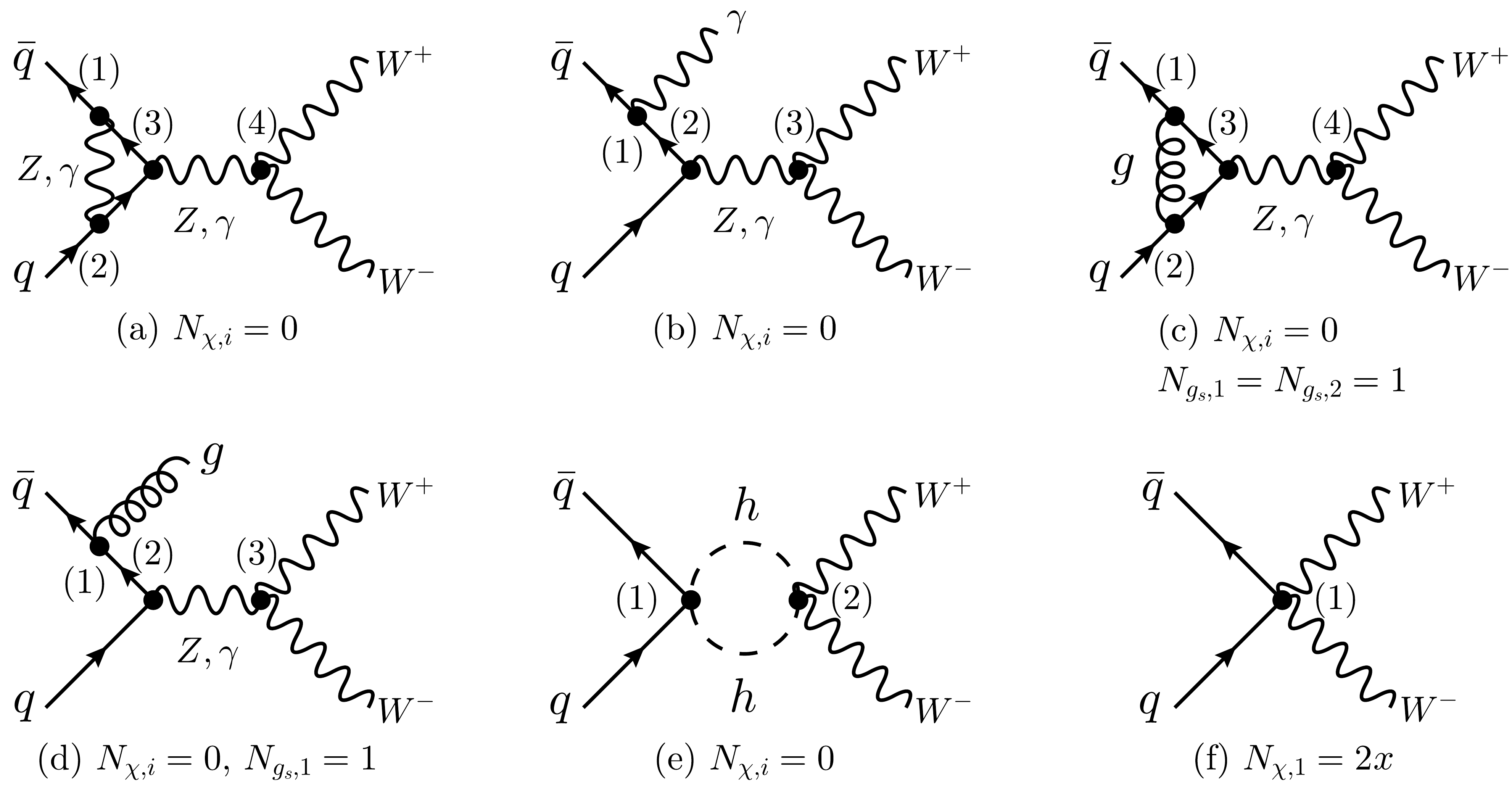}
\caption{Higher-order Feynman diagrams for $q\bar{q}\to W^+ W^-$. We display one-loop topologies, real radiation, and an insertion of the dipole operator having $N_{\chi} = 2x$.}
\label{fig:FDqqWWNLO}
\end{figure}

Fig.~\ref{fig:FDqqWWNLO} shows representative diagrams that arise either at higher order in perturbation theory or from NLO operators.  The corresponding countings are also reported in Table~\ref{tab.qqWW_diagram_counts}.
Diagrams (a)--(d) only contain $\Lag_{\rm LO}$ vertices, with $N_{\chi}=0$. They scale as
\begin{align}
\meLabel_{\ref{fig:FDqqWWNLO}a} &\sim (4\pi)^2\, 
\left(\frac{g}{4\pi}\right)^4\,,
&
\meLabel_{\ref{fig:FDqqWWNLO}c} &\sim (4\pi)^2\, 
\left(\frac{g_s}{4\pi}\right)^4\,,
\\
\meLabel_{\ref{fig:FDqqWWNLO}b} &\sim p^{-1}(4\pi)^3\, 
\left(\frac{g}{4\pi}\right)^3\,,
\label{eq.M8b}
&
\meLabel_{\ref{fig:FDqqWWNLO}d} &\sim p^{-1}(4\pi)^3\, 
\left(\frac{g}{4\pi}\right)^2\left(\frac{g_s}{4\pi}\right)\,,
\end{align}
such that (a) has $N_{\rm HEFT}^{s,\meLabel}=N_{\rm HEFT}=4$,  (b) has $N_{\rm HEFT}^{s,\meLabel}=N_{\rm HEFT}=3$, while (c) has $N_{\rm HEFT}^{s,\meLabel}=4$, $N_{\rm HEFT}=2$ and (d) has $N_{\rm HEFT}^{s,\meLabel}=3, N_{\rm HEFT}=2$. All four diagrams have $N_{\rm HEFT}^{\xi,\meLabel}=0$.  
Note that the different factors of $p,(4\pi)$ in Eq.~\eqref{eq.M8b} are due to the fact that diagrams $(b), (d)$ have an extra external leg. A comparison between (a),(b) and (c),(d) showcases the difference between  $N_{\rm HEFT}$, that keeps the EFT expansion and the powers of the strong coupling constant orthogonal, and $N_{\rm HEFT}^s$ treats them on the same footing. Adopting the latter requires the inclusion of all four diagrams simultaneously, while expanding in $N_{\rm HEFT}$ allows one retain the QCD corrections while discarding the EW ones, that are higher-order.

At the level of observables (in this case the cross section) we will have that $|\meLabel_{\ref{fig:FDqqWWNLO}b}|^2$ and the interference $|\meLabel_{\ref{fig:FDqqWW}a}\meLabel_{\ref{fig:FDqqWWNLO}a}^\dag|$ both enter at the perturbative order $g^6$.  Analogously, both $|\meLabel_{\ref{fig:FDqqWWNLO}d}|^2$ and  $|\meLabel_{\ref{fig:FDqqWW}a}\meLabel_{\ref{fig:FDqqWWNLO}c}^\dag|$ enter at order $g^4 g_s^2$. It is important that the two contributions in each pair are classified equally by the power counting, because both are required to cancel IR divergences in the diagrams with photons and gluons respectively. This is indeed the case for all the counting options considered: $|\meLabel_{\ref{fig:FDqqWWNLO}b}|^2$, $|\meLabel_{\ref{fig:FDqqWW}a}\meLabel_{\ref{fig:FDqqWWNLO}a}^\dag|$ have $\alpha_{\rm HEFT} = \alpha_{\rm HEFT}^s = 6$, 
$|\meLabel_{\ref{fig:FDqqWWNLO}d}|^2$,  $|\meLabel_{\ref{fig:FDqqWW}a}\meLabel_{\ref{fig:FDqqWWNLO}c}^\dag|$ have $\alpha_{\rm HEFT} = 4, \alpha_{\rm HEFT}^s=6$.

Notably, the consistency in the counting of loop and real-emission corrections is ensured by the presence of the $n$-term in Eqs.~\eqref{eq.NHEFT_to_Nchi},~\eqref{eq.NHEFT_to_NchiminGs}: in terms of the chiral dimension $[\mathcal{D}]_c=N_{\rm HEFT}^{s,\meLabel}+4-n$ defined in~\cite{Weinberg:1978kz,Buchalla:2013eza}, $|\meLabel_{\ref{fig:FDqqWWNLO}b}|^2$ appears to be 2 orders lower than $|\meLabel_{\ref{fig:FDqqWW}a}\meLabel_{\ref{fig:FDqqWWNLO}a}^\dag|$. As discussed in the previous sections, the dependence on the number of external legs is automatically accounted for if one expresses $N_{\rm HEFT}^s$ (or analogously $N_{\rm HEFT}, N_{\rm HEFT}^\xi$), as a function of $N_{\Lambda}^p$ and the numbers of couplings. Indeed, as reported in Table~\ref{tab.qqWW_diagram_counts}, all diagrams (a)--(d) have $N_{\Lambda,\meLabel}^p=0$, making the consistency check trivial.

\begin{table}[t]\centering
\renewcommand{\arraystretch}{1.2}
\begin{tabular}{c|*2{>{$}c<{$}}|*3{>{$}c<{$}}|*2{>{$}c<{$}}}
\hline
diagram&
N_{\rm HEFT}^\meLabel& N_{\rm HEFT}^{s,\meLabel}&
N_{\Lambda,\meLabel}^p & N_{g+g'+y+\lambda,\meLabel}&  N_{g_s,\meLabel}&
N_v&
N_{\rm HEFT}^{\xi,\meLabel}
\\\hline
\ref{fig:FDqqWW}.(a)& 
2& 2&
0& 2& 0& 0&
0
\\
\ref{fig:FDqqWW}.(b)& 
2& 2&
-1& 3& 0& 1&
0
\\
\ref{fig:FDqqWW}.(c)& 
2& 2&
0& 2& 0& 0&
0
\\
\hdashline
\ref{fig:FDqqWW}.(a, $W^3$)& 
3+x& 3+x&
2& 1+x& 0& 0&
2
\\
\hdashline
\ref{fig:FDqqWWNLO}.(a)& 
4& 4& 
0& 4& 0& 0&
0
\\
\ref{fig:FDqqWWNLO}.(b)& 
3& 3&
0& 3& 0& 0&
0
\\
\ref{fig:FDqqWWNLO}.(c)& 
2& 4&
0& 2& 2& 0&
0
\\
\ref{fig:FDqqWWNLO}.(d)& 
2& 3&
0& 2& 1& 0&
0
\\
\ref{fig:FDqqWWNLO}.(e)& 
4& 4&
1& 3& 0& 0&
1
\\
\ref{fig:FDqqWWNLO}.(f)& 
2+2x& 2+2x&
2& 2x& 0& 0&
2
\\\hline
\end{tabular}
\caption{Values of the relevant countings for the Feynman diagrams for $\bar qq \to W^+ W^-$ shown in Figs.~\ref{fig:FDqqWW} and~\ref{fig:FDqqWWNLO}. the label (a,$W^3$) refers to diagram (a), with the insertion of the operator in Eq.~\eqref{eq.op_W3} in the triple gauge vertex. The column $N_v$ is computed for the Lagrangian defined as in Eq.~\eqref{eq.YV_xi_new},~\eqref{eq.DU_xi}, which is relevant for the counting in $N_{\rm HEFT}^\xi$.}\label{tab.qqWW_diagram_counts}
\end{table}

Figs.~\ref{fig:FDqqWWNLO}~(e) and (f) show diagrams containing non-SM interactions, namely a coupling of two quarks and two Higgses and a coupling of two quarks and two gauge bosons. 
The former interaction appears in $\Lag_{\rm LO}$ in HEFT, and it has $N_\chi=0.$ Therefore the $L=1$ diagram (e) has $N_{\rm HEFT}^{\mathcal{M}}=N_{\rm HEFT}^{s,\mathcal{M}}=4$. Indeed, it scales as
\begin{equation}
\mathcal{M}_{\ref{fig:FDqqWWNLO}e} \sim (4 \pi)^2\, 
\left(\frac{p}{\Lambda}\right)
\left( \frac{y_q}{4\pi}\right) 
\left( \frac{g}{4\pi}\right)^2
\left( \frac{4\pi v}{\Lambda} \right)^{-1}\,.
\end{equation}
This diagram will also contribute to the NLO cross section (at order $y_qg^3$) once interfered with the LO diagrams of Fig.~\ref{fig:FDqqWW}. This diagram is UV divergent, having $P=0$.
The positive power of $(p/\Lambda)$ appearing in its scaling is related to the fact that, at the level of mass dimensions, $\bar \psi\psi h^2$ is a $d=5$ interaction. Therefore, the diagram would naively have a $\Lambda^{-1}$ scaling. In the HEFT Lagrangian defined with insertions of $h/v$ as in Eq.~\eqref{eq.lagrangian_HEFT_LO}, though, the suppression role is taken by $v$, which introduces the $(4\pi v/\Lambda)^{-1}$ factor.  If $\Lag_{\rm HEFT}$ is defined with two separate scales $f>v$, this changes in general. Taking the normalization choice in Eq.~\eqref{eq.YV_xi_new} gives 
\begin{equation}
\mathcal{M}_{\ref{fig:FDqqWWNLO}e} \sim (4 \pi)^2\, 
\left(\frac{p}{\Lambda}\right)
\left( \frac{y_q}{4\pi}\right) 
\left( \frac{g}{4\pi}\right)^2
\left( \frac{4\pi f}{\Lambda} \right)^{-2}
\,,
\end{equation}
which is a conservative choice that corresponds to $N_{\rm HEFT}^{\xi,\meLabel} = 1$. 

Finally, the diagram in Fig.~\ref{fig:FDqqWWNLO} (f) contains an insertion of the dipole operator, that we normalize as
\begin{equation}
\label{eq.dipole_ex}
\left(\frac{g y_q}{16\pi^2}\right)^x  \frac{4\pi}{\Lambda}\, \bar{Q}_L \sigma^{\mu\nu} \mathbf{U} \sigma^I Q_R W^I_{\mu\nu}\,,
\end{equation}
leaving the exponent $x$ as a free quantity. 
This operator induces vertices with $N_{\chi}=2x$. 
Diagram (f) has $N_{\rm HEFT}^{\mathcal{M}}=N_{\rm HEFT}^{s,\mathcal{M}}=2+2x$, and it scales as 
\begin{equation}
\mathcal{M}_{\ref{fig:FDqqWWNLO}f} \sim (4 \pi)^2\, 
\left(\frac{p}{\Lambda}\right)^2
\left( \frac{y_q}{4\pi}\right)^x
\left( \frac{g}{4\pi}\right)^x
\,.
\end{equation}
If the operator is introduced as naked, this diagram will have the same $N_{\rm HEFT}^{(s)}$ order as the SM ones in Fig.~\ref{fig:FDqqWW}, as $(p/\Lambda)^2\sim (g/4\pi)^2$\,. If the dressing by $y_q g$ is retained, instead, the diagram contributes at the same order as (a), (b), and (e) in Fig.~\ref{fig:FDqqWWNLO}. In this case, it can also act as a counterterm for UV divergences at $N_{\rm HEFT}^{\mathcal{M}}=N_{\rm HEFT}^{s,\mathcal{M}}=4$.

Normalizing the operator  by $g y_q$ (see \eg\ the SMEFT discussion in~\cite{Contino:2013kra}) can be motivated, for instance, as reflecting a UV bias in which such a structure would arise from new physics coupling via Yukawas and weak interactions. Removing the factors of $(4\pi)$ from the normalization in Eq.~\eqref{eq.dipole_ex} will simply introduce spurious $(4\pi)$'s in the scaling of $\meLabel_{\ref{fig:FDqqWWNLO}f}$ without altering any conclusions, see Section~\ref{sec.heft_other_normalizations}. Finally, if the introduction of two scales $f>v$ does not add powers of $v$ to the operator's normalization, diagram (f) has $N_{\rm HEFT}^{\xi,\meLabel}=2$.

We conclude this example by considering the operator
\begin{equation}
\label{eq.op_W3}
\left(\frac{g}{4\pi}\right)^x \frac{4\pi}{\Lambda^2} \, \epsilon_{IJK}\, W^I_{\mu\nu} W^{J\mu}_{\;\lambda} W^{K\nu\lambda}\,,
\end{equation}
which has chiral dimension $N_\chi = 1+x$. Introducing the coupling $g$ (with $x=1$) has the advantage of yielding an even $N_\chi$, which can better align to the loop expansion. This operator modifies the gauge boson self-interactions and it can be inserted into diagram (a) of Fig.~\ref{fig:FDqqWW}. Doing so, we would obtain a diagram scaling with 
\begin{equation}
\mathcal{M}_{\ref{fig:FDqqWW}a,W^3} \sim (4 \pi)^2\, 
\left(\frac{p}{\Lambda}\right)^2
\left( \frac{g}{4\pi}\right)^{x+1}
\,,
\end{equation}
corresponding to $N_{\rm HEFT}^{\mathcal{M}} = N_{\rm HEFT}^{s,\mathcal{M}}=3+x$, \ie\ at NLO in the EFT expansion. Thus we see that choosing $x=1$ puts this diagram at the same NLO order as the 1-loop corrections in diagrams (a), (c) of Fig.~\ref{fig:FDqqWWNLO}. We also have that $N_{\rm HEFT}^{\xi,\meLabel}=2$, consistent with this effect being equivalent to a dimension-6 SMEFT operator.

\subsection{Example two: Four-top production}

As a second example, we consider four-top production to illustrate how four-fermion operators can be counted in HEFT. Beyond serving as an example of operator counting, this process is also of experimental interest~\cite{ATLAS:2023ajo, CMS:2023ftu}, since it allows one to constrain four-top operators, which so far are rather weakly bounded~\cite{DiNoi:2025uhu}.  

\begin{figure}[t] \centering
\includegraphics[width=\textwidth]{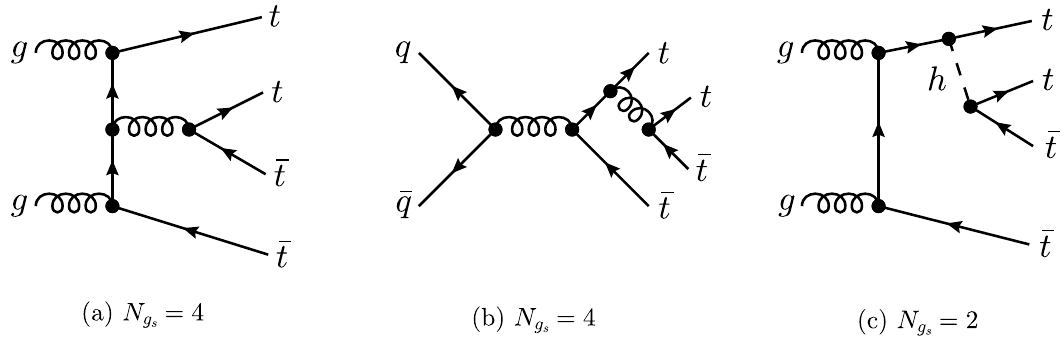}
\caption{SM-like tree-level diagrams for four-top production at the LHC. All vertices represent $N_\chi=0$ insertions.}
\label{fig:4topSM}
\end{figure}

Fig.~\ref{fig:4topSM} displays representative SM diagrams.  
Diagrams (a) and (b) are purely QCD-like, with $N_{g_s}=4$, $N_{\rm HEFT}^{\mathcal{M}}=0$ and $N_{\rm HEFT}^{s,\mathcal{M}}=4$.  
Diagram (c), instead, involves an electroweak exchange, yielding $N_{\rm HEFT}^{\mathcal{M}}=2$ and $N_{\rm HEFT}^{s,\mathcal{M}}=4$. Explicitly:
\begin{align}
\mathcal{M}_{\ref{fig:4topSM}a,b} &\sim p^{-2}(4 \pi)^4\, 
\left( \frac{g_s}{4\pi}\right)^{4}\,,
&
\mathcal{M}_{\ref{fig:4topSM}c} &\sim p^{-2}(4 \pi)^4\, 
\left( \frac{y_t}{4\pi}\right)^{2}
\left( \frac{g_s}{4\pi}\right)^{2}\,.
\end{align}
As already noted in the $\bar qq \to WW$ example, unlike $N_{\rm HEFT}^s$, the $N_{\rm HEFT}$ counting captures the electroweak suppressions, assigning \eg\ a higher order to diagram (c) compared to (a), (b). This allows one to make a selection, if desired, and retain only the QCD contributions. Note, however, that although numerically smaller, in SMEFT such electroweak contributions can be phenomenologically relevant, due to interference effects between the SM electroweak diagrams and the four-top operators~\cite{Aoude:2022deh}.  

\begin{figure}
\includegraphics[width=14cm, angle=0]{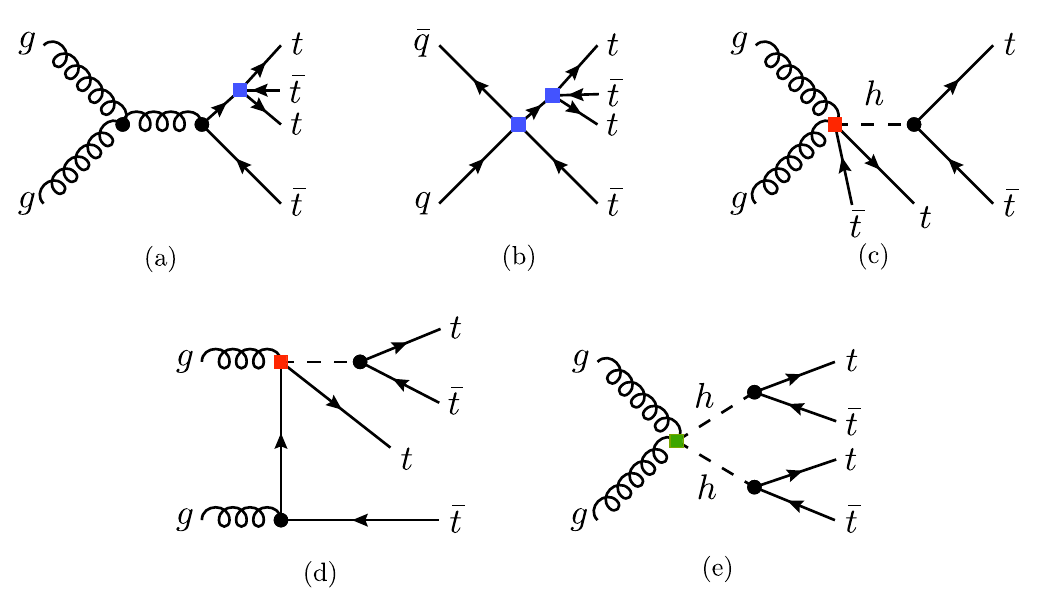}
\caption{Tree-level diagrams for four-top production with HEFT operator insertions. Blue dots indicate insertions of a $\psi^4$ operator ($N_\chi=0$), red dots are insertions of the chromomagnetic operator in Eq.~\eqref{eq.OtG} ($N_\chi = 1+2x$), and green dots insertions of the operator in Eq.~\eqref{eq.OHG} ($N_\chi=2x$).}
\label{fig:4topHEFT} 
\end{figure}

Fig.~\ref{fig:4topHEFT}~(a) and (b) show two diagrams with insertions of four-fermion operators
\begin{equation}
\frac{(4\pi)^2}{\Lambda^2}\, (\bar{\psi}\psi)(\bar{\psi}\psi)
\end{equation}
that, in their "naked" version, are assigned a chiral dimension $N_{\chi}=0$. 
In this case we have that both (a) and (b) have $N_{\rm HEFT}^{s,\mathcal{M}}=4$;  (a) has $N_{\rm HEFT}^{\mathcal{M}}=N_{\rm HEFT}^{\xi,\meLabel}=2$ and (b) $N_{\rm HEFT}^{\mathcal{M}}=N_{\rm HEFT}^{\xi,\meLabel}=4$. 
The explicit scalings are 
\begin{align}
\mathcal{M}_{\ref{fig:4topHEFT}a} &\sim p^{-2}(4 \pi)^4\, 
\left(\frac{p}{\Lambda}\right)^2
\left( \frac{g_s}{4\pi}\right)^{2}\,,
&
\mathcal{M}_{\ref{fig:4topHEFT}b} &\sim p^{-2}(4 \pi)^4\, 
\left(\frac{p}{\Lambda}\right)^4\,.
\end{align}
Here we see that each $\psi^4$ insertion brings a factor $(p/\Lambda)^2$, which is consistent with these being dimension-6 operators in SMEFT. Counting $(p/\Lambda)\sim (g/4\pi)$ identifies diagram (a) as exactly of the same order as diagram (c) of Fig.~\ref{fig:4topSM}. Moreover, we have that the $N_{\rm HEFT}^{s}$ counting interprets both as LO, whereas $N_{\rm HEFT}$ assigns them a suppression compared to the SM QCD-like diagrams of Fig.~\ref{fig:4topSM}, with (a) entering at lower order than~(b).  

Depending on their chirality structure and on UV-matching assumptions, powers of the SM Yukawa or gauge couplings could be added to the normalization of four-fermion operators. In this case, the $N_\chi$ of the operator would increase, pushing diagrams (a), (b) to higher orders and introducing a relative suppression of (b) compared to (a) in $N_{\rm HEFT}^s$. 

Figs.~\ref{fig:4topHEFT}~(c), (d) show diagrams with insertions of the chromo-magnetic dipole operator, that we normalize as
\begin{equation}
\label{eq.OtG}
\left(\frac{g_s}{4\pi}\right)^x \frac{4\pi}{\Lambda}\,\bar{Q}_L\sigma^{\mu\nu} \U  T^a Q_R G^a_{\mu \nu}\, \F_{qG}(h)\,,
\end{equation}
leaving again $x$ free in order to trace the contributions of the prefactor. Note that here we only introduce $g_s$, different from what is done for the electroweak dipole in Eq.~\eqref{eq.dipole_ex}. This operator produces a number of interactions among top quarks, gluons and Higgs bosons, whose counting orders were reported in Table~\ref{tab.operator_counting_examples} (for the case $x=0$).
Both diagrams (c) and (d) scale as
\begin{align}
\mathcal{M}_{\ref{fig:4topHEFT}c,d} &\sim p^{-2}(4 \pi)^4\, 
\left(\frac{p}{\Lambda}\right)^2
\left( \frac{y_t}{4\pi}\right)
\left( \frac{g_s}{4\pi}\right)^{1+x}
\left( \frac{4\pi v}{\Lambda}\right)^{-1}
\,,
\label{eq.M10cd}
\end{align}
corresponding to $N_{\rm HEFT}^{s,\mathcal{M}}=4+x$ and $N_{\rm HEFT}^{\mathcal{M}}=3$. 
The momentum scalings can be understood as follows:
in both diagrams we have a factor $p^2$ from the external spinors, diagram (c) has a $(1/p^2)$ from the Higgs propagator, while diagram (d) has $1/p^3$ from the three propagators, and $p$ from the momentum-dependent $\bar tt (\de G)h$ vertex.
The powers of $g_s$ are also the same, because (c) contains a power of the coupling in the dipole vertex, while diagram (d) contains one in the $G\bar tt$ interaction.

If $\Lag_{\rm HEFT}$ is defined with two scales $f>v$, Higgs insertions would be typically suppressed by $f$, which would yield a scaling analogous to Eq.~\eqref{eq.M10cd}, with $v$ replaced by $f$. In the absence of further $v$ insertions, both diagrams would have $N_{\rm HEFT}^{\xi,\meLabel} = 2$.

Finally, Fig.~\ref{fig:4topHEFT}~(e) shows a diagram with an insertion of the operator
\begin{equation}
\label{eq.OHG}
\left(\frac{g_s^2}{16\pi^2}\right)^x\, G_{\mu\nu}G^{\mu\nu} \F_{HG}(h)\,,    
\end{equation}
that has $N_{\chi}=2x$ and whose interactions were listed in Table~\ref{tab.operator_counting_examples}.
The diagram scales as 
\begin{align}
\mathcal{M}_{\ref{fig:4topHEFT}e} &\sim p^{-2}(4 \pi)^4\, 
\left(\frac{p}{\Lambda}\right)^2
\left( \frac{y_t}{4\pi}\right)^2
\left( \frac{g_s}{4\pi}\right)^{2x}
\left( \frac{4\pi v}{\Lambda}\right)^{-2}
\,,
\end{align}
and therefore it has $N_{\rm HEFT}^{\mathcal{M}}=4$ and $N_{\rm HEFT}^{s,\mathcal{M}}=4+2x$. If $v$ is replaced with $f$ in $\F_{HG}(h)$, the diagram has $N_{\rm HEFT}^{\xi,\meLabel}=2$\,.

\begin{table}[t]\centering
\renewcommand{\arraystretch}{1.2}
\begin{tabular}{c|*2{>{$}c<{$}}|*3{>{$}c<{$}}|*2{>{$}c<{$}}}
\hline
diagram&
N_{\rm HEFT}^\meLabel& N_{\rm HEFT}^{s,\meLabel}&
N_{\Lambda,\meLabel}^p & N_{g+g'+y+\lambda,\meLabel}&  N_{g_s,\meLabel}&
N_v&
N_{\rm HEFT}^{\xi,\meLabel}
\\\hline
\ref{fig:4topSM}.(a)& 
0& 4&
0& 0& 4& 0&
0
\\
\ref{fig:4topSM}.(b)& 
0& 4&
0& 0& 4& 0&
0
\\
\ref{fig:4topSM}.(c)& 
2& 4&
0& 2& 2& 0&
0
\\
\hdashline
\ref{fig:4topHEFT}.(a)& 
2& 4& 
2& 0& 2& 0&
2
\\
\ref{fig:4topHEFT}.(b)& 
4& 4&
4& 0& 0& 0&
4
\\
\ref{fig:4topHEFT}.(c)& 
3& 4+x&
2& 1& 1+x& 0&
2
\\
\ref{fig:4topHEFT}.(d)& 
3& 4+x&
2& 1& 1+x& 0&
2
\\
\ref{fig:4topHEFT}.(e)& 
4& 4+2x&
2& 2& 2x& 0&
2
\\\hline
\end{tabular}
\caption{Values of the relevant countings for the Feynman diagrams for $pp\to \bar tt\bar tt$ shown in Figs.~\ref{fig:4topSM} and~\ref{fig:4topHEFT}. 
The column $N_v$ is computed for the Lagrangian defined as in Eq.~\eqref{eq.YV_xi_new},~\eqref{eq.DU_xi}, which is relevant for the counting in $N_{\rm HEFT}^\xi$.}\label{tab.4top_diagram_counts}
\end{table}

\section{Conclusions \label{sec.con}}
We revisited the power counting of the Higgs EFT, with the aim of improving some unsatisfactory aspects that characterize existing prescriptions for the organization of the operator series. Addressing these issues is important for the potential development of a systematic program of HEFT searches, that could include for instance the automation of HEFT predictions and BSM-HEFT matching. 

We addressed the task from first principles, by requiring the power counting to reflect a series expansion in dimensionless, physical quantities at the level of observable predictions. We then worked backwards through the calculation process, to infer coherent prescriptions for the organization of HEFT diagrams and operators.  
This approach was introduced through a pedagogical review of dimensional analysis in general EFTs, in which we paid particular attention to identifying how the relevant suppression factors propagate from the Lagrangian to amplitudes and observables.

In the spirit of relying as much as possible on genuine EFT considerations, minimizing UV assumptions and theoretical prejudice, we only required HEFT predictions to follow an expansion in $(p/\Lambda)$ and that SM mass insertions are weighted as momentum insertions $m\sim p$. These simple rules identify two alternative power counting rationales, depending on whether HEFT is formulated in terms of a unique low-energy scale $v$ (corresponding to the Higgs VEV), or in terms of two scales $v<f$. In the former case, HEFT diagrams are assigned a suppression order $N_{\rm HEFT}^\meLabel$ (or alternatively $N_{\rm HEFT}^{s,\meLabel}$, depending on the treatment of the QCD expansion), which depends on the "chiral order" $N_\chi$ of the HEFT operators inserted in the vertices. The latter can then be adopted as a sorting rule for the Lagrangian, see Table~\ref{tab.HEFT_operators_orders}.
If a separation between the EW scale $v$ and the scalars suppression scale $f$ is allowed in the HEFT formulation, then the amplitudes can be assigned an order $N_{\rm HEFT}^{\xi,\meLabel}$ which, to some extent, is reminiscent of the SMEFT expansion. 
$N_{\rm HEFT}^{\xi,\meLabel}$ depends on properties of the contributing vertices that do not take homogeneous values over all the interactions produced by a given HEFT operator. As a consequence, it is difficult to organize the Lagrangian in a way that reflects properly the expected suppressions. Both the chiral order $N_\chi$ and the primary dimension $d_p$ of an operator are informative in this sense, but neither provides an unambiguous criterion.

For both of the proposed HEFT power countings, we provided quantitative prescriptions that allow the computation of the  suppression order of a given (squared) amplitude, based on its topology and on the properties of the participating interactions, see App.~\ref{app.nutshell} for a self-contained summary. The definition of $N_{\rm HEFT}^\meLabel$ is very similar to the chiral order $\mathcal{D}$ typically defined for Feynman diagrams in chiral perturbation theory. At variance with that prescription, though, $N_{\rm HEFT}^\meLabel$ depends on the number of external legs in a diagram, and introduces an explicit distinction in the treatment of QCD vs. EW corrections. The former feature partially restores a dependence on the canonical dimension of the effective interactions and it allows, for instance, to assign equal orders to the contributions from loop and real-emission corrections at observables level.

The expressions we found are general: they can be applied to all orders in the HEFT and perturbative expansions, and they do not rely on specific normalization choices for the effective operators, nor on UV-dependent assumptions. By construction, the proposed $N_{\rm HEFT}^\meLabel$, $N_{\rm HEFT}^{\xi,\meLabel}$ orders have a direct interpretation in terms of $(p/\Lambda), (4\pi v/\Lambda), (g/4\pi)\dots$ suppressions, which was demonstrated for a number of diagrams contributing to diboson and four-top production at the LHC. The systematic application of the power counting to a specific process, namely $gg\to hh$, is presented in a companion publication~\cite{Brivio:2025sib}.

We hope that our results can remove some obstacles towards a proper deployment of HEFT in phenomenological analyses and experimental measurements, and help clarify relevant aspects for the comparison of SMEFT and HEFT. A better understanding of the HEFT expansion is relevant for the matching to BSM models s well: a more detailed investigation of the implications of our results in this direction is left for future work.

\acknowledgments
The shown Feynman diagrams have been created with \texttt{FeynGame} \cite{Harlander:2020cyh, Harlander:2024qbn}.
\par
IB thanks
Aneesh Manohar, Ben Stefanek and Peter Stangl for useful discussions and CERN for hospitality during the completion of this project.
This work received funding by the INFN Iniziativa Specifica APINE and by the University of Padua under the 2023 STARS Grants@Unipd programme (Acronym and title of the project: HiggsPairs – Precise Theoretical Predictions for Higgs pair production at the LHC). This work was also partially supported by the Italian MUR Departments of Excellence grant 2023-2027 “Quantum Frontiers” and by SNF through the PRIMA grant no. 201508.  We acknowledge support by the COST Action COMETA CA22130.

\appendix 
\section{How to organize a HEFT calculation in a nutshell}
\label{app.nutshell}

This appendix provides a concise and user-friendly summary of the minimal set of power counting formulas required to organize a HEFT calculation. More details on the derivation of these results can be found in the main text, particularly Sections~\ref{sec.pc} and~\ref{sec.HEFT_PC}. 

Two main power counting rationales apply to the HEFT, depending on whether it is defined using the EW scale $v$ or a BSM scale $f>v$ to suppress Goldstone and Higgs insertions. In the former case, we define prescriptions called $N_{\rm HEFT}$ (which can be taken as a baseline) and $N_{\rm HEFT}^s$. In the latter we define a prescription called $N_{\rm HEFT}^\xi$. 

\subsubsection*{Effective operators and interaction terms}
We assume that the HEFT Lagrangian is normalized using the NDA prescription, see Section~\ref{sec.pc_general}. 
A generic HEFT operator can be characterized by its chiral order, Eq.~\eqref{eq.Nchi_amplitude}
\begin{align}
    N_{\chi} &= q + \frac{f}{2} -2 + N_\chi^\kappa\geq 0\,,
\end{align}
where $q$ is the total number of $D_\mu,\de_\mu,\V_\mu$ or $X_{\mu\nu}$ in the operator's definition, $f$ is the number of fermions and $N_{\chi}^\kappa$ counts the chiral dimension of the SM couplings: gauge couplings $g,g',g_s$ and Yukawa couplings  $y$ have $N_\chi=1$ while the scalar quartic coupling $\lambda$ has $N_\chi=2$:
\begin{align}
 N_{\chi}^\kappa &= N_{g}+N_{g'} +N_{g_s}+ N_{y} + 2N_{\lambda}\,.
\end{align}
This relation implies that, if an operator is multiplied by a SM coupling constant, its $N_\chi$ becomes higher. The minimum $N_\chi$ that can be assigned to the operator is the one obtained when it is "naked", \ie\ without prefactors.
The HEFT Lagrangian can be organized based on the chiral order, \ie 
\begin{align}
\Lag_{\rm HEFT} = \Lag_0 + \Lag_1 + \Lag_2 + \dots
\end{align}
where $\Lag_k$ contains a complete basis of operators with $N_\chi=k$. 
The order $N_\chi$ for a number of HEFT operator classes is shown in Table~\ref{tab.HEFT_operators_orders}. 
\\
In some cases, it is convenient to characterize the individual interaction terms obtained when opening the covariant derivatives and gauge field strengths of an operator. A generic interaction, labeled with $i$, has the form 
\begin{equation}
\gCoupl_i \; \de_\mu^{q_i}\,\phi^{s_i}\, \psi^{f_i}\,,
\end{equation}
where $\phi$ represents either a scalar or a gauge boson: $\phi=\{h,\pi^I,W^I_\mu,B_\mu,G_\mu^a\}$, $\de_\mu$ is a derivative, $\psi$ a fermion and $\kappa$ a product of coupling constants. The relevant quantities are:
\begin{align}
N_{\Lambda,i}^p &= q_i +s_i +\frac{3}{2}f_i -4
&
N_{\chi,i}^p &=  q_i + \frac{f_i}{2} -2 \,.
\end{align}
and they were given in Eqs.~\eqref{eq.Nip_lag},~\eqref{eq.Nchi_amplitude}.
$N_{\Lambda,i}^p$ is essentially the canonical dimension of the interaction $-4$. The orders are such that $N_{\chi,i}^p + N_\chi(\kappa_i) = N_{\chi,i}$ is always equal to the $N_\chi$ of the operator generating the interactions.

\subsubsection*{Amplitudes} 
The HEFT order of a Feynman diagram $\meLabel$ can be computed as
\begin{align}
    N_{\rm HEFT}^{\meLabel} & = n - 2 + 2L + \sum_{i\in{\rm vert}} (N_{\chi,i}-N_{g_s,i}) = N_{\rm HEFT}^\meLabel-N_{g_s},
    \\
    N_{\rm HEFT}^{s,\meLabel} &= n-2 + 2L + \sum_{i\in{\rm vert}} N_{\chi,i} ,
    \\
    N_{\rm HEFT}^{\xi,\meLabel} &= 
    \sum_{i\in\text{vert}}(N_{\Lambda,i}^p + N_{v,i})
    = n-2 + 2L + \sum_{i\in{\rm vert}} (N_{\chi,i}^p+N_{v,i}),
\end{align}
in the three power countings considered, that were given in Eqs.~\eqref{eq.NHEFT_to_NchiminGs},~\eqref{eq.NHEFT_to_Nchi},~\eqref{eq.NHEFTxi_to_Nchi} respectively. The sum runs over all vertices in the diagram, $N_v$ counts the powers of the EW scale $v$, and all orders only take values $\geq 0$.
\\\\
$N_{\rm HEFT}^\meLabel$ gives the total powers of 
\begin{equation}
\label{eq.NH_suppressions}
    \frac{p}{\Lambda}\sim \frac{g,g',y,\sqrt{\lambda}}{4\pi}
\end{equation}
in the amplitude. The $\sim$ sign indicates that both ratios count equally, as 1. This is the most general power counting that applies when HEFT is written using $v$ as a suppression scale. It allows to treat QCD and EW differently. The fact that it does not count the number of $g_s$ insertions $N_{g_s}$, causes QCD operators to enter at lower order compared to their EW counterparts, for the same $N_\chi$. Table~\ref{tab.HEFT_operators_orders_noGs} classifies operator classes keeping this into account.
\\\\
$N_{\rm HEFT}^{s,\meLabel}$ gives the total powers of 
\begin{equation}
    \frac{p}{\Lambda}\sim \frac{g,g',g_s,y,\sqrt{\lambda}}{4\pi}\,,
\end{equation}
\ie\ it treats QCD corrections on the same footing as EW ones, and it counts both as increasing the HEFT suppression order. This counting also applies when HEFT is defined with $v$ as a suppression scale. It makes the calculation of the HEFT orders simplest, as $N_\chi$ (interaction vertex) = $N_\chi$ (operator). However, it is less general and more constraining than $N_{\rm HEFT}^{s,\meLabel}$.  $N_{\rm HEFT}^{\meLabel}$ and $N_{\rm HEFT}^{s,\meLabel}$ are discussed in Section~\ref{sec.HEFT_PC_a}.
\\\\
Finally, $N_{\rm HEFT}^{\xi,\meLabel}$ gives the total powers of
\begin{equation}
\label{eq.Nxi_suppressions}
    \frac{p}{\Lambda}\sim \frac{4\pi v}{\Lambda}\,.
\end{equation}
This counting applies when the suppression scale for scalar insertions is $f\neq v$, and therefore $v$ can be inserted in numerators of the effective interactions, for instance through factors of $\xi=v^2/f^2$. Its drawback is that it can only be written as a function of $N_{\Lambda,i}^p$ or $N_{\chi,i}^p$, which are interaction-dependent, \ie\ $N_\chi^p$ (interaction) $\neq$ $N_\chi^p$ (operator) and the same for $N_\Lambda^p$. $N_{\rm HEFT}^{\xi,\meLabel}$ is discussed in Section~\ref{sec.xi}.

\subsubsection*{Observables} 
Finally, when calculating an observable, such as a cross section $\sigma$ or a decay rate $\Gamma$, one needs to consider the products $\mathcal{M}_a \mathcal{M}_b^{\dagger}$ of two matrix elements. The HEFT order of the product is simply given by
\begin{align}
    N_{\rm HEFT} = N_{\rm HEFT}^{\meLabel_a} + N_{\rm HEFT}^{\meLabel_b}.
\end{align}
and analogously for $N_{\rm HEFT}^s$, $N_{\rm HEFT}^\xi$.
These quantities give the net powers of the suppression factors defined in Eqs.~\eqref{eq.NH_suppressions}--\eqref{eq.Nxi_suppressions} that appear in the observable contribution.

In practice, computations should be truncated by choosing one of the three power countings, and taking a maximal order $N_{\mathrm{max}}$ at which the \emph{observable} should be computed. This in turn determines the range of amplitude orders $N_{\rm HEFT}^{\meLabel}$ that have to be calculated and informs the selection of the operators that need to be retained. Formulas for the truncation of the operators series are given in Section~\ref{sec.HEFT_counting_operators}. 

\section{Diagrammatic relations}\label{app.diagrammatics}
In this appendix we list some useful diagrammatic relations and derive some results that were used in the main text.

We consider a generic connected Feynman diagram with $L$ loops and $n$ external legs, out of which $S$ scalar or gauge fields (we do not distinguish between them as they are dimensionally identical) and $F$ fermions, $n=S+F$. The diagram will contain $I_S$ internal scalar/gauge propagators and $I_F$ internal fermion propagators, and $V$ vertices.
Each vertex is labeled with an index $i$ and treated with the notation of Eq.~\eqref{eq.generic_interaction}, \ie\ it has $n_i$ legs, out of which $s_i$ scalar/gauge ones and $f_i$ fermionic ones, and $q_i$ derivatives.

Some of these relations are well-known. Our main source is Ref.~\cite{Gavela:2016bzc}.

\subsubsection*{General relations}
In any connected Feynman diagram, the number of loops $L$ is related to the numbers of vertices $V$ and internal propagators $I=I_S+I_F$:
\begin{align}
\label{graph_theory_identity}
    L = I - V + 1.
\end{align}
The numbers of external fields $S,F$ and internal propagators $I_S,I_F$ are related to the number of fields in each vertex:
\begin{align}
\label{contractions_propagator}
    2 I_S +S&=\sum_{i \in \mathrm{vert}} s_i\,, 
    &
    2 I_F +F&= \sum_{i \in \mathrm{vert}} f_i \,,
\end{align}
and also
\begin{align}
\label{eq.I-to-n}
2I + n =  \sum_{i \in \mathrm{vert}} n_i\,.
\end{align}
Overall, the diagram scales with $P$ powers of the momenta that, excluding external spinors, are given by 
\begin{align}
\label{diagram_momenta}
    P = 4L - 2I_S - I_F + \sum_{i \in \mathrm{vert}} q_i\,,
\end{align}
which accounts, respectively, for the $p^4$ powers brought by each loop integration, the contributions $\sim p^{-2},p^{-1}$ from each internal scalar/gauge and fermionic propagator, and the powers of momenta from the $q_i$ derivatives in the $i$-th vertex.

\subsubsection*{Maximum number of vertices in a diagram}
Consider a diagram with fixed numbers of external legs $n$ and loops $L$. 
\\
Using  the fact that each vertex contains at least 3 legs ($n_i\geq 3$) we can obtain
\begin{align}\label{eq.Vmax_3}
 V &\stackrel{\eqref{graph_theory_identity}}{=} I-L+1 
 \stackrel{\eqref{eq.I-to-n}}{=} \frac{1}{2}\sum_{i \in \mathrm{vert}}   n_i - \frac{n}{2}-L+1 
 \quad\geq\quad \frac{3}{2}\,V -\frac{n}{2}- L+1 \,.
\end{align}
Rearranging the inequality gives the maximum number of vertices
\begin{align}
V \leq V_{\max} = 2L+n-2\,.
\end{align}
\\
Let us now generalize to the case where $k_4$ of the vertices have 4 legs, $k_5$ have 5 legs and $k_6$ have 6 legs. Then
\begin{align}
\sum_{i \in \mathrm{vert}}  n_i \geq 4k_4 + 5 k_5 + 6k_6+3(V-k_4 - k_5 -k_6) = 3V+k_4 + 2 k_5 + 3k_6\,.
\end{align}
Following a reasoning analogous to Eq.~\eqref{eq.Vmax_3}, this implies
\begin{align}
\label{eq.Vmax_4}
 V = \frac{1}{2}\sum_{i \in \mathrm{vert}}   n_i -\frac{n}{2}-L+1 
 \quad\geq\quad \frac{3}{2}V + \frac{k_4+2k_5+3k_6}{2}-\frac{n}{2}-L+1\,,
\end{align}
and therefore
\begin{align}
    V \leq 2L+n-2-k_4-2k_5-3k_6\,,
\end{align}
which can be easily generalized to higher-point interactions.
Note that the $k$'s satisfy
\begin{align}
    0\leq k_4+2k_5+3k_6\leq {\rm floor}\left( \frac{n}{2}+L-1\right)\,.
\end{align}
For instance, for $k_5=k_6=0$, the maximum is reached when all vertices have 4 legs ($V=k_4$, if $\sum n_i=4k_4$ is even) or all vertices have 4 legs except one who has 3 ($V=k_4+1$, if $\sum n_i=4k_4+3$ is odd). Inserting $V=k_4$ or $V=k_4+1$ in Eq.~\eqref{eq.Vmax_4} gives the same result
\begin{align}
k_{4,\max} &= {\rm floor}\left(\frac{n}{2}+L-1\right)\,.
\end{align}

\subsubsection*{Derivation of counting rules for scattering amplitudes}
The diagram scales with $\meCount_{\Lambda,\meLabel}$ powers of $\Lambda$ and $\meCount_{4\pi,\meLabel}$ powers of $4\pi$. As indicated in the main text, we can split these between powers $\meCount^\gCoupl_{\Lambda/4\pi,\meLabel}$ brought by the coupling constants, that are composed trivially as in Eq.~\eqref{eq.me_NG_def}, and powers $\meCount^p_{\Lambda/4\pi,\meLabel}$ brought by fields and derivatives, which we compute here.
\\
Powers of $\Lambda$ can only arise in vertices, therefore:
\begin{align}
\meCount_{\Lambda, \meLabel}^p 
&= \sum_{i \in \mathrm{vert}} N_{\Lambda,i}^p
\\
{\scriptsize \eqref{eq.Nip_lag}}\qquad\qquad
&=\sum_{i \in \mathrm{vert}} (-4 + q_i + s_i + 3f_i/2) 
\\
&=-4V + \sum_{i \in \mathrm{vert}} q_i + \sum_{i \in \mathrm{vert}} s_i + \frac{3}{2} \sum_{i \in \mathrm{vert}} f_i
\\
{\scriptsize \eqref{contractions_propagator},\eqref{diagram_momenta}}\qquad\qquad
&= -4 (L+V-I) + P + S + \frac{3}{2} F \\
{\scriptsize \eqref{graph_theory_identity}}\qquad\qquad
&= - 4 + P + S + \frac{3}{2} F\,.
\label{eq.NLambdap_derivation}
\end{align}
Powers of $4\pi$ can arise from both vertices and loop factors, therefore: 
\begin{align}
\meCount_{4 \pi, \meLabel}^p &= 2L + \sum_{i \in \mathrm{vert}} N_{4 \pi,i}^p 
\\
{\scriptsize \eqref{eq.Nip_lag}}\qquad\qquad
&= 2L + \sum_{i \in \mathrm{vert}} (2-s_i - f_i)
\\
{\scriptsize \eqref{contractions_propagator} } \qquad\qquad
& = 2(L + V - I)  - (S+F)
\\
{\scriptsize \eqref{graph_theory_identity}}\qquad\qquad
& = 2 - n\,.
\label{eq.N4pi_derivation}
\end{align}
The corresponding quantity for the chiral dimension counting is then just 
\begin{align}
\meCount_{\chi, \meLabel}^p=\meCount_{\Lambda, \meLabel}^p + \meCount_{4 \pi, \meLabel}^p = -2+P +\frac{1}{2}F\,.
\end{align}

Once the momentum dependence from external spinors $u\sim\sqrt{p}$ is included, the overall scaling of a diagram is
\begin{align}
 \mathcal{M} &\sim p^{P+F/2} \Lambda^{-\meCount_{\Lambda,\meLabel}}  (4\pi)^{-\meCount_{4\pi,\meLabel}}   
\end{align}
In particular, the overall powers of momenta can be cast as
\begin{align}
P +\frac{F}{2} &\stackrel{\eqref{eq.NLambdap_derivation}}{=}\meCount_{\Lambda,\meLabel}^p  +4-S-\frac{3}{2}F+\frac{1}{2}F = \meCount_{\Lambda,\meLabel}^p  +4-n\,,\\
&\quad\!\!=\quad\!\!\meCount_{\chi,\meLabel}^p + 2
\end{align}
which, together with Eq.~\eqref{eq.N4pi_derivation}, allows to rewrite 
\begin{align}
 \mathcal{M} &\sim p^{4-n}  (4\pi)^{2-n} \left(\frac{p}{\Lambda}\right)^{\meCount_{\Lambda,\meLabel}^p}\, \left(\frac{1}{\Lambda}\right)^{\meCount_{\Lambda,\meLabel}^\gCoupl} \left(\frac{1}{4\pi}\right)^{\meCount_{4\pi,\meLabel}^\gCoupl}\,.
\end{align}

\subsubsection*{Self-consistency of non-negative powers of the SM coupling constants}
When discussing the HEFT power counting at the Lagrangian level and examining possible algorithms for the construction of complete HEFT operator bases consistent with the $N_{\rm HEFT}^{(s)}$ countings in Section~\ref{sec.HEFT_PC}, we made the crucial assumption that the coupling constants $g,g',y_\psi,\lambda (,g_s)$ only appear in $\Lag_{\rm HEFT}$ with positive powers.
To be more precise, we assumed that one can safely choose to write $\Lag_{\rm HEFT}$ with the SM-like terms as in Eq.~\eqref{eq.lagrangian_HEFT_LO} and introducing all other operators as either "naked" or dressed with constant pre-factors $\gCoupl_i$ that \emph{only contain positive overall powers} of the coupling constants. Dimensionless ratios of the constants (such as $g'/g'$) leave the power counting unaffected and are therefore allowed, but they represent the only situation in which a coupling constant could appear at a denominator. 

The adjective "safely" in the previous paragraph refers to the fact that this convention on $\Lag_{\rm HEFT}$ is not spoiled by renormalization or by the matching to UV models. 
The argument for UV matching is trivially that,  if HEFT is an expansion in $(p/\Lambda)\sim (g',g,y_\psi,\sqrt{\lambda} ) /4\pi$, then by definition any matching expression will be polynomial in these quantities, that will only be featured in numerators. 

The argument concerning HEFT renormalization can be verified explicitly by checking the 1-loop RGEs provided in Refs.~\cite{Alonso:2017tdy,Buchalla:2017jlu,Buchalla:2020kdh,Morales:2025jyu} (see also~\cite{Bijnens:1999hw,Jenkins:2023bls,Banerjee:2024rbc} for higher-loop corrections to similar theories and~\cite{Fonseca:2025zjb,Misiak:2025xzq,Aebischer:2025zxg} for the 1-loop renormalization of general EFTs), but it can be useful to show in general with a diagrammatic argument:
the total powers of coupling constants appearing in a loop diagram is given by 
\begin{equation}
N_{g} + N_{g'} +  N_{y_\psi} + 2 N_{\lambda}\; (+N_{g_s}) + P_{m}
\end{equation}
where $P_{m}$ are the total powers of SM masses appearing in the diagram, that in principle could be negative due to propagators. We would like to prove that, if we start from a Lagrangian where  coupling constants only appear in positive powers, then, in any counterterm, $N_g+N_{g'}+\dots + P_{m_{SM}}\geq0$. The positivity of the first terms in the sum follows trivially from the starting assumption. The proof is completed by showing that necessarily $P_{m}\geq 0$: it is always positive for super-renormalizable counterterms and vanishing in all other cases, which means that, in RG expressions, SM masses can only appear at numerators or in dimensionless ratios that preserve $P_{m}=0$.

To see that $P_{m}\geq 0$, consider a generic HEFT operator with $q$ derivatives, that plays the role of counterterm for the renormalization of a set of divergent diagrams with $L$ loops. 
We know that, overall, the loop diagrams must have exactly the same mass and $\hbar$ dimensions as the counterterm, the same fields/external legs and momentum scaling. When the diagram is evaluated, its overall powers of momenta $P$ can be separated into $P=P_\de + P_{m}$, \ie\ the powers of kinematic momenta $p$ and SM masses $m$. Then it must be that $P_\de =q$. We also know that $P\geq 0$, because the diagram diverges.
If the counterterm is a super-renormalizable operator, then it has $q=0$,\footnote{An operator has at least two fields, which give a minimum mass dimension 2. To keep it super-renormalizable one could add one more dimensionful object, but Lorentz invariance doesn't allow it to be a single derivative. So derivatives must be absent.} and therefore $P = P_m \geq0$: all the powers of momenta from the diagram must turn into masses, which appear in positive powers. If the counterterm is a marginal or non-renormalizable operator, then the only way to accommodate the momentum dependence is that  $P = q = P_\de$, and therefore $P_{m}=0$.  
Another way of seeing this is that, were $P_m<0$, we would have a divergent loop diagram scaling  as $p^q (p/m)^r$ for some power $r$, and therefore explicit momenta would appear in the counterterms, which would render renormalization over the whole energy range impossible. Whenever divergencies are accompanied with momenta they are associated to wave functions or to operators with derivatives: in this case the counterterm should be a $(q+r)$-derivatives operator, contradicting the initial hypothesis.

\section{HEFT basis reduction with field redefinitions}\label{app.redefinitions}
The construction of non-redundant operator bases for a generic EFT requires removing all parameters that would give vanishing contributions to on-shell scattering amplitudes. Redundancies originating from the invariance of the effective action under field redefinitions require careful treatment, and they can have a non-trivial interplay with the power counting~\cite{Alonso:2025jvv}. 
In this appendix we revisit the procedure for the construction of HEFT bases consistent with expansions in $N_{\rm HEFT}^{(s)}$ employing field redefinitions, instead of naively applying EOMs as in Section~\ref{sec.HEFT_basis}.

Let us start by sorting HEFT interactions by their chiral order $N_\chi$, computed as in Eq.~\eqref{eq.Nchi_amplitude} and given in Table~\ref{tab.HEFT_operators_orders} for the leading operator classes: the HEFT Lagrangian can be written as
\begin{align}
\label{eq.LHEFT_orders_appendix}
 \Lag_{\rm HEFT} = \sum_{n=0}^\infty \Lag_n   
\end{align}
where $\Lag_n$ contains a sum over all HEFT operators with $N_\chi=n$, that we will assume to be NDA-normalized and  defined as "naked", except for the SM-like terms in $\Lag_{\rm LO}\subset \Lag_0$, which are defined as in Eq.~\eqref{eq.lagrangian_HEFT_LO}.

Following Ref.~\cite{Criado:2018sdb}, in order to obtain a non-redundant basis at each order $k>0$, it is sufficient to remove all operators of the form
\begin{align}
 \slashed{\mathcal{O}}_k^i (\phi) &=   F_k^i(\phi)\, \frac{\delta \mathcal{K}(\phi)}{\delta \phi^i}\,,
 &
 N_\chi (\slashed{\mathcal{O}}_k^i) &= k
 \label{Eq.Otoremove_redefinitions}
\end{align}
where $\phi^i$ denotes a generic HEFT field, $F_k^i(\phi)$ is a combination of fields, derivatives and constant factors with the appropriate dimensions and quantum numbers and  $\mathcal{K}(\phi)$ is a linear combination of terms appearing in $\Lag_0$, which is conventionally taken to be the same for all orders $k$, and equal to the sum of all kinetic terms, such that
\begin{align}
\label{eq.dKdphi_examples}
  \frac{\delta \mathcal{K}}{\delta \bar \psi_L} &= i \slashed{D}\psi_L  \,,
  &
  \frac{\delta \mathcal{K}}{\delta G^A_\mu} &= (D_\nu G^{\mu\nu})^A\,,
  &
  \frac{\delta \mathcal{K}}{\delta h} &= \square h\,,
\end{align}
and analogously for the other fields.
We will keep this choice here, as it ensures that all redundancies are removed~\cite{Criado:2018sdb} and it is consistent with the rule that field redefinitions should be used to remove the operators with the highest $N_\chi$, as discussed in Section~\ref{sec.HEFT_basis}. We will come back to this point later in this appendix.

The operator $\slashed{\mathcal{O}}_k^i$ in Eq.~\eqref{Eq.Otoremove_redefinitions} is removed by performing the redefinition
\begin{align}
\label{eq.redefinition_phi}
\phi^i \mapsto \phi^i - F_k^i(\phi)
\end{align}
over the whole $\Lag_{\rm HEFT}$. This has the effect of shifting each order $\Lag_n$ by:
\begin{align}
\label{eq.Ln_shift}
    \Lag_n\mapsto \Lag_n - F_k^i(\phi) \frac{\delta \Lag_n}{\delta \phi^i} + F_k^{i_1} (\phi) F_k^{i_2} (\phi) \frac{\delta^2\Lag_n}{\delta \phi^{i_1}\delta\phi^{i_2}} + \dots
\end{align}
where the Lagrangian term with $r$ derivatives has $N_\chi=n+rk$. Collecting the various terms order by order, one finds the redefined Lagrangian at order $n$
\begin{align}
\Lag_n^\prime &= \Lag_n + \sum_{r=1}^{{\rm floor}(n/k)}\;
(-1)^{r} \,F_k^{i_1}\dots F_k^{i_r}
\,\frac{\delta^r\Lag_{n-rk}}{\delta \phi^{i_1}\dots \delta \phi^{i_r}}\,.
\label{eq.redefinition_Lm}
\end{align}
This equation shows that each order $\Lag_n$ only receives corrections from lower orders $\Lag_m$ with $m=n-rk \in[0,n-k]$. Also, removing an operator of order $k$ only impacts Lagrangians at orders $n=m+rk \geq k$. 
For $n=k$, $\Lag_0$ is the only one contributing and one has
\begin{align}
    \Lag_k^\prime 
    \;= \;
    \Lag_k - F_k^i\, \frac{\delta\Lag_0}{\delta \phi^i}
    \;= \;
    \Lag_k - \slashed{\mathcal{O}}_k^i - F_k^i\, \frac{\delta (\Lag_0-\mathcal{K})}{\delta\phi^i}\,,
    \label{eq.Lk_redefinition_from_L0}
\end{align}
without further corrections. In the last step of the equation we used $\Lag_0 = \mathcal{K} + (\Lag_0 - \mathcal{K})$ to  illustrate explicitly how the unwanted operator is removed and replaced by a different structure.

These results allow us to refine the prescription for the construction of HEFT operator bases given in Section~\ref{sec.HEFT_basis}. One can work iteratively, order by order in $N_\chi$: at a given $\Lag_k$, and the set of naked operators can be reduced first using IBP and algebraic relations. Among the latter, those containing dimensionful coupling constants, such as Eq.~\eqref{eq.DD_X}, should be applied only to remove $\Lag_k$ operators containing the highest-$N_\chi$ structure appearing in the relation, \eg\ $[D_\mu,D_\nu]$. This will trade $\Lag_k$ operators for contributions to lower-$N_\chi$ naked operators, that are relevant \eg\ for matching calculations, but can be ignored in the basis construction. 

As a last step, operators of the form~\eqref{Eq.Otoremove_redefinitions} can be removed via shifts of the form~\eqref{eq.redefinition_phi} for any of the HEFT fields. This operation results in the Lagrangian shifts given by Eq.~\eqref{eq.Ln_shift}, whose effect must be unphysical. This is technically equivalent to an equation containing infinite terms: 
\begin{equation}
\label{eq.infinite_EOM}
F_k^i\frac{\delta\Lag_0}{\delta\phi^i} 
+F_k^{i_1}F_k^{i_2}\frac{\delta^2\Lag_0}{\delta\phi^{i_1}\delta\phi^{i_2}} 
+\dots 
+ F_k^i \frac{\delta\Lag_{1}}{\delta\phi^i} 
+F_k^{i_1} F_k^{i_2}\frac{\delta^2\Lag_{1}}{\delta\phi^{i_1}\delta\phi^{i_2}}
+\dots =0 
\end{equation}
where the first term contains operators with $N_\chi=k$, while the subsequent ones contain operators with $N_\chi>k$. 
However, when building a HEFT basis order by order in $N_\chi$, one is implicitly fixing the form of HEFT amplitudes order by order in $N_{\rm HEFT}^s$. In particular, an operator basis for $N_\chi\leq k$ is complete and non-redundant \emph{iff} it can produce -- without redundancies -- the complete set of allowed contributions to amplitudes with $N_{\rm HEFT}^s\leq n-2+k$. Since all contributions to higher-$N_\chi$ operators can only impact amplitudes with $N_{\rm HEFT}^s > n-2+k$, they play no role in establishing the basis completeness at  $N_\chi=k$. 
At the end of the day, Eq.~\eqref{eq.infinite_EOM} can be truncated at the first order, \ie\ we can reduce the basis at order $N_\chi=k$ using only equations of the form $F_k^i\,(\delta \Lag_0/\delta \phi^i)=0$. Following the prescriptions from Section~\ref{sec.HEFT_basis}, the value of $\min_i(N_{\chi,i}-N_{g_s,i})$ can be disregarded when making operators selections, and $\delta \mathcal{K}/\delta\phi^i$ should be selected among the terms in $\delta \Lag_0/\delta \phi^i$ that are not accompanied by a dimensionful constant. The choice in Eq.~\eqref{eq.dKdphi_examples}, that we consider the "default" option, satisfies this requirement. 

\subsection*{Action of field redefinitions on the $N_\chi=0$ HEFT Lagrangian}
In the remainder of this appendix we look in more detail at the expressions of $\delta \Lag_0/\delta \phi^i$. We have 
\begin{align}
\label{eq.HEFT_L0}
\Lag_0 &= \Lag_{\rm LO} + \Lag_{d^2} + \Lag_{d\psi^2} +  \Lag_{\psi^4}\,,
\\[2mm]
\Lag_{d^2} &= \Lag_{\V^2} + \Lag_{D\V} + \Lag_{X^2} {\color{black!50}\,+ \Lag_{D^2}}\,,
\\
\Lag_{d\psi^2} &= \Lag_{\V\psi^2} + \Lag_{X\psi^2} {\color{black!50} \,+ \Lag_{D\psi^2}}\,,
\end{align}
where $\Lag_{\rm LO}$ contains the SM-like terms as given in Eq.~\eqref{eq.lagrangian_HEFT_LO}, and the last two rows show breakdowns in subclasses obtained specializing the $d$-structures into derivatives ($D$), $\V_\mu$ insertions ($\V$) and gauge field strengths ($X$). 
Each of the Lagrangian terms contains a sum over an independent set of operators:
\begin{align}
\Lag_\alpha &=  w_{\alpha} \sum_{i\in[\alpha]} c_i \cO_{i}\,, 
\end{align}
with $w_\alpha$ the NDA weight: $w_{\psi^4} = (4\pi/\Lambda)^2, w_{X\psi^2}=(4\pi/\Lambda)$ and $w_{\alpha}=1$ otherwise.

The classes in gray are actually empty, in the sense that all the allowed operators are equivalent to SM-like terms already contained in $\Lag_{\rm LO}$. For some of these operators, this redundancy is resolved via field rescalings:
this is required for instance to "remove the~$\F$" from  $\bar\psi_Li\slashed{D}\psi_L \F$ and $\de_\mu h\de^\mu h\F$ structures~\cite{Brivio:2016fzo}.
It is interesting to note that, as scalar insertions do not increase the chiral order, field redefinitions in HEFT can in principle involve non-linear functions of the $h$ field, as in $\psi\to\psi/\sqrt{1+\F(h)}$.

Operators in the remaining classes can be reduced using only integration by parts and algebraic relations. Classes $\V^2, D\V$ contain only one operator each:
\begin{align}
\Lag_{\V^2} = \cO_T &= \frac{v^2}{4}\tr[\T\V_\mu]^2 \F_T(h)
    &
\Lag_{D\V}=  \cO_{2D} &= \frac{v^2}{4} i\tr[\T\V_\mu]\de^\mu\F_{2D}(h)\,,
\end{align}
where $\cO_T$ is CP even while $\cO_{2D}$ is CP odd and both violate the custodial symmetry. Class $X^2$, on the other hand, contains the operators~\cite{Buchalla:2013rka,Alonso:2012px,Gavela:2014vra}
\begin{align}
\cO_B&= B_{\mu\nu} B^{\mu\nu} \bar\F_B(h)
&
\cO_{\tilde B}&= B_{\mu\nu} \tilde B^{\mu\nu} \bar\F_{\tilde B}(h)  
\\
\cO_W&= W^I_{\mu\nu} W^{I \mu\nu} \bar \F_W(h)
&
\cO_{\tilde W}&= W^I_{\mu\nu} \tilde W^{I\mu\nu}\bar \F_{\tilde W}(h)
\\
\cO_{WB}&= B_{\mu\nu} \tr(\T W^{\mu\nu})\F_{WB}(h) 
&
\cO_{W\tilde B}&= \tilde B_{\mu\nu} \tr(\T W^{\mu\nu})\F_{W\tilde B}(h)
\\
\cO_{\T W}&= \tr(\T W^{\mu\nu})^2 \F_{\T W}(h)   
&
\cO_{\T\tilde W}&= \tr(\T W^{\mu\nu})\tr(\T \tilde W^{\mu\nu})\F_{\T\tilde W}(h)
\\
\cO_{G}&= G^A_{\mu\nu} G^{A\mu\nu}\bar \F_G(h)
&
\cO_{\tilde G}&= G^A_{\mu\nu} \tilde G^{A\mu\nu}\bar \F_{\tilde G}(h)\,.
\end{align}
Operators in the two columns are respectively CP-even and CP-odd and $\bar \F_i = \F_i - 1$ indicates that only $h$-dependent terms are retained. In the case of $O_{B,W,G}$ this choice avoids redundancy with the SM kinetic terms in $\Lag_{\rm LO}$, while in the case of $O_{\tilde B,\tilde W,\tilde G}$ it avoids introducing topological terms in $\Lag_{\rm HEFT}$.

Complete, non-redundant sets of operators in the remaining classes $\psi^4, \V\psi^2, X\psi^2$ can be found in Refs.~\cite{Buchalla:2013rka,Brivio:2016fzo,Sun:2022ssa} and won't be reported here. 

Once $\Lag_0$ has been fixed, $\delta\Lag_0/\delta\phi^i$ can be computed expanding the functional derivative and integrating by parts, which gives the usual:
{\allowdisplaybreaks
\begin{align}
\frac{\delta \Lag_n}{\delta\phi^i}
&=
\frac{\de\Lag_n}{\de \phi^i } - D_\mu\frac{\de \Lag_n}{\de(D_\mu\phi^i)} + D_\mu D_\nu \frac{\de \Lag_n}{\de(D_\mu D_\nu\phi^i)}+ \dots
\end{align}
for any order $\Lag_n$.

We point out that computing $\delta\Lag_0/\delta\phi^i$ for the full $\Lag_0$ in Eq.~\eqref{eq.HEFT_L0} is \emph{not} the same as deriving naive EOMs from $\Lag_{\rm LO}$ as done \eg\ in~\cite{Brivio:2016fzo} (App.~D), as several additional terms must be included in the equations. One has for instance
\begin{align}
\label{eq.dL0_eoms_first}
\frac{\delta \Lag_0}{\delta \bar \psi_{Lp}} =&\,
i\slashed{D}\psi_{Lp} - \frac{v}{\sqrt 2}\U\mathcal{Y}_\psi(h)_{pr}\psi_{Rr}
+ \frac{\delta \Lag_{\V\psi^2}}{\delta\bar\psi_{Lp}}
+ \frac{\delta \Lag_{X\psi^2}}{\delta\bar\psi_{Lp}}
+ \frac{\delta \Lag_{\psi^4}}{\delta\bar\psi_{Lp}}
\\[2mm] 
\frac{\delta \Lag_0}{\delta B_\nu} =&\, 
\de_\mu B^{\mu\nu} 
 - \sum_{\psi=Q_{L,R},L_{L,R}} g' \left( \bar\psi\,\mathbf{h}_{\psi}\gamma^\nu \psi \right) 
\nonumber\\
&
+ \frac{i g' v^2}{4}\tr[\T\V^\nu]\F_C(h) 
-  \frac{i g' v^2}{2}C_T\,\tr[\T\V^\nu]\F_T(h) + \frac{g'v^2}{4} C_{2D}\, \de^\nu\F_{2D}(h)
\nonumber\\
& 
+ \frac{\delta \Lag_{X^2}}{\delta B_\nu}
+ \frac{\delta \Lag_{\V\psi^2}}{\delta B_\nu}
+ \frac{\delta \Lag_{X\psi^2}}{\delta B_\nu}
\label{eq.dL0_eoms_B}
\\[2mm]
\frac{\delta \Lag_0}{\delta G^A_\nu} =&\,
(D_\mu G^{\mu\nu})^A 
- g_s \left(\bar Q_L T^A \gamma^\nu Q_L + \bar Q_R T^A \gamma^\nu Q_R\right)
+ \frac{\delta \Lag_{X^2}}{\delta G_\nu^A}
+ \frac{\delta \Lag_{X\psi^2}}{\delta G_\nu^A}
 \\[2mm]
\frac{\delta \Lag_0}{\delta h} =&\,
-\square h -\lambda v^4\mathcal{V}'(h) 
- \frac{v}{\sqrt2}\sum_{\psi=Q,L}\left(\bar\psi_L \U \mathcal{Y}^\prime_\psi(h) \psi_R+\text{h.c.}\right)
\nonumber\\
&- \frac{v^2}{4}\tr[\V_\mu\V^\mu]\F'_C(h)
+ \frac{v^2}{4}C_T\,\tr[\T\V^\mu]^2\F'_T(h)
+\frac{i v^2}{4}C_{2D}\, \tr[\T\V_\mu] \de^\mu \F'_{2D}(h)
\nonumber\\
&
+ \frac{\delta \Lag_{X^2}}{\delta h}
+ \frac{\delta \Lag_{\V\psi^2}}{\delta h}
+ \frac{\delta \Lag_{X\psi^2}}{\delta h}
+ \frac{\delta \Lag_{\psi^4}}{\delta h}
\\[2mm]
\frac{\delta \Lag_0}{\delta \U_{ij}} =&\,
\bigg[\U^\dag \bigg(
\frac{v^2}{2} D_\mu\left(\V^\mu\F_C(h)\right)
-\frac{v^2}{2}C_T D_\mu \left( \T \,\tr[\T\V^\mu]\F_T(h)\right)
-\frac{iv^2}{4}C_{2D} D_\mu\left(\T\, \de^\mu\F_{2D}(h)\right)
\bigg)
\bigg]_{ji}
\nonumber\\
&
-\frac{v}{\sqrt2}\sum_{\psi=Q,L}\left[\bar\psi_{L,i}\,\left(\mathcal{Y}_\psi(h) \psi_{R}\right)_j - (\bar\psi_{R}\mathcal{Y}^\dag_\psi(h)\U^\dag)_i (\U^\dag\psi_{L})_j
\right]
\nonumber\\
&
+ \frac{\delta \Lag_{X^2}}{\delta \U_{ij}}
+ \frac{\delta \Lag_{\V\psi^2}}{\delta \U_{ij}}
+ \frac{\delta \Lag_{X\psi^2}}{\delta \U_{ij}}
+ \frac{\delta \Lag_{\psi^4}}{\delta \U_{ij}}
\label{eq.dL0_eoms_last}
\end{align}
where, for the $\U$ field, the functional variation was obtained as
\begin{align}
 \frac{\delta \Lag}{\delta \U_{ij}} &= 
 \left[\frac{\de\Lag}{\de\U_{ij}} - D_\mu \frac{\de\Lag}{\de(D_\mu \U)_{ij}}+\dots\right] - (\U^\dag)_{ki}(\U^\dag)_{jl} 
 \left[\frac{\de\Lag}{\de(\U^\dag)_{kl}} - D_\mu \frac{\de\Lag}{\de(D_\mu \U^\dag)_{kl}}+\dots\right]
\end{align}
to ensure that the unitarity condition $\U^\dag\U=\mathbbm{1}$ is preserved, and one can further simplify the expressions using
\begin{align}
    D_\mu\V^\mu &= \frac{1}{2}\left(\square \U \U^\dag - \U \square \U^\dag\right)\,,
    &
    D_\mu\T &= [\V_\mu,\T]\,.
\end{align}
In Eqs.~\eqref{eq.dL0_eoms_first}--\eqref{eq.dL0_eoms_last} we have made the contributions from $\Lag_{\rm LO} + \Lag_{\V^2} + \Lag_{D\V}$ explicit, while we left those from other classes as implicit functional derivatives,  as they are lengthy, basis dependent expressions.  
In Eq.~\eqref{eq.dL0_eoms_B} $\mathbf{h}_\psi$ denotes the hypercharges
\begin{align}
\mathbf{h}_{Q_L} &= {\rm diag}\left(1/6,\,1/6\right)  \,,  
&
\mathbf{h}_{L_L} &= {\rm diag}\left(-1/2,\,-1/2\right) \,,  
\\
\mathbf{h}_{Q_R} &= {\rm diag}\left(2/3,\,-1/3\right) \,,   
&
\mathbf{h}_{L_R} &= {\rm diag}\left(0,\,-1\right) \,. 
\end{align}
Moreover, it can be checked that all the equations are homogeneous in the chiral dimension.

The contributions from $\Lag_{\rm LO}$ are consistent with previous results, see \eg\ Ref.~\cite{Brivio:2016fzo}.
The contributions from other classes introduce significant complications compared to the naive LO EOMs. For instance, expliciting $\delta \Lag_{X^2}/\delta B_\nu$ using the basis given above, we would find:
\begin{align}
\frac{\delta\Lag_{X^2}}{\delta B_\nu} &= 
-4C_B\de_\mu B^{\mu\nu} \bar \F_B
-4C_B B^{\mu\nu}\de_\mu\bar\F_B
-4C_{\tilde B}\de_\mu \tilde B^{\mu\nu}\bar \F_{\tilde B}
-4C_{\tilde B} \tilde B^{\mu\nu}\de_\mu\bar\F_{\tilde B}
\\
&
-2C_{WB} \tr[[\V_\mu,\T] W^{\mu\nu}] \F_{WB}
-2C_{WB} \tr[\T D_\mu W^{\mu\nu}] \F_{WB}
-2C_{WB} \tr[\T W^{\mu\nu}] \de_\mu\F_{WB}
\nonumber\\
&
-2C_{W\tilde B} \tr[[\V_\mu,\T] \tilde W^{\mu\nu}] \F_{W\tilde B}
-2C_{W\tilde B} \tr[\T D_\mu \tilde W^{\mu\nu}] \F_{W\tilde B}
-2C_{W\tilde B} \tr[\T \tilde W^{\mu\nu}] \de_\mu\F_{W\tilde B}\,.
\nonumber
\end{align}
The "equation of motion" for the $W^I_\nu$ gauge fields has a similar structure. Note, in particular, that due to the operator $\cO_{WB}$, $\delta\Lag_0/\delta B_\nu$ contains terms in $D_\mu W^{\mu\nu}$ and, conversely, $\delta\Lag_0/\delta W_\nu^I$ will contain terms in $\de_\mu B^{\mu\nu}$.  Moreover, in general the "EOMs" derived in this way can contain multiple occurrences of the structure $\de_\mu B^{\mu\nu}, D_\mu\V^\mu$, etc. that one wishes to remove. This makes the removal of "EOM operators" less trivial than expected.  Imagine, for instance, that we wish to remove an $N_\chi=k$ operator of the generic form $C_XA_\nu\de_\mu B^{\mu\nu}\F$. The required field redefinition is $B_\nu\to B_\nu - C_XA_\nu \F/(1-4C_B\bar\F_B)$, which yields
\begin{align}
\Delta\Lag_k &= - C_X A_\nu \de_\mu B^{\mu\nu}\F  - C_XA_\nu\de_\mu \tilde B^{\mu\nu}\frac{4C_{\tilde B}\F\bar \F_{\tilde B}}{1-4C_B\bar\F_B}
\nonumber\\
&
-C_X A_\nu B^{\mu\nu}\frac{4C_B\F\de_\mu \bar \F_{B}}{1-4C_B\bar\F_B}
- C_XA_\nu\Tr[\T D_\mu W^{\mu\nu}]\frac{2C_{WB}\F\F_{WB}}{1-4C_B\bar\F_B}+\dots
\label{eq.dB_removal_example}
\end{align}
which subtracts the desired term, but introduces new operators with unwanted structures. Operators containing $D_\mu \V^\mu$ are also interesting: consider the removal of a generic $C'_X A \tr[B D_\mu \V^\mu]\F'$, with $A$ some singlet structure and $B$ a bidoublet. Performing the field redefinition $\U_{ij}\to \U_{ij} - (2C_X'/v^2) A (B\U)_{ij}(\F/\F_C)$ yields
\begin{align}
 \Delta\Lag_k &= 
 -C_X'A \Tr[B D_\mu \V^\mu]\F' -C_X' A \Tr[B\V_\mu]\frac{\F'\de^\mu \F_C}{\F_C}
\nonumber\\
&+ C_X'C_T A \tr[\T\V_\mu]\Tr[B[\V^\mu,\T]] \frac{\F'\F_T}{\F_C}
+ C_X'C_T A \tr[\T B]\Tr[\T D_\mu\V^\mu] \frac{\F'\F_T}{\F_C}
\nonumber\\
& - C_X'C_T A \tr[\T\V_\mu]\Tr[\T\V_\mu] \frac{\F'\de^\mu\F_T}{\F_C}
+\frac{i}{2}C_{2D} A \tr[B[\V_\mu,\T]]\frac{\F'\de^\mu\F_{2D}}{\F_C}
\nonumber\\
&
+\frac{i}{2}C_{2D}A\Tr[\T B] \frac{\F'\square \F_{2D}}{\F_C}
+ \dots
\end{align}
which again removes the unwanted term, but introduces a new operator with $D_\mu\V^\mu$, that we can parameterize as $C_X''A'\tr[\T D_\mu \V^\mu]\F''$, and a term  $\sim\square h$. To remove the former, we can apply the redefinition $\U_{ij}\to \U_{ij} - (2C_X''/v^2)A' (\T\U)_{ij}\, \F''/\F_C/(1-2C_T \F_T/\F_C)$\,. One can check that this fully removes $D_\mu\V^\mu$ from the Lagrangian shifts, but introduces a $A \F''\square \F_{2D}/\F_C$ term, that would be removed via a redefinition of the Higgs field $h$.

Investigating in detail the consequences of Eqs.~\eqref{eq.dL0_eoms_first}--\eqref{eq.dL0_eoms_last} for the basis reduction in HEFT is beyond the scope of the present paper. We merely note that accounting for the correct structure of the "EOMs" given above is certainly important for a correct treatment of operator basis conversions and matching to BSM models.
Based on the examples given above, we expect that all the structures $\de_\mu B^{\mu\nu}, D_\mu W^{\mu\nu},D_\mu \V^\mu$ etc. can still be fully removed from higher-$N_\chi$ operators, and therefore that the existing operator bases are indeed complete. However, the algorithmic procedure to realize this condition is clearly more involved than the simple application of LO EOMs. 
Finally, as the operator classes $\V\psi^2,X\psi^2$ give contributions to $\delta\Lag_0/\delta\phi^i$, it is not obvious that using LO EOMs for their reduction leads to a correct identification of the corresponding operator bases. However, we do not investigate this aspect here.

\bibliographystyle{JHEP.bst}
\bibliography{bibliography}

\end{document}